\documentclass[twocolumn,traditabstract,longauth]{aa}
\usepackage{graphicx,amsmath}
\usepackage{epsf}
\usepackage{url}
\usepackage{txfonts}
\usepackage{natbib}
\bibpunct{(}{)}{;}{a}{}{,} 

\def\setsymbol#1#2{\expandafter\def\csname #1\endcsname{#2}}
\def\getsymbol#1{\csname #1\endcsname}

\def\Planck{{\it Planck\/}}

\def\allearlypapers{\nocite{planck2011-1.1, planck2011-1.3, planck2011-1.4, planck2011-1.5, planck2011-1.6, planck2011-1.7, planck2011-1.10, planck2011-1.10sup, planck2011-5.1a, planck2011-5.1b, planck2011-5.2a, planck2011-5.2b, planck2011-5.2c, planck2011-6.1, planck2011-6.2, planck2011-6.3a, planck2011-6.4a, planck2011-6.4b, planck2011-6.6, planck2011-7.0, planck2011-7.2, planck2011-7.3, planck2011-7.7a, planck2011-7.7b, planck2011-7.12, planck2011-7.13}}

\newbox\tablebox    \newdimen\tablewidth
\def\leaderfil{\leaders\hbox to 5pt{\hss.\hss}\hfil}
\def\tablenote#1 #2\par{\begingroup \parindent=0.8em
    \abovedisplayshortskip=0pt\belowdisplayshortskip=0pt
    \noindent
    $$\hss\vbox{\hsize\tablewidth \hangindent=\parindent \hangafter=1 \noindent
    \hbox to \parindent{\sup{\rm #1}\hss}\strut#2\strut\par}\hss$$
    \endgroup}

\def\L2{\ifmmode L_2\else $L_2$\fi}

\def\DeltaT{\ifmmode \Delta T\else $\Delta T$\fi}
\def\deltat{\ifmmode \Delta t\else $\Delta t$\fi}
\def\fknee{\ifmmode f_{\rm knee}\else $f_{\rm knee}$\fi}
\def\Fmax{\ifmmode F_{\rm max}\else $F_{\rm max}$\fi}
\def\solar{\ifmmode{\rm M}_{\mathord\odot}\else${\rm M}_{\mathord\odot}$\fi}

\def\inv{\ifmmode^{-1}\else$^{-1}$\fi}
\def\mo{\ifmmode^{-1}\else$^{-1}$\fi}
\def\sup#1{\ifmmode ^{\rm #1}\else $^{\rm #1}$\fi}
\def\expo#1{\ifmmode \times 10^{#1}\else $\times 10^{#1}$\fi}
\def\,{\thinspace}
\def\lsim{\mathrel{\raise .4ex\hbox{\rlap{$<$}\lower 1.2ex\hbox{$\sim$}}}}
\def\gsim{\mathrel{\raise .4ex\hbox{\rlap{$>$}\lower 1.2ex\hbox{$\sim$}}}}

\def\simprop{\mathrel{\raise .4ex\hbox{\rlap{$\propto$}\lower 1.2ex\hbox{$\sim$}}}}
\def\deg{\ifmmode^\circ\else$^\circ$\fi}
\def\pdeg{\ifmmode $\setbox0=\hbox{$^{\circ}$}\rlap{\hskip.11\wd0 .}$^{\circ}
          \else \setbox0=\hbox{$^{\circ}$}\rlap{\hskip.11\wd0 .}$^{\circ}$\fi}
\def\arcs{\ifmmode {^{\scriptstyle\prime\prime}}
          \else $^{\scriptstyle\prime\prime}$\fi}
\def\arcm{\ifmmode {^{\scriptstyle\prime}}
          \else $^{\scriptstyle\prime}$\fi}
\newdimen\sa  \newdimen\sb
\def\parcs{\sa=.07em \sb=.03em
     \ifmmode \hbox{\rlap{.}}^{\scriptstyle\prime\kern -\sb\prime}\hbox{\kern -\sa}
     \else \rlap{.}$^{\scriptstyle\prime\kern -\sb\prime}$\kern -\sa\fi}
\def\parcm{\sa=.08em \sb=.03em
     \ifmmode \hbox{\rlap{.}\kern\sa}^{\scriptstyle\prime}\hbox{\kern-\sb}
     \else \rlap{.}\kern\sa$^{\scriptstyle\prime}$\kern-\sb\fi}
\def\ra[#1 #2 #3.#4]{#1\sup{h}#2\sup{m}#3\sup{s}\llap.#4}
\def\dec[#1 #2 #3.#4]{#1\deg#2\arcm#3\arcs\llap.#4}
\def\deco[#1 #2 #3]{#1\deg#2\arcm#3\arcs}
\def\rra[#1 #2]{#1\sup{h}#2\sup{m}}

\def\dots{\relax\ifmmode \ldots\else $\ldots$\fi}
\def\WHzsr{\ifmmode $W\,Hz\mo\,sr\mo$\else W\,Hz\mo\,sr\mo\fi}
\def\mHz{\ifmmode $\,mHz$\else \,mHz\fi}
\def\GHz{\ifmmode $\,GHz$\else \,GHz\fi}
\def\mKs{\ifmmode $\,mK\,s$^{1/2}\else \,mK\,s$^{1/2}$\fi}
\def\muKs{\ifmmode \,\mu$K\,s$^{1/2}\else \,$\mu$K\,s$^{1/2}$\fi}
\def\muKRJs{\ifmmode \,\mu$K$_{\rm RJ}$\,s$^{1/2}\else \,$\mu$K$_{\rm RJ}$\,s$^{1/2}$\fi}
\def\muKHz{\ifmmode \,\mu$K\,Hz$^{-1/2}\else \,$\mu$K\,Hz$^{-1/2}$\fi}
\def\MJysr{\ifmmode \,$MJy\,sr\mo$\else \,MJy\,sr\mo\fi}
\def\MJysrmK{\ifmmode \,$MJy\,sr\mo$\,mK$_{\rm CMB}\mo\else \,MJy\,sr\mo\,mK$_{\rm CMB}\mo$\fi}
\def\microns{\ifmmode \,\mu$m$\else \,$\mu$m\fi}
\def\micron{\microns}
\def\muK{\ifmmode \,\mu$K$\else \,$\mu$\hbox{K}\fi}
\def\microK{\ifmmode \,\mu$K$\else \,$\mu$\hbox{K}\fi}
\def\muW{\ifmmode \,\mu$W$\else \,$\mu$\hbox{W}\fi}
\def\kms{\ifmmode $\,km\,s$^{-1}\else \,km\,s$^{-1}$\fi}
\def\kmsMpc{\ifmmode $\,\kms\,Mpc\mo$\else \,\kms\,Mpc\mo\fi}
\setsymbol{LFI:center:frequency:70GHz:units}{70.3\,GHz}
\setsymbol{LFI:center:frequency:44GHz:units}{44.1\,GHz}
\setsymbol{LFI:center:frequency:30GHz:units}{28.5\,GHz}

\setsymbol{LFI:center:frequency:70GHz}{70.3}
\setsymbol{LFI:center:frequency:44GHz}{44.1}
\setsymbol{LFI:center:frequency:30GHz}{28.5}

\setsymbol{LFI:center:frequency:LFI18:Rad:M:units}{71.7\GHz}
\setsymbol{LFI:center:frequency:LFI19:Rad:M:units}{67.5\GHz}
\setsymbol{LFI:center:frequency:LFI20:Rad:M:units}{69.2\GHz}
\setsymbol{LFI:center:frequency:LFI21:Rad:M:units}{70.4\GHz}
\setsymbol{LFI:center:frequency:LFI22:Rad:M:units}{71.5\GHz}
\setsymbol{LFI:center:frequency:LFI23:Rad:M:units}{70.8\GHz}
\setsymbol{LFI:center:frequency:LFI24:Rad:M:units}{44.4\GHz}
\setsymbol{LFI:center:frequency:LFI25:Rad:M:units}{44.0\GHz}
\setsymbol{LFI:center:frequency:LFI26:Rad:M:units}{43.9\GHz}
\setsymbol{LFI:center:frequency:LFI27:Rad:M:units}{28.3\GHz}
\setsymbol{LFI:center:frequency:LFI28:Rad:M:units}{28.8\GHz}
\setsymbol{LFI:center:frequency:LFI18:Rad:S:units}{70.1\GHz}
\setsymbol{LFI:center:frequency:LFI19:Rad:S:units}{69.6\GHz}
\setsymbol{LFI:center:frequency:LFI20:Rad:S:units}{69.5\GHz}
\setsymbol{LFI:center:frequency:LFI21:Rad:S:units}{69.5\GHz}
\setsymbol{LFI:center:frequency:LFI22:Rad:S:units}{72.8\GHz}
\setsymbol{LFI:center:frequency:LFI23:Rad:S:units}{71.3\GHz}
\setsymbol{LFI:center:frequency:LFI24:Rad:S:units}{44.1\GHz}
\setsymbol{LFI:center:frequency:LFI25:Rad:S:units}{44.1\GHz}
\setsymbol{LFI:center:frequency:LFI26:Rad:S:units}{44.1\GHz}
\setsymbol{LFI:center:frequency:LFI27:Rad:S:units}{28.5\GHz}
\setsymbol{LFI:center:frequency:LFI28:Rad:S:units}{28.2\GHz}

\setsymbol{LFI:center:frequency:LFI18:Rad:M}{71.7}
\setsymbol{LFI:center:frequency:LFI19:Rad:M}{67.5}
\setsymbol{LFI:center:frequency:LFI20:Rad:M}{69.2}
\setsymbol{LFI:center:frequency:LFI21:Rad:M}{70.4}
\setsymbol{LFI:center:frequency:LFI22:Rad:M}{71.5}
\setsymbol{LFI:center:frequency:LFI23:Rad:M}{70.8}
\setsymbol{LFI:center:frequency:LFI24:Rad:M}{44.4}
\setsymbol{LFI:center:frequency:LFI25:Rad:M}{44.0}
\setsymbol{LFI:center:frequency:LFI26:Rad:M}{43.9}
\setsymbol{LFI:center:frequency:LFI27:Rad:M}{28.3}
\setsymbol{LFI:center:frequency:LFI28:Rad:M}{28.8}
\setsymbol{LFI:center:frequency:LFI18:Rad:S}{70.1}
\setsymbol{LFI:center:frequency:LFI19:Rad:S}{69.6}
\setsymbol{LFI:center:frequency:LFI20:Rad:S}{69.5}
\setsymbol{LFI:center:frequency:LFI21:Rad:S}{69.5}
\setsymbol{LFI:center:frequency:LFI22:Rad:S}{72.8}
\setsymbol{LFI:center:frequency:LFI23:Rad:S}{71.3}
\setsymbol{LFI:center:frequency:LFI24:Rad:S}{44.1}
\setsymbol{LFI:center:frequency:LFI25:Rad:S}{44.1}
\setsymbol{LFI:center:frequency:LFI26:Rad:S}{44.1}
\setsymbol{LFI:center:frequency:LFI27:Rad:S}{28.5}
\setsymbol{LFI:center:frequency:LFI28:Rad:S}{28.2}

\setsymbol{LFI:white:noise:sensitivity:70GHz:units}{152.6\muKs}
\setsymbol{LFI:white:noise:sensitivity:44GHz:units}{173.1\muKs}
\setsymbol{LFI:white:noise:sensitivity:30GHz:units}{146.8\muKs}

\setsymbol{LFI:white:noise:sensitivity:70GHz}{152.6}
\setsymbol{LFI:white:noise:sensitivity:44GHz}{173.1}
\setsymbol{LFI:white:noise:sensitivity:30GHz}{146.8}

\setsymbol{LFI:white:noise:sensitivity:LFI18:Rad:M:units}{512.0\muKs}
\setsymbol{LFI:white:noise:sensitivity:LFI19:Rad:M:units}{581.4\muKs}
\setsymbol{LFI:white:noise:sensitivity:LFI20:Rad:M:units}{590.8\muKs}
\setsymbol{LFI:white:noise:sensitivity:LFI21:Rad:M:units}{455.2\muKs}
\setsymbol{LFI:white:noise:sensitivity:LFI22:Rad:M:units}{492.0\muKs}
\setsymbol{LFI:white:noise:sensitivity:LFI23:Rad:M:units}{507.7\muKs}
\setsymbol{LFI:white:noise:sensitivity:LFI24:Rad:M:units}{462.2\muKs}
\setsymbol{LFI:white:noise:sensitivity:LFI25:Rad:M:units}{413.6\muKs}
\setsymbol{LFI:white:noise:sensitivity:LFI26:Rad:M:units}{478.6\muKs}
\setsymbol{LFI:white:noise:sensitivity:LFI27:Rad:M:units}{277.7\muKs}
\setsymbol{LFI:white:noise:sensitivity:LFI28:Rad:M:units}{312.3\muKs}
\setsymbol{LFI:white:noise:sensitivity:LFI18:Rad:S:units}{465.7\muKs}
\setsymbol{LFI:white:noise:sensitivity:LFI19:Rad:S:units}{555.6\muKs}
\setsymbol{LFI:white:noise:sensitivity:LFI20:Rad:S:units}{623.2\muKs}
\setsymbol{LFI:white:noise:sensitivity:LFI21:Rad:S:units}{564.1\muKs}
\setsymbol{LFI:white:noise:sensitivity:LFI22:Rad:S:units}{534.4\muKs}
\setsymbol{LFI:white:noise:sensitivity:LFI23:Rad:S:units}{542.4\muKs}
\setsymbol{LFI:white:noise:sensitivity:LFI24:Rad:S:units}{399.2\muKs}
\setsymbol{LFI:white:noise:sensitivity:LFI25:Rad:S:units}{392.6\muKs}
\setsymbol{LFI:white:noise:sensitivity:LFI26:Rad:S:units}{418.6\muKs}
\setsymbol{LFI:white:noise:sensitivity:LFI27:Rad:S:units}{302.9\muKs}
\setsymbol{LFI:white:noise:sensitivity:LFI28:Rad:S:units}{285.3\muKs}

\setsymbol{LFI:white:noise:sensitivity:LFI18:Rad:M}{512.0}
\setsymbol{LFI:white:noise:sensitivity:LFI19:Rad:M}{581.4}
\setsymbol{LFI:white:noise:sensitivity:LFI20:Rad:M}{590.8}
\setsymbol{LFI:white:noise:sensitivity:LFI21:Rad:M}{455.2}
\setsymbol{LFI:white:noise:sensitivity:LFI22:Rad:M}{492.0}
\setsymbol{LFI:white:noise:sensitivity:LFI23:Rad:M}{507.7}
\setsymbol{LFI:white:noise:sensitivity:LFI24:Rad:M}{462.2}
\setsymbol{LFI:white:noise:sensitivity:LFI25:Rad:M}{413.6}
\setsymbol{LFI:white:noise:sensitivity:LFI26:Rad:M}{478.6}
\setsymbol{LFI:white:noise:sensitivity:LFI27:Rad:M}{277.7}
\setsymbol{LFI:white:noise:sensitivity:LFI28:Rad:M}{312.3}
\setsymbol{LFI:white:noise:sensitivity:LFI18:Rad:S}{465.7}
\setsymbol{LFI:white:noise:sensitivity:LFI19:Rad:S}{555.6}
\setsymbol{LFI:white:noise:sensitivity:LFI20:Rad:S}{623.2}
\setsymbol{LFI:white:noise:sensitivity:LFI21:Rad:S}{564.1}
\setsymbol{LFI:white:noise:sensitivity:LFI22:Rad:S}{534.4}
\setsymbol{LFI:white:noise:sensitivity:LFI23:Rad:S}{542.4}
\setsymbol{LFI:white:noise:sensitivity:LFI24:Rad:S}{399.2}
\setsymbol{LFI:white:noise:sensitivity:LFI25:Rad:S}{392.6}
\setsymbol{LFI:white:noise:sensitivity:LFI26:Rad:S}{418.6}
\setsymbol{LFI:white:noise:sensitivity:LFI27:Rad:S}{302.9}
\setsymbol{LFI:white:noise:sensitivity:LFI28:Rad:S}{285.3}

\setsymbol{LFI:knee:frequency:70GHz:units}{29.5\mHz}
\setsymbol{LFI:knee:frequency:44GHz:units}{56.2\mHz}
\setsymbol{LFI:knee:frequency:30GHz:units}{113.7\mHz}

\setsymbol{LFI:knee:frequency:70GHz}{29.5}
\setsymbol{LFI:knee:frequency:44GHz}{56.2}
\setsymbol{LFI:knee:frequency:30GHz}{113.7}

\setsymbol{LFI:knee:frequency:LFI18:Rad:M:units}{16.3\mHz}
\setsymbol{LFI:knee:frequency:LFI19:Rad:M:units}{15.1\mHz}
\setsymbol{LFI:knee:frequency:LFI20:Rad:M:units}{18.7\mHz}
\setsymbol{LFI:knee:frequency:LFI21:Rad:M:units}{37.2\mHz}
\setsymbol{LFI:knee:frequency:LFI22:Rad:M:units}{12.7\mHz}
\setsymbol{LFI:knee:frequency:LFI23:Rad:M:units}{34.6\mHz}
\setsymbol{LFI:knee:frequency:LFI24:Rad:M:units}{46.2\mHz}
\setsymbol{LFI:knee:frequency:LFI25:Rad:M:units}{24.9\mHz}
\setsymbol{LFI:knee:frequency:LFI26:Rad:M:units}{67.6\mHz}
\setsymbol{LFI:knee:frequency:LFI27:Rad:M:units}{187.4\mHz}
\setsymbol{LFI:knee:frequency:LFI28:Rad:M:units}{122.2\mHz}
\setsymbol{LFI:knee:frequency:LFI18:Rad:S:units}{17.7\mHz}
\setsymbol{LFI:knee:frequency:LFI19:Rad:S:units}{22.0\mHz}
\setsymbol{LFI:knee:frequency:LFI20:Rad:S:units}{8.7\mHz}
\setsymbol{LFI:knee:frequency:LFI21:Rad:S:units}{25.9\mHz}
\setsymbol{LFI:knee:frequency:LFI22:Rad:S:units}{15.8\mHz}
\setsymbol{LFI:knee:frequency:LFI23:Rad:S:units}{129.8\mHz}
\setsymbol{LFI:knee:frequency:LFI24:Rad:S:units}{100.9\mHz}
\setsymbol{LFI:knee:frequency:LFI25:Rad:S:units}{38.9\mHz}
\setsymbol{LFI:knee:frequency:LFI26:Rad:S:units}{58.9\mHz}
\setsymbol{LFI:knee:frequency:LFI27:Rad:S:units}{104.4\mHz}
\setsymbol{LFI:knee:frequency:LFI28:Rad:S:units}{40.7\mHz}

\setsymbol{LFI:knee:frequency:LFI18:Rad:M}{16.3}
\setsymbol{LFI:knee:frequency:LFI19:Rad:M}{15.1}
\setsymbol{LFI:knee:frequency:LFI20:Rad:M}{18.7}
\setsymbol{LFI:knee:frequency:LFI21:Rad:M}{37.2}
\setsymbol{LFI:knee:frequency:LFI22:Rad:M}{12.7}
\setsymbol{LFI:knee:frequency:LFI23:Rad:M}{34.6}
\setsymbol{LFI:knee:frequency:LFI24:Rad:M}{46.2}
\setsymbol{LFI:knee:frequency:LFI25:Rad:M}{24.9}
\setsymbol{LFI:knee:frequency:LFI26:Rad:M}{67.6}
\setsymbol{LFI:knee:frequency:LFI27:Rad:M}{187.4}
\setsymbol{LFI:knee:frequency:LFI28:Rad:M}{122.2}
\setsymbol{LFI:knee:frequency:LFI18:Rad:S}{17.7}
\setsymbol{LFI:knee:frequency:LFI19:Rad:S}{22.0}
\setsymbol{LFI:knee:frequency:LFI20:Rad:S}{8.7}
\setsymbol{LFI:knee:frequency:LFI21:Rad:S}{25.9}
\setsymbol{LFI:knee:frequency:LFI22:Rad:S}{15.8}
\setsymbol{LFI:knee:frequency:LFI23:Rad:S}{129.8}
\setsymbol{LFI:knee:frequency:LFI24:Rad:S}{100.9}
\setsymbol{LFI:knee:frequency:LFI25:Rad:S}{38.9}
\setsymbol{LFI:knee:frequency:LFI26:Rad:S}{58.9}
\setsymbol{LFI:knee:frequency:LFI27:Rad:S}{104.4}
\setsymbol{LFI:knee:frequency:LFI28:Rad:S}{40.7}

\setsymbol{LFI:slope:70GHz:units}{$-1.03$\mHz}
\setsymbol{LFI:slope:44GHz:units}{$-0.89$\mHz}
\setsymbol{LFI:slope:30GHz:units}{$-0.87$\mHz}

\setsymbol{LFI:slope:70GHz}{$-1.03$}
\setsymbol{LFI:slope:44GHz}{$-0.89$}
\setsymbol{LFI:slope:30GHz}{$-0.87$}

\setsymbol{LFI:slope:LFI18:Rad:M:units}{$-1.04$\mHz}
\setsymbol{LFI:slope:LFI19:Rad:M:units}{$-1.09$\mHz}
\setsymbol{LFI:slope:LFI20:Rad:M:units}{$-0.69$\mHz}
\setsymbol{LFI:slope:LFI21:Rad:M:units}{$-1.56$\mHz}
\setsymbol{LFI:slope:LFI22:Rad:M:units}{$-1.01$\mHz}
\setsymbol{LFI:slope:LFI23:Rad:M:units}{$-0.96$\mHz}
\setsymbol{LFI:slope:LFI24:Rad:M:units}{$-0.83$\mHz}
\setsymbol{LFI:slope:LFI25:Rad:M:units}{$-0.91$\mHz}
\setsymbol{LFI:slope:LFI26:Rad:M:units}{$-0.95$\mHz}
\setsymbol{LFI:slope:LFI27:Rad:M:units}{$-0.87$\mHz}
\setsymbol{LFI:slope:LFI28:Rad:M:units}{$-0.88$\mHz}
\setsymbol{LFI:slope:LFI18:Rad:S:units}{$-1.15$\mHz}
\setsymbol{LFI:slope:LFI19:Rad:S:units}{$-1.00$\mHz}
\setsymbol{LFI:slope:LFI20:Rad:S:units}{$-0.95$\mHz}
\setsymbol{LFI:slope:LFI21:Rad:S:units}{$-0.92$\mHz}
\setsymbol{LFI:slope:LFI22:Rad:S:units}{$-1.01$\mHz}
\setsymbol{LFI:slope:LFI23:Rad:S:units}{$-0.95$\mHz}
\setsymbol{LFI:slope:LFI24:Rad:S:units}{$-0.73$\mHz}
\setsymbol{LFI:slope:LFI25:Rad:S:units}{$-1.16$\mHz}
\setsymbol{LFI:slope:LFI26:Rad:S:units}{$-0.79$\mHz}
\setsymbol{LFI:slope:LFI27:Rad:S:units}{$-0.82$\mHz}
\setsymbol{LFI:slope:LFI28:Rad:S:units}{$-0.91$\mHz}

\setsymbol{LFI:slope:LFI18:Rad:M}{$-1.04$}
\setsymbol{LFI:slope:LFI19:Rad:M}{$-1.09$}
\setsymbol{LFI:slope:LFI20:Rad:M}{$-0.69$}
\setsymbol{LFI:slope:LFI21:Rad:M}{$-1.56$}
\setsymbol{LFI:slope:LFI22:Rad:M}{$-1.01$}
\setsymbol{LFI:slope:LFI23:Rad:M}{$-0.96$}
\setsymbol{LFI:slope:LFI24:Rad:M}{$-0.83$}
\setsymbol{LFI:slope:LFI25:Rad:M}{$-0.91$}
\setsymbol{LFI:slope:LFI26:Rad:M}{$-0.95$}
\setsymbol{LFI:slope:LFI27:Rad:M}{$-0.87$}
\setsymbol{LFI:slope:LFI28:Rad:M}{$-0.88$}
\setsymbol{LFI:slope:LFI18:Rad:S}{$-1.15$}
\setsymbol{LFI:slope:LFI19:Rad:S}{$-1.00$}
\setsymbol{LFI:slope:LFI20:Rad:S}{$-0.95$}
\setsymbol{LFI:slope:LFI21:Rad:S}{$-0.92$}
\setsymbol{LFI:slope:LFI22:Rad:S}{$-1.01$}
\setsymbol{LFI:slope:LFI23:Rad:S}{$-0.95$}
\setsymbol{LFI:slope:LFI24:Rad:S}{$-0.73$}
\setsymbol{LFI:slope:LFI25:Rad:S}{$-1.16$}
\setsymbol{LFI:slope:LFI26:Rad:S}{$-0.79$}
\setsymbol{LFI:slope:LFI27:Rad:S}{$-0.82$}
\setsymbol{LFI:slope:LFI28:Rad:S}{$-0.91$}

\setsymbol{LFI:FWHM:70GHz:units}{13\parcm01}
\setsymbol{LFI:FWHM:44GHz:units}{27\parcm92}
\setsymbol{LFI:FWHM:30GHz:units}{32\parcm65}

\setsymbol{LFI:FWHM:70GHz}{13.01}
\setsymbol{LFI:FWHM:44GHz}{27.92}
\setsymbol{LFI:FWHM:30GHz}{32.65}

\setsymbol{LFI:FWHM:LFI18:units}{13\parcm39}
\setsymbol{LFI:FWHM:LFI19:units}{13\parcm01}
\setsymbol{LFI:FWHM:LFI20:units}{12\parcm75}
\setsymbol{LFI:FWHM:LFI21:units}{12\parcm74}
\setsymbol{LFI:FWHM:LFI22:units}{12\parcm87}
\setsymbol{LFI:FWHM:LFI23:units}{13\parcm27}
\setsymbol{LFI:FWHM:LFI24:units}{22\parcm98}
\setsymbol{LFI:FWHM:LFI25:units}{30\parcm46}
\setsymbol{LFI:FWHM:LFI26:units}{30\parcm31}
\setsymbol{LFI:FWHM:LFI27:units}{32\parcm65}
\setsymbol{LFI:FWHM:LFI28:units}{32\parcm66}

\setsymbol{LFI:FWHM:LFI18}{13.39}
\setsymbol{LFI:FWHM:LFI19}{13.01}
\setsymbol{LFI:FWHM:LFI20}{12.75}
\setsymbol{LFI:FWHM:LFI21}{12.74}
\setsymbol{LFI:FWHM:LFI22}{12.87}
\setsymbol{LFI:FWHM:LFI23}{13.27}
\setsymbol{LFI:FWHM:LFI24}{22.98}
\setsymbol{LFI:FWHM:LFI25}{30.46}
\setsymbol{LFI:FWHM:LFI26}{30.31}
\setsymbol{LFI:FWHM:LFI27}{32.65}
\setsymbol{LFI:FWHM:LFI28}{32.66}

\setsymbol{LFI:FWHM:uncertainty:LFI18:units}{0.170\arcm}
\setsymbol{LFI:FWHM:uncertainty:LFI19:units}{0.174\arcm}
\setsymbol{LFI:FWHM:uncertainty:LFI20:units}{0.170\arcm}
\setsymbol{LFI:FWHM:uncertainty:LFI21:units}{0.156\arcm}
\setsymbol{LFI:FWHM:uncertainty:LFI22:units}{0.164\arcm}
\setsymbol{LFI:FWHM:uncertainty:LFI23:units}{0.171\arcm}
\setsymbol{LFI:FWHM:uncertainty:LFI24:units}{0.652\arcm}
\setsymbol{LFI:FWHM:uncertainty:LFI25:units}{1.075\arcm}
\setsymbol{LFI:FWHM:uncertainty:LFI26:units}{1.131\arcm}
\setsymbol{LFI:FWHM:uncertainty:LFI27:units}{1.266\arcm}
\setsymbol{LFI:FWHM:uncertainty:LFI28:units}{1.287\arcm}

\setsymbol{LFI:FWHM:uncertainty:LFI18}{0.170}
\setsymbol{LFI:FWHM:uncertainty:LFI19}{0.174}
\setsymbol{LFI:FWHM:uncertainty:LFI20}{0.170}
\setsymbol{LFI:FWHM:uncertainty:LFI21}{0.156}
\setsymbol{LFI:FWHM:uncertainty:LFI22}{0.164}
\setsymbol{LFI:FWHM:uncertainty:LFI23}{0.171}
\setsymbol{LFI:FWHM:uncertainty:LFI24}{0.652}
\setsymbol{LFI:FWHM:uncertainty:LFI25}{1.075}
\setsymbol{LFI:FWHM:uncertainty:LFI26}{1.131}
\setsymbol{LFI:FWHM:uncertainty:LFI27}{1.266}
\setsymbol{LFI:FWHM:uncertainty:LFI28}{1.287}

\setsymbol{HFI:center:frequency:100GHz:units}{100\,GHz}
\setsymbol{HFI:center:frequency:143GHz:units}{143\,GHz}
\setsymbol{HFI:center:frequency:217GHz:units}{217\,GHz}
\setsymbol{HFI:center:frequency:353GHz:units}{353\,GHz}
\setsymbol{HFI:center:frequency:545GHz:units}{545\,GHz}
\setsymbol{HFI:center:frequency:857GHz:units}{857\,GHz}

\setsymbol{HFI:center:frequency:100GHz}{100}
\setsymbol{HFI:center:frequency:143GHz}{143}
\setsymbol{HFI:center:frequency:217GHz}{217}
\setsymbol{HFI:center:frequency:353GHz}{353}
\setsymbol{HFI:center:frequency:545GHz}{545}
\setsymbol{HFI:center:frequency:857GHz}{857}

\setsymbol{HFI:Ndetectors:100GHz}{8}
\setsymbol{HFI:Ndetectors:143GHz}{11}
\setsymbol{HFI:Ndetectors:217GHz}{12}
\setsymbol{HFI:Ndetectors:353GHz}{12}
\setsymbol{HFI:Ndetectors:545GHz}{3}
\setsymbol{HFI:Ndetectors:857GHz}{4}

\setsymbol{HFI:FWHM:Maps:100GHz:units}{9\parcm88}
\setsymbol{HFI:FWHM:Maps:143GHz:units}{7\parcm18}
\setsymbol{HFI:FWHM:Maps:217GHz:units}{4\parcm87}
\setsymbol{HFI:FWHM:Maps:353GHz:units}{4\parcm65}
\setsymbol{HFI:FWHM:Maps:545GHz:units}{4\parcm72}
\setsymbol{HFI:FWHM:Maps:857GHz:units}{4\parcm39}
\setsymbol{HFI:FWHM:Maps:100GHz}{9.88}
\setsymbol{HFI:FWHM:Maps:143GHz}{7.18}
\setsymbol{HFI:FWHM:Maps:217GHz}{4.87}
\setsymbol{HFI:FWHM:Maps:353GHz}{4.65}
\setsymbol{HFI:FWHM:Maps:545GHz}{4.72}
\setsymbol{HFI:FWHM:Maps:857GHz}{4.39}

\setsymbol{HFI:beam:ellipticity:Maps:100GHz}{1.15}
\setsymbol{HFI:beam:ellipticity:Maps:143GHz}{1.01}
\setsymbol{HFI:beam:ellipticity:Maps:217GHz}{1.06}
\setsymbol{HFI:beam:ellipticity:Maps:353GHz}{1.05}
\setsymbol{HFI:beam:ellipticity:Maps:545GHz}{1.14}
\setsymbol{HFI:beam:ellipticity:Maps:857GHz}{1.19}

\setsymbol{HFI:FWHM:Mars:100GHz:units}{9\parcm37}
\setsymbol{HFI:FWHM:Mars:143GHz:units}{7\parcm04}
\setsymbol{HFI:FWHM:Mars:217GHz:units}{4\parcm68}
\setsymbol{HFI:FWHM:Mars:353GHz:units}{4\parcm43}
\setsymbol{HFI:FWHM:Mars:545GHz:units}{3\parcm80}
\setsymbol{HFI:FWHM:Mars:857GHz:units}{3\parcm67}

\setsymbol{HFI:FWHM:Mars:100GHz}{9.37}
\setsymbol{HFI:FWHM:Mars:143GHz}{7.04}
\setsymbol{HFI:FWHM:Mars:217GHz}{4.68}
\setsymbol{HFI:FWHM:Mars:353GHz}{4.43}
\setsymbol{HFI:FWHM:Mars:545GHz}{3.80}
\setsymbol{HFI:FWHM:Mars:857GHz}{3.67}

\setsymbol{HFI:beam:ellipticity:Mars:100GHz}{1.18}
\setsymbol{HFI:beam:ellipticity:Mars:143GHz}{1.03}
\setsymbol{HFI:beam:ellipticity:Mars:217GHz}{1.14}
\setsymbol{HFI:beam:ellipticity:Mars:353GHz}{1.09}
\setsymbol{HFI:beam:ellipticity:Mars:545GHz}{1.25}
\setsymbol{HFI:beam:ellipticity:Mars:857GHz}{1.03}

\setsymbol{HFI:CMB:relative:calibration:100GHz}{$\lsim 1\%$}
\setsymbol{HFI:CMB:relative:calibration:143GHz}{$\lsim 1\%$}
\setsymbol{HFI:CMB:relative:calibration:217GHz}{$\lsim 1\%$}
\setsymbol{HFI:CMB:relative:calibration:353GHz}{$\lsim 1\%$}
\setsymbol{HFI:CMB:relative:calibration:545GHz}{}
\setsymbol{HFI:CMB:relative:calibration:857GHz}{}

\setsymbol{HFI:CMB:absolute:calibration:100GHz}{$\lsim 2\%$}
\setsymbol{HFI:CMB:absolute:calibration:143GHz}{$\lsim 2\%$}
\setsymbol{HFI:CMB:absolute:calibration:217GHz}{$\lsim 2\%$}
\setsymbol{HFI:CMB:absolute:calibration:353GHz}{$\lsim 2\%$}
\setsymbol{HFI:CMB:absolute:calibration:545GHz}{}
\setsymbol{HFI:CMB:absolute:calibration:857GHz}{}

\setsymbol{HFI:FIRAS:gain:calibration:accuracy:statistical:100GHz}{}
\setsymbol{HFI:FIRAS:gain:calibration:accuracy:statistical:143GHz}{}
\setsymbol{HFI:FIRAS:gain:calibration:accuracy:statistical:217GHz}{}
\setsymbol{HFI:FIRAS:gain:calibration:accuracy:statistical:353GHz}{2.5\%}
\setsymbol{HFI:FIRAS:gain:calibration:accuracy:statistical:545GHz}{1\%}
\setsymbol{HFI:FIRAS:gain:calibration:accuracy:statistical:857GHz}{0.5\%}

\setsymbol{HFI:FIRAS:gain:calibration:accuracy:systematic:100GHz}{}
\setsymbol{HFI:FIRAS:gain:calibration:accuracy:systematic:143GHz}{}
\setsymbol{HFI:FIRAS:gain:calibration:accuracy:systematic:217GHz}{}
\setsymbol{HFI:FIRAS:gain:calibration:accuracy:systematic:353GHz}{}
\setsymbol{HFI:FIRAS:gain:calibration:accuracy:systematic:545GHz}{7\%}
\setsymbol{HFI:FIRAS:gain:calibration:accuracy:systematic:857GHz}{7\%}

\setsymbol{HFI:FIRAS:zero:point:accuracy:100GHz:units}{0.8\MJysr}
\setsymbol{HFI:FIRAS:zero:point:accuracy:143GHz:units}{}
\setsymbol{HFI:FIRAS:zero:point:accuracy:217GHz:units}{}
\setsymbol{HFI:FIRAS:zero:point:accuracy:353GHz:units}{1.4\MJysr}
\setsymbol{HFI:FIRAS:zero:point:accuracy:545GHz:units}{2.2\MJysr}
\setsymbol{HFI:FIRAS:zero:point:accuracy:857GHz:units}{1.7\MJysr}

\setsymbol{HFI:FIRAS:zero:point:accuracy:100GHz}{0.8}
\setsymbol{HFI:FIRAS:zero:point:accuracy:143GHz}{}
\setsymbol{HFI:FIRAS:zero:point:accuracy:217GHz}{}
\setsymbol{HFI:FIRAS:zero:point:accuracy:353GHz}{1.4}
\setsymbol{HFI:FIRAS:zero:point:accuracy:545GHz}{2.2}
\setsymbol{HFI:FIRAS:zero:point:accuracy:857GHz}{1.7}

\setsymbol{HFI:unit:conversion:100GHz:units}{0.2415\MJysrmK}
\setsymbol{HFI:unit:conversion:143GHz:units}{0.3694\MJysrmK}
\setsymbol{HFI:unit:conversion:217GHz:units}{0.4811\MJysrmK}
\setsymbol{HFI:unit:conversion:353GHz:units}{0.2883\MJysrmK}
\setsymbol{HFI:unit:conversion:545GHz:units}{0.05826\MJysrmK}
\setsymbol{HFI:unit:conversion:857GHz:units}{0.002238\MJysrmK}

\setsymbol{HFI:unit:conversion:100GHz}{0.2415}
\setsymbol{HFI:unit:conversion:143GHz}{0.3694}
\setsymbol{HFI:unit:conversion:217GHz}{0.4811}
\setsymbol{HFI:unit:conversion:353GHz}{0.2883}
\setsymbol{HFI:unit:conversion:545GHz}{0.05826}
\setsymbol{HFI:unit:conversion:857GHz}{0.002238}

\setsymbol{HFI:colour:correction:alpha=-2:V1.01:100GHz}{0.9893}
\setsymbol{HFI:colour:correction:alpha=-2:V1.01:143GHz}{0.9759}
\setsymbol{HFI:colour:correction:alpha=-2:V1.01:217GHz}{1.0007}
\setsymbol{HFI:colour:correction:alpha=-2:V1.01:353GHz}{1.0028}
\setsymbol{HFI:colour:correction:alpha=-2:V1.01:545GHz}{1.0019}
\setsymbol{HFI:colour:correction:alpha=-2:V1.01:857GHz}{0.9889}

\setsymbol{HFI:colour:correction:alpha=0:V1.01:100GHz}{1.0008}
\setsymbol{HFI:colour:correction:alpha=0:V1.01:143GHz}{1.0148}
\setsymbol{HFI:colour:correction:alpha=0:V1.01:217GHz}{0.9909}
\setsymbol{HFI:colour:correction:alpha=0:V1.01:353GHz}{0.9888}
\setsymbol{HFI:colour:correction:alpha=0:V1.01:545GHz}{0.9878}
\setsymbol{HFI:colour:correction:alpha=0:V1.01:857GHz}{1.0014}

\begin{document}
\allearlypapers
\title{\textit{Planck} Early Results. VII. The Early Release Compact Source Catalogue}
\author{\small
Planck Collaboration:
P.~A.~R.~Ade\inst{76}
\and
N.~Aghanim\inst{50}
\and
M.~Arnaud\inst{63}
\and
M.~Ashdown\inst{61, 4}
\and
J.~Aumont\inst{50}
\and
C.~Baccigalupi\inst{74}
\and
A.~Balbi\inst{30}
\and
A.~J.~Banday\inst{82, 8, 68}
\and
R.~B.~Barreiro\inst{57}
\and
J.~G.~Bartlett\inst{3, 59}
% ...
}
\institute{\small
Aalto University Mets\"{a}hovi Radio Observatory, Mets\"{a}hovintie 114, FIN-02540 Kylm\"{a}l\"{a}, Finland\\
\and
Agenzia Spaziale Italiana Science Data Center, c/o ESRIN, via Galileo Galilei, Frascati, Italy\\
\and
Astroparticule et Cosmologie, CNRS (UMR7164), Universit\'{e} Denis Diderot Paris 7, B\^{a}timent Condorcet, 10 rue A. Domon et L\'{e}onie Duquet, Paris, France\\
\and
Astrophysics Group, Cavendish Laboratory, University of Cambridge, J J Thomson Avenue, Cambridge CB3 0HE, U.K.\\
\and
Atacama Large Millimeter/submillimeter Array, ALMA Santiago Central Offices, Alonso de Cordova 3107, Vitacura, Casilla 763 0355, Santiago, Chile\\
\and
Australia Telescope National Facility, CSIRO, P.O. Box 76, Epping, NSW 1710, Australia\\
\and
CITA, University of Toronto, 60 St. George St., Toronto, ON M5S 3H8, Canada\\
\and
CNRS, IRAP, 9 Av. colonel Roche, BP 44346, F-31028 Toulouse cedex 4, France\\
\and
California Institute of Technology, Pasadena, California, U.S.A.\\
\and
Centre of Mathematics for Applications, University of Oslo, Blindern, Oslo, Norway\\
\and
DAMTP, University of Cambridge, Centre for Mathematical Sciences, Wilberforce Road, Cambridge CB3 0WA, U.K.\\
\and
DSM/Irfu/SPP, CEA-Saclay, F-91191 Gif-sur-Yvette Cedex, France\\
\and
DTU Space, National Space Institute, Juliane Mariesvej 30, Copenhagen, Denmark\\
\and
Departamento de F\'{\i}sica, Universidad de Oviedo, Avda. Calvo Sotelo s/n, Oviedo, Spain\\
\and
Department of Astronomy and Astrophysics, University of Toronto, 50 Saint George Street, Toronto, Ontario, Canada\\
\and
Department of Physics \& Astronomy, University of British Columbia, 6224 Agricultural Road, Vancouver, British Columbia, Canada\\
\and
Department of Physics and Astronomy, Tufts University, Medford, MA 02155, U.S.A.\\
\and
Department of Physics, Gustaf H\"{a}llstr\"{o}min katu 2a, University of Helsinki, Helsinki, Finland\\
\and
Department of Physics, Princeton University, Princeton, New Jersey, U.S.A.\\
\and
Department of Physics, Purdue University, 525 Northwestern Avenue, West Lafayette, Indiana, U.S.A.\\
\and
Department of Physics, University of California, Berkeley, California, U.S.A.\\
\and
Department of Physics, University of California, One Shields Avenue, Davis, California, U.S.A.\\
\and
Department of Physics, University of California, Santa Barbara, California, U.S.A.\\
\and
Department of Physics, University of Illinois at Urbana-Champaign, 1110 West Green Street, Urbana, Illinois, U.S.A.\\
\and
Dipartimento di Fisica G. Galilei, Universit\`{a} degli Studi di Padova, via Marzolo 8, 35131 Padova, Italy\\
\and
Dipartimento di Fisica, Universit\`{a} La Sapienza, P. le A. Moro 2, Roma, Italy\\
\and
Dipartimento di Fisica, Universit\`{a} degli Studi di Milano, Via Celoria, 16, Milano, Italy\\
\and
Dipartimento di Fisica, Universit\`{a} degli Studi di Trieste, via A. Valerio 2, Trieste, Italy\\
\and
Dipartimento di Fisica, Universit\`{a} di Ferrara, Via Saragat 1, 44122 Ferrara, Italy\\
\and
Dipartimento di Fisica, Universit\`{a} di Roma Tor Vergata, Via della Ricerca Scientifica, 1, Roma, Italy\\
\and
Discovery Center, Niels Bohr Institute, Blegdamsvej 17, Copenhagen, Denmark\\
\and
Dpto. Astrof\'{i}sica, Universidad de La Laguna (ULL), E-38206 La Laguna, Tenerife, Spain\\
\and
European Southern Observatory, ESO Vitacura, Alonso de Cordova 3107, Vitacura, Casilla 19001, Santiago, Chile\\
\and
European Space Agency, ESAC, Planck Science Office, Camino bajo del Castillo, s/n, Urbanizaci\'{o}n Villafranca del Castillo, Villanueva de la Ca\~{n}ada, Madrid, Spain\\
\and
European Space Agency, ESTEC, Keplerlaan 1, 2201 AZ Noordwijk, The Netherlands\\
\and
Haverford College Astronomy Department, 370 Lancaster Avenue, Haverford, Pennsylvania, U.S.A.\\
\and
Helsinki Institute of Physics, Gustaf H\"{a}llstr\"{o}min katu 2, University of Helsinki, Helsinki, Finland\\
\and
INAF - Osservatorio Astrofisico di Catania, Via S. Sofia 78, Catania, Italy\\
\and
INAF - Osservatorio Astronomico di Padova, Vicolo dell'Osservatorio 5, Padova, Italy\\
\and
INAF - Osservatorio Astronomico di Roma, via di Frascati 33, Monte Porzio Catone, Italy\\
\and
INAF - Osservatorio Astronomico di Trieste, Via G.B. Tiepolo 11, Trieste, Italy\\
\and
INAF/IASF Bologna, Via Gobetti 101, Bologna, Italy\\
\and
INAF/IASF Milano, Via E. Bassini 15, Milano, Italy\\
\and
INRIA, Laboratoire de Recherche en Informatique, Universit\'{e} Paris-Sud 11, B\^{a}timent 490, 91405 Orsay Cedex, France\\
\and
IPAG: Institut de Plan\'{e}tologie et d'Astrophysique de Grenoble, Universit\'{e} Joseph Fourier, Grenoble 1 / CNRS-INSU, UMR 5274, Grenoble, F-38041, France\\
\and
ISDC Data Centre for Astrophysics, University of Geneva, ch. d'Ecogia 16, Versoix, Switzerland\\
\and
Imperial College London, Astrophysics group, Blackett Laboratory, Prince Consort Road, London, SW7 2AZ, U.K.\\
\and
Infrared Processing and Analysis Center, California Institute of Technology, Pasadena, CA 91125, U.S.A.\\
\and
Institut N\'{e}el, CNRS, Universit\'{e} Joseph Fourier Grenoble I, 25 rue des Martyrs, Grenoble, France\\
\and
Institut d'Astrophysique Spatiale, CNRS (UMR8617) Universit\'{e} Paris-Sud 11, B\^{a}timent 121, Orsay, France\\
\and
Institut d'Astrophysique de Paris, CNRS UMR7095, Universit\'{e} Pierre \& Marie Curie, 98 bis boulevard Arago, Paris, France\\
\and
Institut de Ci\`{e}ncies de l'Espai, CSIC/IEEC, Facultat de Ci\`{e}ncies, Campus UAB, Torre C5 par-2, Bellaterra 08193, Spain\\
\and
Institute of Astronomy and Astrophysics, Academia Sinica, Taipei, Taiwan\\
\and
Institute of Astronomy, University of Cambridge, Madingley Road, Cambridge CB3 0HA, U.K.\\
\and
Institute of Theoretical Astrophysics, University of Oslo, Blindern, Oslo, Norway\\
\and
Instituto de Astrof\'{\i}sica de Canarias, C/V\'{\i}a L\'{a}ctea s/n, La Laguna, Tenerife, Spain\\
\and
Instituto de F\'{\i}sica de Cantabria (CSIC-Universidad de Cantabria), Avda. de los Castros s/n, Santander, Spain\\
\and
Istituto di Fisica del Plasma, CNR-ENEA-EURATOM Association, Via R. Cozzi 53, Milano, Italy\\
\and
Jet Propulsion Laboratory, California Institute of Technology, 4800 Oak Grove Drive, Pasadena, California, U.S.A.\\
\and
Jodrell Bank Centre for Astrophysics, Alan Turing Building, School of Physics and Astronomy, The University of Manchester, Oxford Road, Manchester, M13 9PL, U.K.\\
\and
Kavli Institute for Cosmology Cambridge, Madingley Road, Cambridge, CB3 0HA, U.K.\\
\and
LERMA, CNRS, Observatoire de Paris, 61 Avenue de l'Observatoire, Paris, France\\
\and
Laboratoire AIM, IRFU/Service d'Astrophysique - CEA/DSM - CNRS - Universit\'{e} Paris Diderot, B\^{a}t. 709, CEA-Saclay, F-91191 Gif-sur-Yvette Cedex, France\\
\and
Laboratoire Traitement et Communication de l'Information, CNRS (UMR 5141) and T\'{e}l\'{e}com ParisTech, 46 rue Barrault F-75634 Paris Cedex 13, France\\
\and
Laboratoire de Physique Subatomique et de Cosmologie, CNRS/IN2P3, Universit\'{e} Joseph Fourier Grenoble I, Institut National Polytechnique de Grenoble, 53 rue des Martyrs, 38026 Grenoble cedex, France\\
\and
Laboratoire de l'Acc\'{e}l\'{e}rateur Lin\'{e}aire, Universit\'{e} Paris-Sud 11, CNRS/IN2P3, Orsay, France\\
\and
Lawrence Berkeley National Laboratory, Berkeley, California, U.S.A.\\
\and
Max-Planck-Institut f\"{u}r Astrophysik, Karl-Schwarzschild-Str. 1, 85741 Garching, Germany\\
\and
MilliLab, VTT Technical Research Centre of Finland, Tietotie 3, Espoo, Finland\\
\and
National University of Ireland, Department of Experimental Physics, Maynooth, Co. Kildare, Ireland\\
\and
Niels Bohr Institute, Blegdamsvej 17, Copenhagen, Denmark\\
\and
Observational Cosmology, Mail Stop 367-17, California Institute of Technology, Pasadena, CA, 91125, U.S.A.\\
\and
Optical Science Laboratory, University College London, Gower Street, London, U.K.\\
\and
SISSA, Astrophysics Sector, via Bonomea 265, 34136, Trieste, Italy\\
\and
SUPA, Institute for Astronomy, University of Edinburgh, Royal Observatory, Blackford Hill, Edinburgh EH9 3HJ, U.K.\\
\and
School of Physics and Astronomy, Cardiff University, Queens Buildings, The Parade, Cardiff, CF24 3AA, U.K.\\
\and
Space Research Institute (IKI), Russian Academy of Sciences, Profsoyuznaya Str, 84/32, Moscow, 117997, Russia\\
\and
Space Sciences Laboratory, University of California, Berkeley, California, U.S.A.\\
\and
Spitzer Science Center, 1200 E. California Blvd., Pasadena, California, U.S.A.\\
\and
Stanford University, Dept of Physics, Varian Physics Bldg, 382 Via Pueblo Mall, Stanford, California, U.S.A.\\
\and
Universit\"{a}t Heidelberg, Institut f\"{u}r Theoretische Astrophysik, Albert-\"{U}berle-Str. 2, 69120, Heidelberg, Germany\\
\and
Universit\'{e} de Toulouse, UPS-OMP, IRAP, F-31028 Toulouse cedex 4, France\\
% ...
}

\titlerunning{ERCSC}
\authorrunning{Planck Collaboration}

\abstract{A brief description of the methodology of construction, contents and usage
of the \Planck\ Early Release Compact Source Catalogue (ERCSC), including
the Early Cold Cores (ECC) and the Early Sunyaev-Zeldovich (ESZ) cluster catalogue is provided. The catalogue is based on data that consist
of mapping the entire sky once and 60\% of the sky a second time by \Planck, thereby comprising the
first high sensitivity radio/submillimetre observations of the entire sky.
Four source detection algorithms were run as part of the ERCSC pipeline.
A Monte-Carlo algorithm based on the injection and extraction of artificial sources into the \Planck\ maps
was implemented to select reliable sources among all extracted candidates 
such that the cumulative reliability of the catalogue is $\geq$90\%. There
is no requirement on completeness for the ERCSC.
As a result of the Monte-Carlo assessment of reliability of sources from the different techniques, 
an implementation of the PowellSnakes source extraction technique was
used at the five frequencies between 30 and 143\,GHz while the SExtractor technique was used between 217 and 857\,GHz. 
The 10$\sigma$ photometric flux density limit of the catalogue at $|b|>30\degr$ is
0.49, 1.0, 0.67, 0.5, 0.33, 0.28, 0.25, 0.47 and 0.82 Jy at each of the nine frequencies between $30$ and $857$\,GHz. Sources 
which are up to a factor of $\sim$2 fainter than this limit, 
and which are present in ``clean'' regions of the Galaxy where the sky background due to emission from the interstellar 
medium is low, are included in the ERCSC if they meet the high
reliability criterion. The \Planck\ ERCSC sources have known associations to
stars with dust shells, stellar cores, radio galaxies, blazars, infrared luminous galaxies and 
Galactic interstellar medium features. 
A significant fraction of unclassified sources are also present in the catalogs.
In addition, two early release catalogs that
contain 915 cold molecular cloud core candidates and 189 SZ cluster candidates that have been
generated using multifrequency algorithms are presented.
The entire source list, with more than 15000 unique sources, 
is ripe
for follow-up characterisation with {\it Herschel}, ATCA, VLA, SOFIA, ALMA and other ground-based observing facilities.}
\maketitle

\section{Introduction}

\Planck\footnote{\Planck\ ({\rm http://www.esa.int/Planck}) is a project of the European
Space Agency (ESA) with instruments provided by two scientific
consortia funded by ESA member states (in particular the lead countries
France and Italy), with contributions from NASA (USA) and telescope
reflectors provided in a collaboration between ESA and a scientific
consortium led and funded by Denmark.}\ \citep{tauber2010a,planck2011-1.1} is the third-generation space mission to measure the anisotropy of the cosmic microwave background (CMB).  It observes the sky in nine frequency bands covering 30--857\,GHz with high sensitivity and angular resolution from 33\,arcmin to 4.2\,arcmin (Table \ref{tab:ascomp}).  The Low Frequency Instrument (LFI; \citealt{Mandolesi2010,Bersanelli2010,planck2011-1.4}) covers the 30, 44, and 70\,GHz bands with amplifiers cooled to 20\,\hbox{K}.  The High Frequency Instrument (HFI; \citealt{Lamarre2010,planck2011-1.5}) covers the 100, 143, 217, 353, 545, and 857\,GHz bands with bolometers cooled to 0.1\,\hbox{K}.  Polarisation is measured in all but the highest two bands \citep{Leahy2010,Rosset2010}.  A combination of radiative cooling and three mechanical coolers produces the temperatures needed for the detectors and optics \citep{planck2011-1.3}.  Two data processing centres (DPCs) check and calibrate the data and make maps of the sky \citep{planck2011-1.7,planck2011-1.6}.  \Planck's sensitivity, angular resolution, and frequency coverage make it a powerful instrument for Galactic and extragalactic astrophysics as well as cosmology.

\begin{figure*}
\centering
\includegraphics[width=8.5cm]{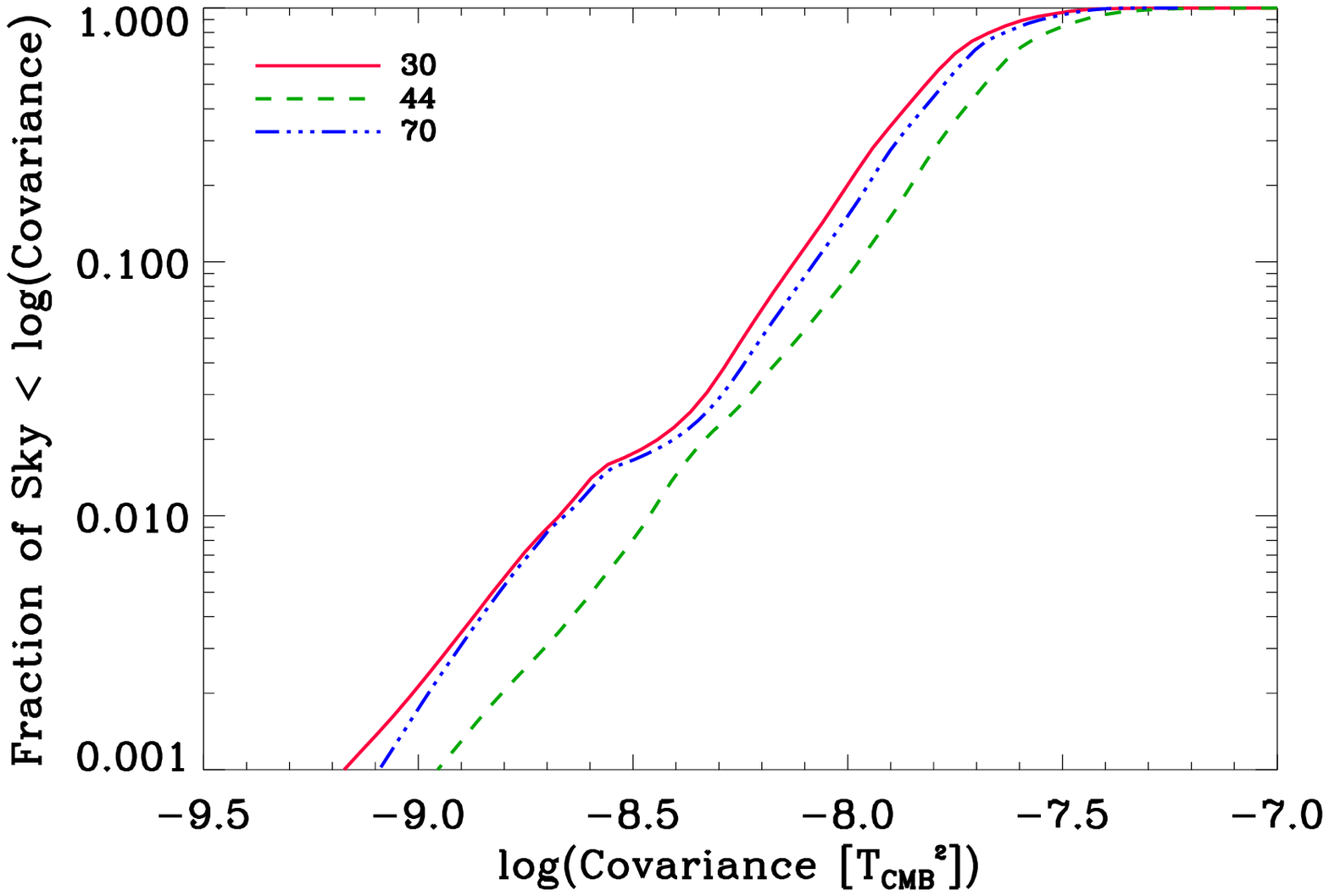}
\includegraphics[width=8.5cm]{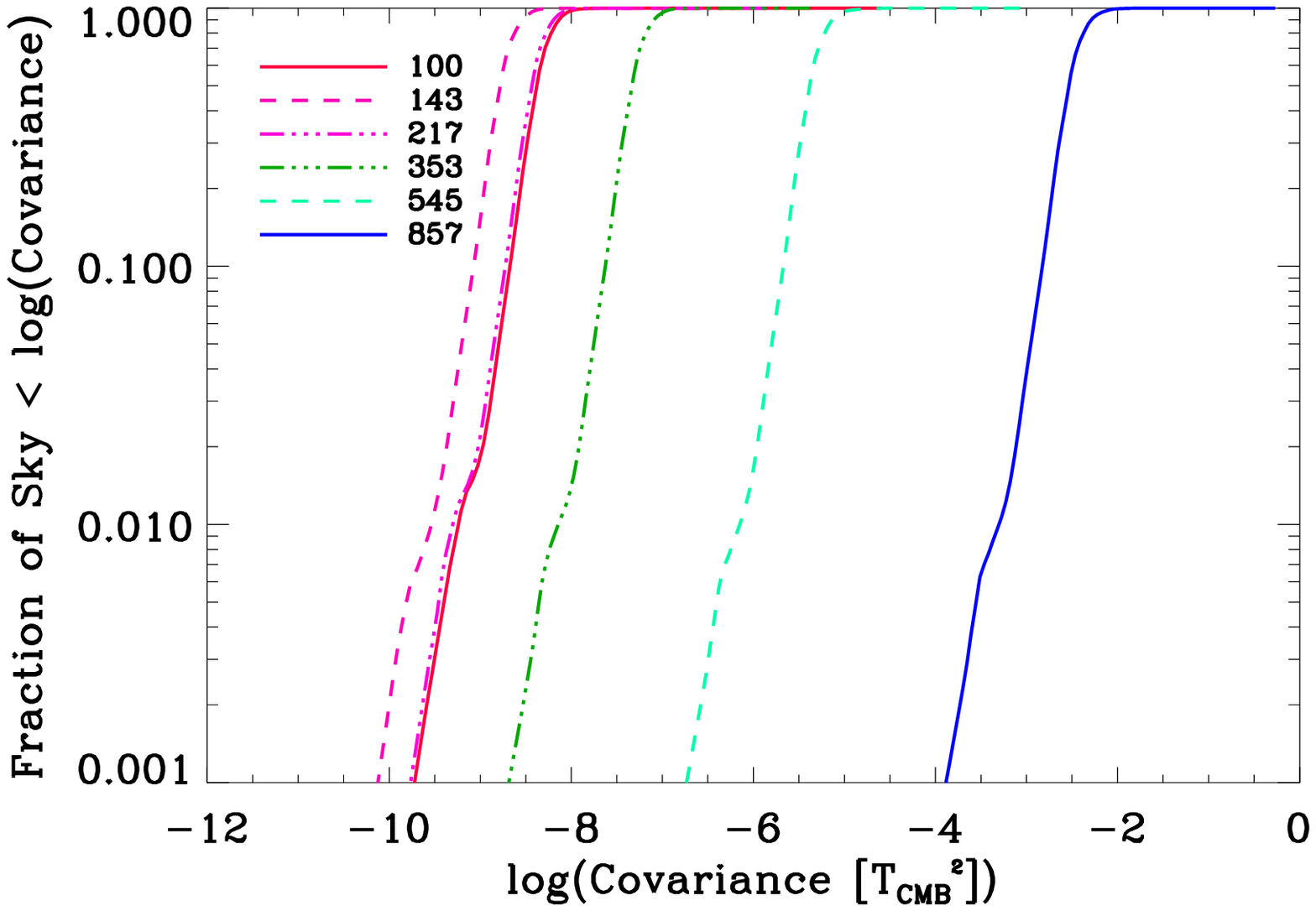}
\caption[\Planck\ Sensitivity]{
The variance in units of Kelvin$^{2}$ across the entire sky, with the left panel showing the variance for the LFI frequencies
and the right panel showing the variance for the HFI frequencies. At any single frequency, the 
variance in the all-sky maps
span almost a factor of 50 over the entire sky.
}
\label{fig:skycov}
\end{figure*}

\Planck\ spins around its axis at a rate of one rotation per minute.
The focal plane is oriented at an angle of 85$\degr$ to the satellite spin axis, which
tracks the direction of the Sun at $\approx$1$\degr$ per day. The effective
\Planck\ scan strategy described in \citet{dupac2005} results in areas near
the ecliptic poles being observed several times more frequently than regions of sky near
the ecliptic plane. 
This implies that there is a range of almost 50 in instrumental noise
between the most- and least-frequently observed areas of the sky (Figure \ref{fig:skycov}).
This scan strategy achieved greater than $99.9$\% coverage of the full sky on 1 April 2010
with gaps predominantly resulting from planet crossings and the masking of associated artefacts. 
A comparison between the all sky imaging capabilities of \Planck, Wilkinson Microwave Anisotropy Probe ({\it WMAP}), Cosmic Background
Explorer/Differential Microwave Radiometer ({\it COBE}/DMR), {\it Akari}, Infrared Astronomy Satellite ({\it IRAS}) and Wide-field Infrared Survey Explorer ({\it WISE}) 
is shown in Table~\ref{tab:ascomp}.
\Planck\ straddles the wavelength range between {\it WMAP} \citep{Bennett2003} at one end and {\it Akari} \citep{Murakami2007} at the other end. 
At its lowest frequencies \Planck\ improves upon the imaging resolution of {\it WMAP}. Although \Planck\ 
does not have the high resolution of {\it Akari} or {\it WISE}\citep{Wright2010} at higher frequencies, it matches the capabilities of {\it IRAS} at
frequencies which are more than a factor of three lower than {\it IRAS}. The consequence of this unprecedented spatial resolution
and wavelength coverage is a unique simultaneous, multiwavelength view of the sky, enabling the study of a broad class of sources, and
facilitating improved separation between Galactic and extragalactic foregrounds and the CMB.

The Early Release Compact Source Catalogue is a catalogue of all high-reliability
sources, both Galactic 
and extragalactic, detected over the entire sky, in the first \Planck\ all-sky survey. This includes
a sample of clusters detected through the Sunyaev-Zeldovich (SZ) effect and a catalogue
of cold, molecular cloud cores with far-infrared colour temperatures cooler than the ambient $T\sim$18K dust in our Galaxy.
No polarisation information is provided for the sources at this time. One of the primary goals of the ERCSC is to provide an early catalogue
of sources for follow-up observations with existing facilities, in particular {\it Herschel}, while they are still in their
cryogenic operational phase. The need for a rapid turnaround (less than nine months) from the end of the first sky coverage to a community-wide release
of source lists is the motivating factor behind the reliability and flux-density accuracy requirements as well as the choice
of algorithms that were adopted for the ERCSC.

The sources of noise vary significantly as a function of location on the sky as well as a function of frequency. 
Apart from the instrumental noise and the Galaxy, at the lowest frequencies the dominant astrophysical source
of noise is the CMB itself. At the highest frequencies, zodiacal dust and emission
from the interstellar medium dominate. As a result, the flux density limits
corresponding to the same reliability vary widely across the sky (Figure~\ref{fig:skycov}), with the
sensitivity typically improving with increasing ecliptic latitude. The areas of deepest coverage are centred
on the ecliptic pole regions due to the scan strategy and individual sources in that vicinity may be observed several times
by \Planck\ in the course of a single sky survey.

The data obtained
from the scans of the sky between 2009 August 13 and 2010 June 6, corresponding to \Planck\ operational days 91--389,
have been processed and converted into all-sky maps at the HFI and LFI Data Processing Centres (DPCs). The data
extend beyond a single sky coverage with 60\% of the second sky coverage included in the maps.
A description of the processing can be found in \citet{planck2011-1.6,planck2011-1.7}.
Four different implementations of source detection algorithms were run on these maps. The performance of these algorithms was
compared and the single implementation which provides superior source statistics at each frequency was selected
for the final catalogue at that frequency. For the early SZ (ESZ) and early cold cores (ECC) catalogues, multifrequency
algorithms described in \citet{melin2006} and \citet{planck2011-7.7b} respectively, have been run to provide a candidate source
list, which has then been culled to maintain the high reliability required for the ERCSC.
This paper describes the methodology through which the ERCSC pipeline generates a high reliability
source catalogue as well as presents the contents and characteristics of the \Planck\ ERCSC data release.

\begin{table*}
\centering
\caption{Comparison between all sky surveys with similar frequencies aligned in rows. The left column for each
mission gives the frequency ($\nu$ in GHz) while the right
column gives the spatial resolution as a full width at half maximum (FWHM) in arcminutes.}
\label{tab:ascomp}
\begin{tabular}{c c c c c c c c c c c c c c c c c}
\hline\hline
\multicolumn{2}{c}{{\it DMR}} &&
\multicolumn{2}{c}{{\it WMAP}} &&
\multicolumn{2}{c}{\Planck} &&
\multicolumn{2}{c}{{\it Akari}} &&
\multicolumn{2}{c}{{\it IRAS}} &&
\multicolumn{2}{c}{{\it WISE}} \\
\cline{1-2} 
\cline{4-5} 
\cline{7-8}
\cline{10-11} 
\cline{13-14} 
\cline{16-17} 
$\nu$ & FWHM && $\nu$ & FWHM && $\nu$ & FWHM && $\nu$ & FWHM && $\nu$ & FWHM && $\nu$ & FWHM  \\
\hline
   & && 23 & 53 \\
       32 & 420 && 33 & 40 && 30 & 32.65 \\
   & && 41 & 31 && 44 & 27.00 && \\
 53 & 420 && 61 & 21 && 70 & 13.01 \\
 90 & 420 && 94 & 13 && 100 & 9.94 \\
   & && & && 143 & 7.04 && \\
 & && & && 217 & 4.66 \\
 & && & && 353 & 4.41 \\
 & && & && 545 & 4.47 \\
 & && & && 857 & 4.23 \\
 & && & && & && 1.9$\times$10$^{3}$ & 0.8 \\
 & && & && & && 2.1$\times$10$^{3}$ & 0.7 \\
 & && & && & && 3.3$\times$10$^{3}$ & 0.45 && 3$\times$10$^{3}$ & 5.2 \\
 & && & && & && 4.6$\times$10$^{3}$ & 0.32 && 5$\times$10$^{3}$ & 3.9 \\
 & && & && & && 16.7$\times$10$^{3}$ & 0.09 && 12$\times$10$^{3}$ & 4.5 && 13.6$\times$10$^{3}$ & 0.2 \\
 & && & && & && 33$\times$10$^{3}$ & 0.05 && 25$\times$10$^{3}$ & 4.7 && 25$\times$10$^{3}$ & 0.11 \\
 & && & && & && & && & && 65$\times$10$^{3}$ & 0.11 \\
 & && & && & && & && & && 88$\times$10$^{3}$ & 0.1 \\

\hline
\hline
\end{tabular}
\end{table*}

\section{The ERCSC Pipeline}
This section summarises the steps involved in progressing
from the initial \Planck\ all-sky maps to the final catalogue. A full description of the entire process
can be found in \citet{planck2011-1.10sup}, which has been released with the ERCSC. 

The intensity maps on which the ERCSC pipeline is run 
are in HEALPix format \citep{gor05} in units of ${\rm K_{RJ}}$ (Kelvin Rayleigh-Jeans, a measure of brightness temperature)
in the Galactic coordinate system. 
Thermodynamic temperature (T$_{\rm CMB}$) is related to the Rayleigh-Jeans brightness temperature by:
\begin{eqnarray}
\centering
K_{\rm RJ} & = & T_{\rm CMB} \times \frac{x^{2} \exp{x}}{(\exp{x}-1)^2} \\
x & = & \frac{h\nu}{k\times2.725}
\end{eqnarray}
\noindent where $h$ is the Planck constant, $k$ is the Boltzmann constant and $\nu$ is the frequency.
The pixel size is 3.4\,arcmin
for the LFI bands (30--70\,GHz) and 1.7\,arcmin for the HFI bands (100--857\,GHz) which corresponds to NSIDE values of
1024 and 2048 in the HEALPix format.
In addition to the intensity map, there is a corresponding covariance map at each frequency, which 
is a measure of the noise in that pixel measured as the standard deviation of all scans
that have gone through that pixel, after removal of an offset from each ring of observations. The covariance maps are in units of ${\rm K_{RJ}}^{2}$.
A full description of the map-making process can be found in \citet{planck2011-1.7}.

The three core processing steps within the ERCSC pipeline are source 
detection, source extraction, and in the case of 857\,GHz, measuring the flux densities of each source at the three lower frequencies, 217, 353, and 545\,GHz, which
is sometimes referred to as band-filling.
These three steps are first run on the intensity maps to obtain catalogues of sources.
The process is then repeated on maps which have a population of artificial point sources
of varying flux densities injected directly into the maps. The performance of the
algorithms are evaluated based on the positions and extracted
flux densities of the artificial sources whose real flux density
and positions are precisely known. Based on the properties of the extracted artificial sources,
signal-to-noise ratio (SNR) cuts are defined such that the properties of the extracted artificial sources
are robust both in terms of position and flux density. The same signal-to-noise cut
is then applied to the real catalogue of sources generated from the intensity maps to obtain a high
reliability catalogues. Secondary quality assessment cuts are applied to the catalogue to eliminate
sources associated with known artefacts in the maps. Additional properties of the reliable sources such as the
dates they were observed, their
presence in CMB subtracted maps, their flux density estimated from point source fitting, and the potential
contribution of Galactic cirrus emission are evaluated in the final stages of the pipeline.

We note that 
the maps used for the ERCSC are affected by uncorrected pointing errors of at least two types. 
The first is due to 
time-dependent, thermally-driven misalignment between the star tracker and the boresight of the telescope 
\citep{planck2011-1.1}.  The second is due to uncorrected stellar aberration across the focal plane. Since 70\% of the sky is 
observed at least twice with different orientations, the effect of stellar aberration for the majority of sources is 
negligible. However, for the remaining 30\% of the sky, the effect would result in an offset ranging from 
21$\arcsec$ near the ecliptic poles to a value close to 0$\arcsec$ in the ecliptic plane.  We do not see this effect clearly 
in the centroids of sources in the ERCSC, likely because the intrinsic uncertainty in the positions is comparable 
to the maximum offset induced by these pointing errors.  
Furthermore, since these pointing errors are not factored into the injection of the
artificial sources into the maps, the positional uncertainty from the Monte-Carlo analysis is an underestimate. 
The effect on the flux density of the catalogue sources is 
negligibly small, $\lesssim$2\%. Both types of pointing error will be corrected in the maps used for future Planck catalogues.

\subsection{Source Detection Algorithms}

Four specific implementations of source detection algorithms were run as part of the  
ERCSC pipeline. These are the Paris Matched Filter \citep[PMF;][]{melin2006}, IFCA Mexican hat wavelet filter \citep[IFCAMex;][]{lop07}, PowellSnakes \citep[PwS; Carvalho et al. 2011, in preparation;][]{car09}
and SExtractor \citep{ber96}. Some of the algorithms are still in development and the results depended on version of algorithms used
and their implementation methodology in the ERCSC pipeline.
Of these, only two were selected to generate the final catalogues.
The two algorithms were selected to provide the largest numbers of high reliability
sources at high Galactic latitude in the \Planck\ maps.
These are PwS v2.0 for frequencies
30--143\,GHz, and SExtractor for frequencies 217--857\,GHz. 

\subsubsection{PowellSnakes}
PowellSnakes is a fast Bayesian method for the detection of discrete objects
immersed in a diffuse background. The application of Bayesian model
selection and the Bayesian information criterion to source detection
and extraction have been reviewed by \cite{hob03} and
\cite{savage07}. PwS builds on these ideas and
incorporates them in a fast implementation.

The all-sky map is resampled onto a set of overlapping flat patches
using a gnomonic (tangent plane) projection. Each patch
is modelled as a set of discrete objects, of known shape,
embedded in a stochastic background, with added instrumental
noise. The object shape is chosen to be a circular Gaussian
approximation to the effective point spread function (PSF), and the background and
instrumental noise are modelled as a Gaussian random field with power
spectrum to be estimated from the data. Because both the PSF and
background vary with sky position, the analysis is performed on
overlapping sky patches within which the properties are assumed to
be uniform. At high latitudes and low frequencies the background is
dominated by the CMB, so the Gaussian assumption is a good one; near
the Galactic Plane, however, the background is dominated by emission
from the ISM and the assumption breaks down. 
In practice, however,
PwS gives good results in these
cases. At the highest frequencies SExtractor was found to perform
better than PwS, probably because the model of the background
statistics is poor, and also because many of the sources are diffuse peaks
in the ISM emission and are not well represented by the PSF
model.

Given these assumptions, PwS estimates source parameters by
maximising the posterior probability (i.e., the product of the
likelihood and an assumed prior), using
a simultaneous multiple maximisation code based on Powell's direction
set algorithm (hence the name) to rapidly locate local maxima in
the posterior. This novel feature makes PwS substantially
faster than Monte-Carlo Markov chain methods used by
\cite{hob03}. Whether or not a posterior peak corresponds
to a source is determined by Bayesian model selection using an
approximate evidence value based on a local Gaussian approximation to
the peak. In this step, PwS minimises the average loss matrix
rather than maximising either reliability or completeness: that is, it
treats spurious detections and missing detections as equally
undesirable.

For detection of sources with high signal-to-noise ratio, PwS 
is fairly insensitive to the choice of priors. For the version of the algorithm
which was used for ERCSC, a flat distribution of priors was adopted with
the distribution of priors on the intrinsic source radius being uniform 
between 0 and $3\parcm4$ for all frequencies. Since this is smaller than
\Planck's spatial resolution at any frequency, the effect of the priors is to 
favour point sources.

After merging the results from each patch, the output of PwS 
is a set of source positions with estimated flux densities. 
The ERCSC pipeline photometry algorithms are then applied at each
position to obtain other measures of flux density and size, taking
into account the instrumental noise in each pixel.

\subsubsection{SExtractor}

SExtractor \citep{ber96}, as for PwS, requires local flat patches
created from gnomonic projections. 
Each map is pre-filtered with a Gaussian kernel the same size as the beam at each frequency (the built-in 
filtering step within SExtractor is not used as it uses a digitised filtering grid). Typically, a Mexican 
hat filter gives slightly more reliable detections of point sources in the presence of noise and background, although bright extended sources are often 
missed. However, a Gaussian filter is adopted
because simulations show that it performed almost as well as the Mexican hat for high-latitude compact sources and 
is still sensitive to sources that are extended. 
The algorithm then finds objects by isolating connected groups of pixels above a certain $n-$sigma threshold. 
Sources which are extremely close to each other are deblended if a saddle point is found in the intensity 
distribution. Spurious detections due to neighbouring bright objects are cleaned, and finally the algorithm determines the
centroids of each source and performs photometry in an elliptical Kron aperture \citep{ber96, kron1980}. 

The performance of SExtractor's own adaptive aperture photometry (MAG\_AUTO) is good at high latitudes for all 
\Planck\ frequencies,
 providing flux densities to within 10$\%$ accuracy, and errors typically 1--5\%. Nevertheless, at low Galactic latitudes, particularly at the highest frequencies, the photometric accuracy is significantly degraded. This is because it uses a variable Kron radius, which becomes unstable in crowded fields with strong residual background fluctuations. To ensure homogeneous flux density estimates, the primary flux density estimate
is obtained from an external source extraction code, as was done for PwS.

\subsubsection{Flux Density Estimation} \label{sec:flux_density_estimation}
Each source that is extracted has four different measures of flux density associated with it. These are
based on aperture photometry, PSF fitting, Gaussian fitting and a measure native to the source detection algorithm 
(Table \ref{tab:ercsccon}). Each of these flux density estimates has a local background subtracted but they have not
been colour corrected. Colour corrections are available in \citet{planck2011-1.10sup}.

\begin{enumerate}
\item The FLUX and FLUX\_ERR columns in the ERCSC FITS files give the flux densities measured
in a circular aperture of radius given by the nominal sky-averaged FWHM.
Appropriate corrections have been applied for the flux density outside the aperture,
assuming that the source profile is a point source.

\item The PSFFLUX and PSFFLUX\_ERR columns give flux densities
estimated by fitting the source with the \Planck\ point spread function
at the location of the source \citep{Mitra2010}. The \Planck\ point spread function is estimated at each point
on the sky by combining the
individual horn beams with the scan strategy, where the individual beams are derived from when the scans cross planet
positions. A 2D Gaussian is fit to the PSF and the derived parameters of the Gaussian are used to perform a constrained Gaussian
fit to the source. 

\item The GAUFLUX and GAUFLUX\_ERR columns give flux densities
estimated by fitting the source with an elliptical Gaussian model whose parameters are free. 

\item The FLUXDET and FLUXDET\_ERR columns gives the flux densities estimated by the native detection algorithm. For the
frequencies at which PwS is used, this is estimated by utilising the
mean of the posterior distribution of all parameters, while
for frequencies at which SExtractor is used, it is the flux density in
an elliptical Kron aperture, i.e., FLUX\_AUTO. The FLUXDET values at the
frequencies where PwS is used have been corrected for an average bias
that was seen in the difference between the extracted and input flux
density of Monte-Carlo sources that were injected into the maps.  This is most likely due to an
inaccurate representation of the true beam inside the PwS detection
algorithm. For faint
extended sources in the upper HFI frequencies, the SExtractor FLUXDET
values might be useful.
\end{enumerate}

Once the initial pass of the algorithm generates the list of all sources in the map, the next
step is to identify the ones which are highly reliable, i.e., those that have accurate positions as well
as flux density uncertainties which are less than 30\%\footnote{Spurious sources can be classified as those which have
an intrinsic flux density of zero but with 
some arbitrary extracted flux density, corresponding to a flux density error of 100\%. The presence of such sources would decrease the reliability at the corresponding extracted flux density.}. 
In the absence of a ``truth'' catalogue
for the sky, it is not possible to definitively identify reliable
sources. The significant frequency
difference between {\it Akari}, {\it IRAS} and \Planck\ at submillimetre frequencies implies that uncertain
extrapolations of the thermal dust spectral energy distribution (SED) need to be made to force associations between 
far-infrared sources and \Planck\ sources. 
At radio frequencies,
deeper surveys such as those with the Green Bank Telescope, Parkes and ATCA have been undertaken \citep[e.g.][]{greg, griffith}.
However, the flat-spectrum radio sources that dominate the source population vary significantly even
on short time scales. 
In addition, the high source density of those surveys requires assumptions about the thermal and non-thermal spectral indices in order to identify possible associations between the \Planck\ sources and the radio sources. 
Although these ancillary external
catalogues are used for cross-validation of the final ERCSC, the primary measure of reliability
for the sources uses a Monte-Carlo Quality Assessment (MCQA) analysis that is described in the next section. This is the first
application of a Monte-Carlo source characterisation algorithm at these frequencies, although the practise
is fairly commonplace at higher frequencies \citep[][ and references therein]{cha}. The process is described below.

\begin{table*}
\centering
\caption{ERCSC Catalogue Columns
\label{tab:ercsccon}}
\begin{tabular}{l l}
\hline\hline
Column Name &
Description \\
\hline
\multicolumn{2}{c}{Identification} \\
\hline
NAME & Source name \\
FLUX & Flux density (mJy)\\
FLUX\_ERR & Flux density error (mJy)\\
CMBSUBTRACT & Flag indicating detection of source in CMB subtracted maps\\
EXTENDED & Flag indicating that source is extended \\
DATESOBS & UTC dates at which this source was observed \\
NUMOBS & Number of days this source observed \\
CIRRUS & Cirrus flag based on 857 GHz source counts \\
\hline
\multicolumn{2}{c}{Source Position} \\
\hline 
GLON  & Galactic longitude (deg) based on extraction algorithm \\
GLAT & Galactic latitude (deg) based on extraction algorithm \\
POS\_ERR & Standard deviation of positional offsets for sources with this SNR (arcminute) \\
RA & Right Ascension (J2000) in degrees transformed from (GLON,GLAT) \\
DEC & Declination (J2000) in degrees transformed from (GLON,GLAT) \\
\hline 
\multicolumn{2}{c}{Effective beam} \\
\hline
BEAM\_FWHMMAJ & Elliptical Gaussian beam FWHM along major axis (arcmin) \\
BEAM\_FWHMMIN & Elliptical Gaussian beam FWHM along minor axis (arcmin) \\
BEAM\_THETA & Orientation of Elliptical Gaussian major axis (measured East 
of Galactic North) \\
\hline
\multicolumn{2}{c}{Morphology} \\
\hline 
ELONGATION & Ratio of major to minor axis lengths \\
\hline
\multicolumn{2}{c}{Source Extraction Results} \\
\hline
FLUXDET & Flux density of source as determined by detection method  (mJy) \\
FLUXDET\_ERR & Uncertainty (1 sigma) of FLUXDET (mJy)\\
MX1 & First moment in X (arcmin)\\
MY1 & First moment in Y (arcmin) \\
MX2 & Second moment in X (arcmin$^2$) \\
MXY & Cross moment in X and Y (arcmin$^2$)\\
MY2 & Second moment in Y (arcmin$^2$) \\
PSFFLUX & Flux density of source as determined from PSF fitting (mJy) \\
PSFFLUX\_ERR & Uncertainty (1 sigma) of PSFFLUX (mJy) \\
GAUFLUX & Flux density of source as determined from 2-D Gaussian fitting (mJy) \\
GAUFLUX\_ERR & Uncertainty (1 sigma) of GAUFLUX (mJy) \\
GAU\_FWHMMAJ & Gaussian fit FWHM along major axis (arcmin) \\
GAU\_FWHMMIN & Gaussian fit FWHM along minor axis (arcmin) \\
GAU\_THETA & Orientation of Gaussian fit major axis \\
\hline
\multicolumn{2}{c}{Quality Assurance} \\
\hline
RELIABILITY & Fraction of MC sources that are matched and have photometric errors $<$ 30\% \\
RELIABILITY\_ERR & Uncertainty (1 sigma) in reliabiliy based on Poisson statistics \\
MCQA\_FLUX\_ERR & Standard deviation of photometric error for 
sources with this SNR \\ 
MCQA\_FLUX\_BIAS & Median photometric error for sources with this SNR \\
BACKGROUND\_RMS & Background point source RMS obtained from threshold  maps (mJy) \\
\hline
\multicolumn{2}{c}{Bandfilling (857 GHz catalogue only)} \\
\hline
BANDFILL217 & 217 GHz Aperture Photometry Flux Density at 857 GHz Source Position (mJy) \\
BANDFILL217\_ERR & Uncertainty in BANDFILL217 \\
BANDFILL353 & 353  GHz Aperture Photometry Flux Density at 857 GHz Source Position (mJy) \\
BANDFILL353\_ERR & Uncertainty in BANDFILL353 \\
BANDFILL545 & 545 GHz Aperture Photometry Flux Density at 857 GHz Source Position (mJy) \\
BANDFILL54\_5ERR & Uncertainty in BANDFILL545 \\
\hline
\end{tabular}
\end{table*}

\subsection{Primary Reliability Selection: Monte-Carlo Analysis}

Quality assessment (QA) is an integral step in the validation of a catalogue.
It helps quantify flux-density biases and flux-density uncertainties, positional errors,
completeness and reliability in a catalogue.
QA metrics based on external (``truth'') catalogues suffer at the brightest flux densities since source numbers
are sparse and resultant QA metrics are dominated by Poisson noise. In addition, generating such a truth
catalogue for the sky from past observational priors,
requires uncertain assumptions about the behaviour of sources across a wide range
of frequencies.
As a result, the Monte-Carlo QA approach is adopted as the primary criterion for selecting high reliability sources.

The goals of the Monte-Carlo QA system are:
\begin{enumerate}
\item To quantify flux-density biases and flux-density uncertainties as a function of background.
\item To quantify completeness in extracted sources as a function of flux density.
\item To quantify contamination or ``spurious sources'' as a function of flux density.
\item To assess positional offsets between extracted and input sources.
\item To assess systematic uncertainties associated with beam shape, gaps in coverage, scan strategy, etc.
\end{enumerate}

The first step of the MC QA analysis is to run the ERCSC pipeline on
the input maps to generate a source
catalogue for the true sky. 
Unresolved point sources, 
convolved with a circularly symmetric
Gaussian with full-width at half maximum identical to that of the 
derived effective beam, are injected into the maps at random positions and with random flux densities ($S_{\nu}$) and re-run the main ERCSC pipeline.
The typical run parameters are 1000 sources per iteration, uniformly distributed
across the sky. In order to minimise Poisson $\sqrt{N}$ variation in our estimates of QA parameters,
while keeping confusion low,
we execute 10 iterations. The present set of runs uses a flat $dN/d\log{S}$ distribution 
at all flux densities ranging from 100 mJy to 100 Jy. We have previously tested Monte-Carlo 
runs where the injected
sources follow a flux-density distribution that is similar to the \Planck\ Sky Model. No
significant differences due to the choice of flux density distribution have been found, particularly because
source extraction in the \Planck\ maps are not significantly affected by source confusion.

We note that to precisely assess the performance of the pipeline, including systematic effects associated
with the generation of the all-sky maps, the artificial sources should be injected into the time-ordered data stream and
processed through each each of the data-processing steps outlined in \citet{planck2011-1.6,planck2011-1.7}. This however, is 
prohibitively expensive in terms of computational resources 
and cannot be accomplished at the present time given the rapid turnaround required for the ERCSC.

At the end of the Monte-Carlo runs, we have one catalogue which only comprises the sources
detected in the original map and 10 catalogues which have the original sources in addition
to the detected fraction of the fake sources that were injected into the maps.
We first match the sources in the original map to each of the remaining 10 catalogues
with a matching threshold of twice the FWHM. This leaves only the artificial and spurious sources in the catalogues,
whose
properties can then be compared to the known flux densities and positions of the injected sources.

Reliability specifies the
fraction of extracted sources that differ from their input flux densities to within 30\%. 
This is based on the flux density accuracy requirement for the ERCSC. 
Imposing the requirement implies
that the catalogue is equivalent to a catalogue with a $>$5$\sigma$ cut if the noise were Gaussian. That is, a typical 5$\sigma$ source
would have a flux density error that is smaller than 20\%, 68\% of the time, which translates to a flux density error of $<$30\% for 90\% of the sources,
for a Gaussian distribution of errors. It is well known that the 
contribution from the Galaxy and the CMB results in a non-Gaussian distribution
for the background RMS, at least on large spatial scales. Future work will attempt to build upon our increased knowledge 
of the foregrounds from the \Planck\ maps and undertake a more precise characterization of the noise.

The reliability
is measured as a function of root-mean-square (RMS) signal-to-background where the signal is a measure of the
flux density of the source and therefore either FLUX or FLUXDET. The background RMS
is derived from the RMS measured in a 2$\degr$ radius annulus on the maps after individual detected sources are masked.
The choice of 2$\degr$ was made empirically. 
It was found that
if the outer radius were too small, i.e., tens of minutes of arc, the RMS was similar to the RMS returned by the detection algorithms
which detect
sources as peaks above the local RMS. These RMS returned by the codes are typically lower than the RMS measured in the 
larger annulus used here.
If the outer radius were too large (several degrees), background structure
gets smoothed out. A radius of 2$\degr$ represents a trade-off between these two extremes and yields a background RMS which is a combination
 of substructure in the background and the instrumental noise in the maps. The RMS is converted to a 1$\sigma$ background
RMS for a point source, by integrating over the \Planck\ beam.

\begin{figure*}
\includegraphics[width=9cm]{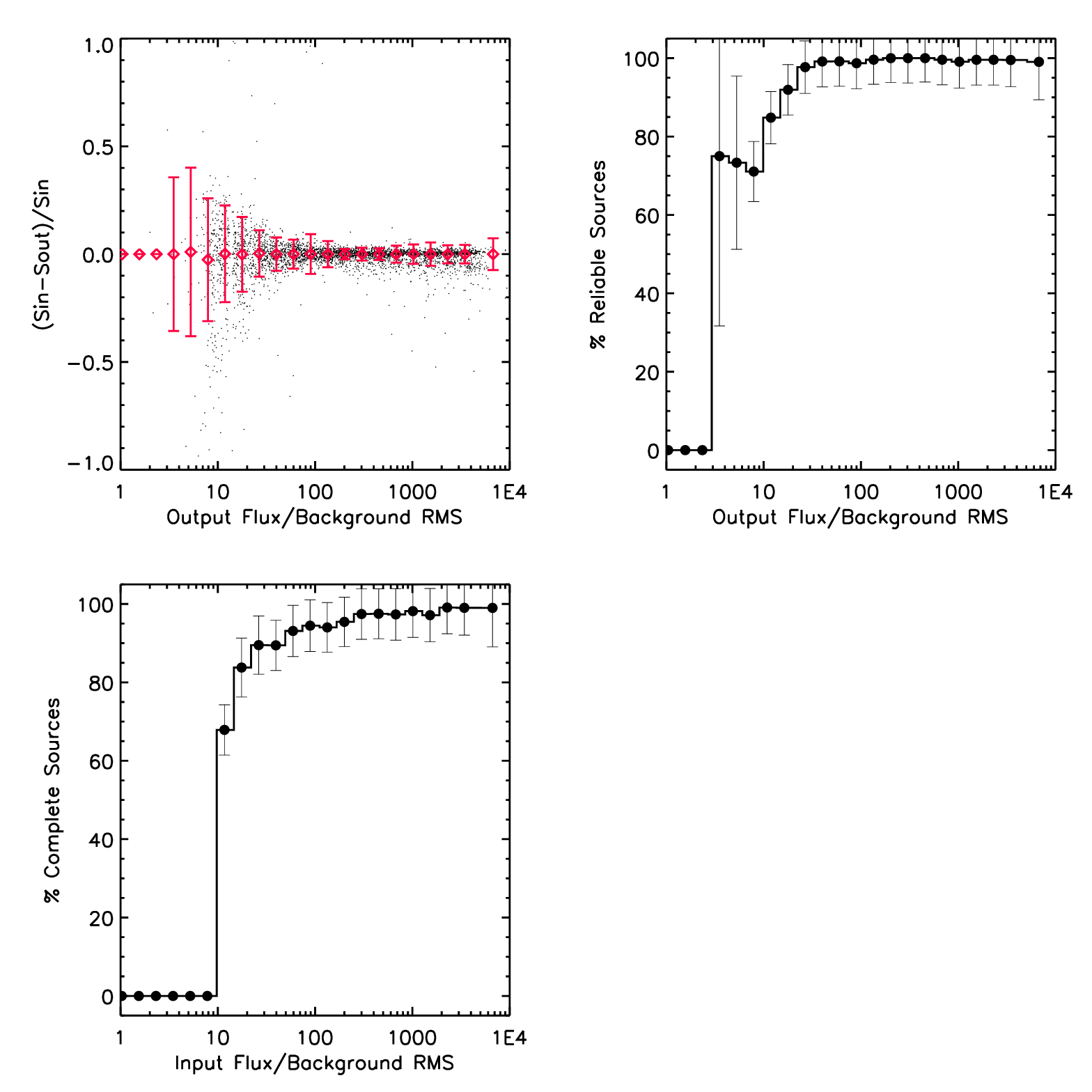}
\includegraphics[width=9cm]{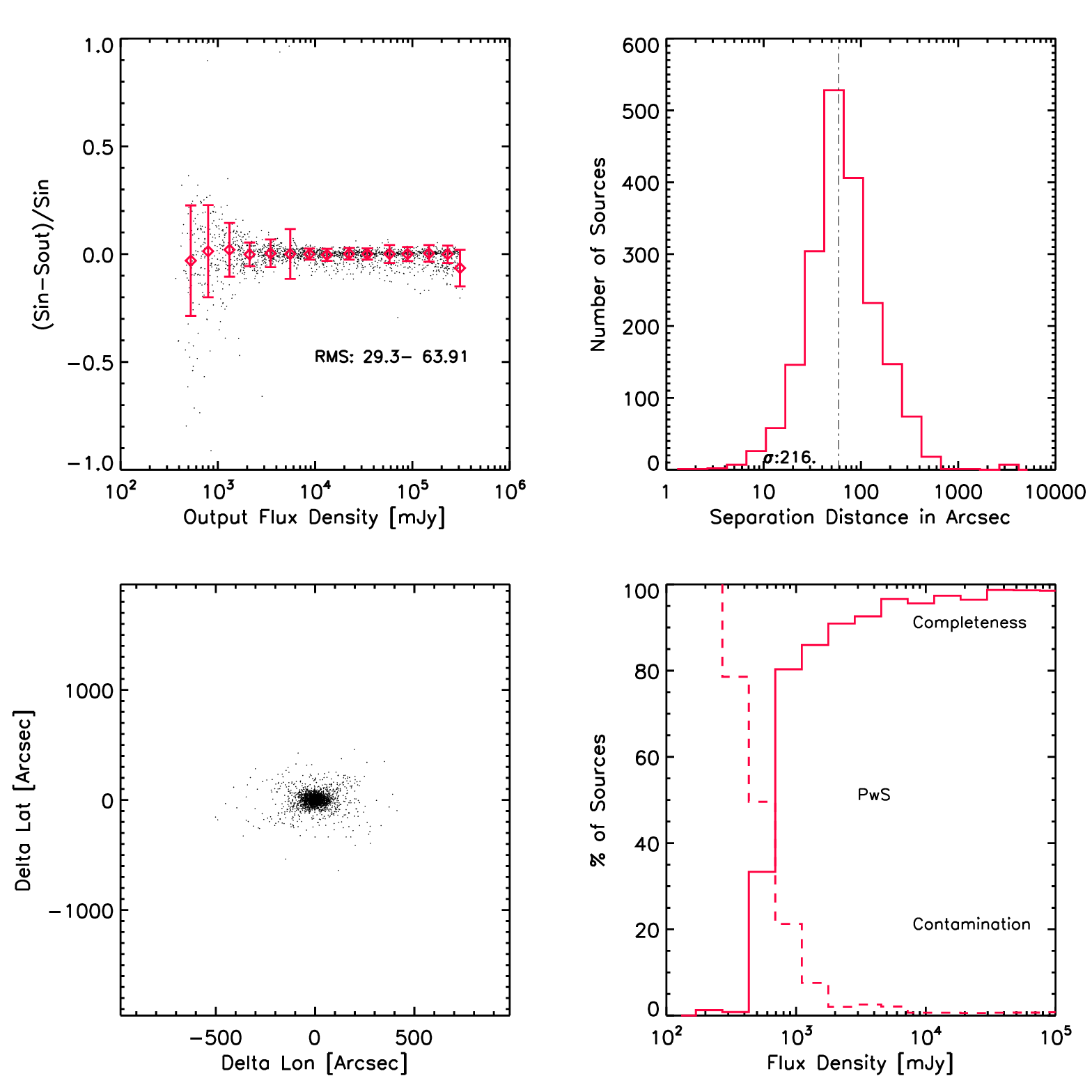}
\includegraphics[width=9cm]{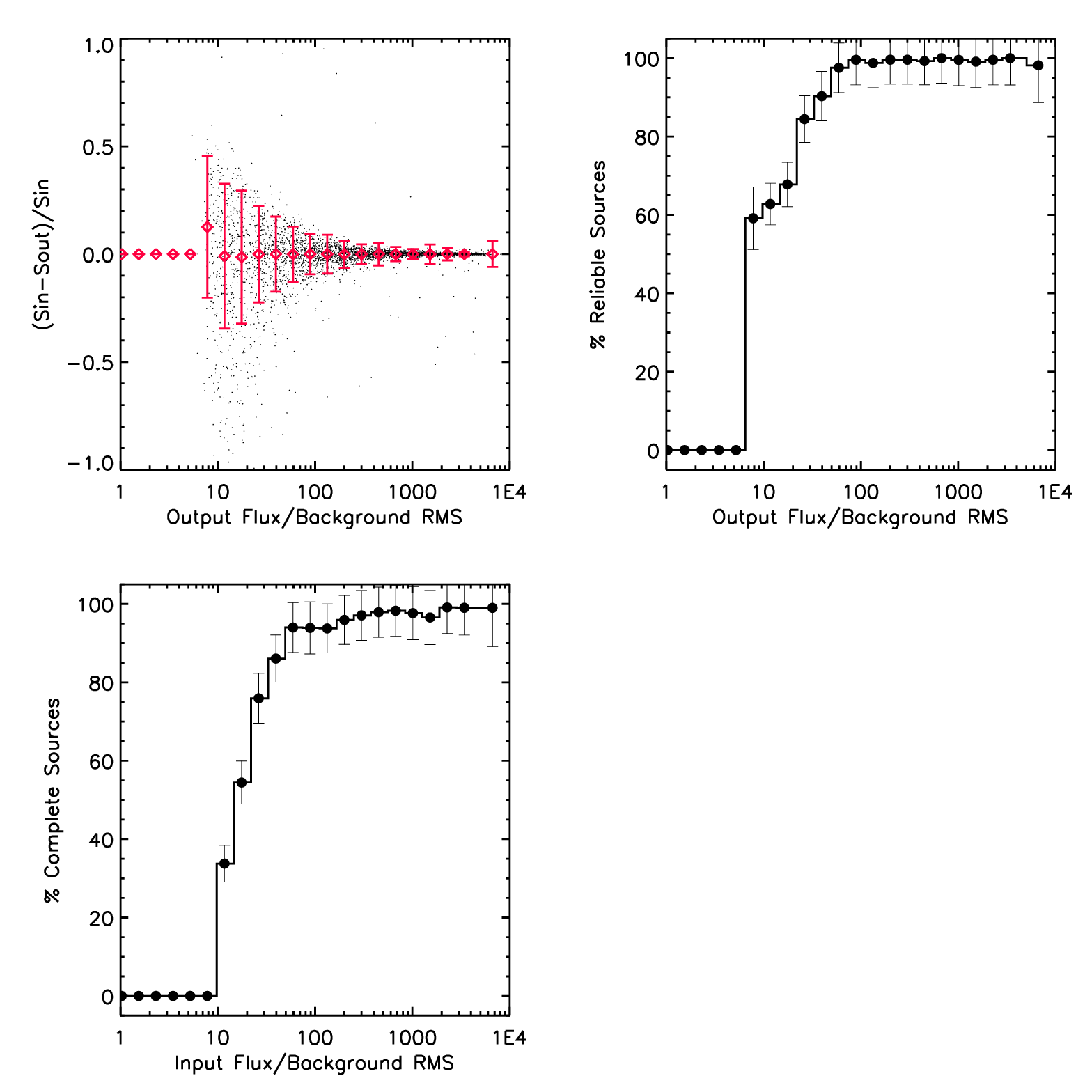}
\includegraphics[width=9cm]{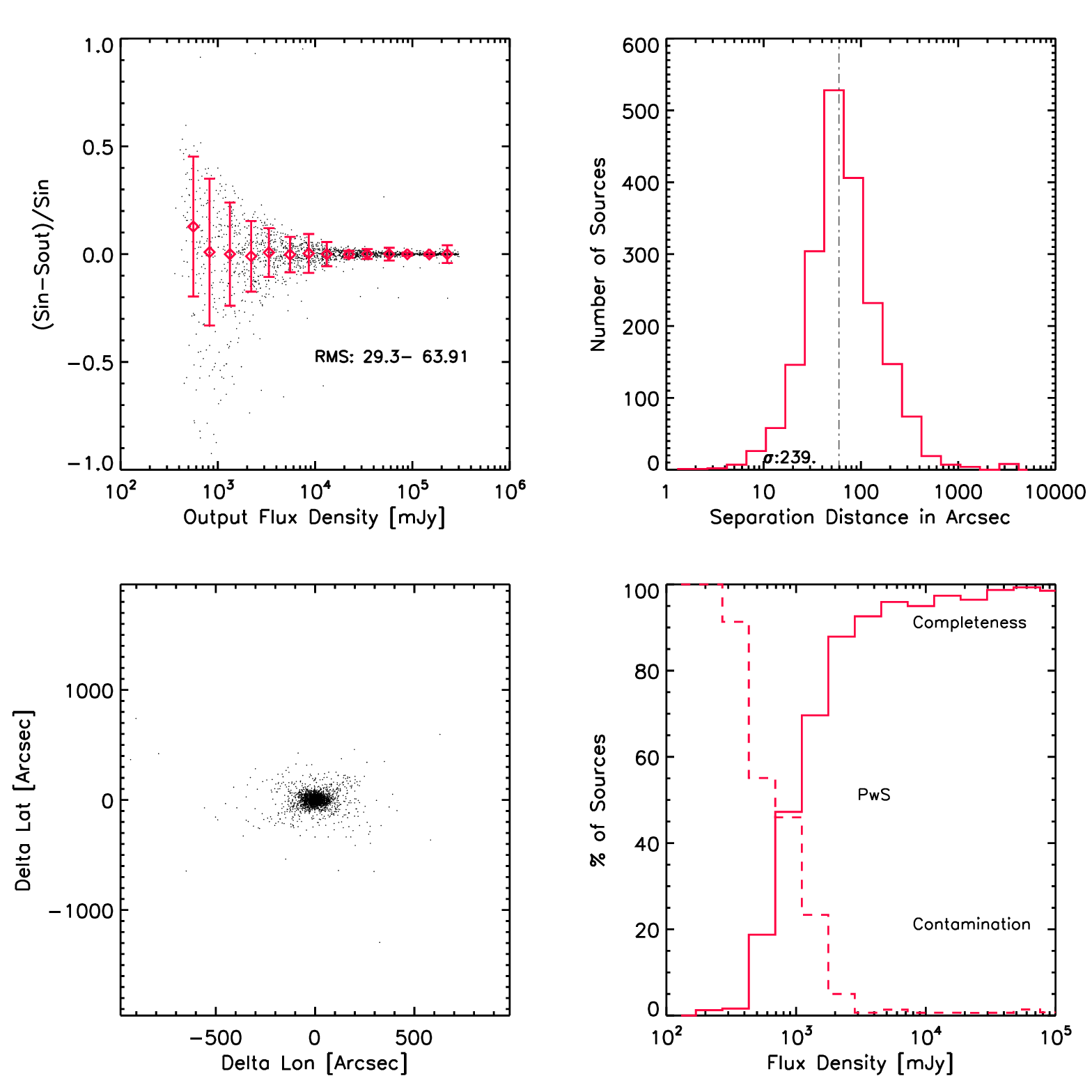}
\caption[Monte-Carlo analysis at 30\,GHz]{
Plots showing the results of the Monte-Carlo analysis at 30\,GHz with the PwS algorithm.
The upper two rows shows the results when the flux density of sources is from FLUXDET while
the lower two rows shows the results when the flux density of sources is from FLUX.
The set of 3 plots in the top-left and bottom-left corner
show the all-sky flux-density uncertainty, differential reliability and differential
completeness of the Monte-Carlo sources as a function of SNR where signal may be FLUX or FLUXDET and the noise
is the background RMS. The set of four plots at the top right and bottom right
show (left to right, top to bottom) the fractional flux density uncertainty, $(S_\textrm{in}-S_\textrm{out})/S_\textrm{in}$ (see Sect.~\ref{sec:catalogue}), the distribution of the absolute positional offset, 
differential positional offset, as well as completeness and contamination ($1-$reliability converted to a percentage) as a function of flux density for the half of 
the sky with the lowest sky background RMS. The range of sky background RMS converted to a point
source flux density uncertainty, is shown in the inset in mJy.
The primary source selections in the catalogues are based on the reliability vs output flux density/background RMS plots such
that the cumulative reliability (integral of the differential reliability) is greater than 90\%. 
}
\label{fig:mcres}
\end{figure*}

Figure~\ref{fig:mcres} shows the flux-density accuracy, the positional accuracy and
the differential reliability as a function of SNR at 30\,GHz based on the artificially
injected Monte-Carlo sources. 
The differential reliability is simply the reliability in each bin of SNR while the cumulative reliability
is the integral of the differential reliability above a particular SNR value.
Also shown in Figure~\ref{fig:mcres} is the flux density accuracy, positional uncertainty,
completeness and reliability as a function of flux density for the half of the sky with the lowest sky background RMS.
Due to the differences between how FLUX and FLUXDET are estimated, the top panels and lower panels are not the same.
First, FLUXDET flux densities have a larger scatter at almost all SNR ratios
compared to FLUX. This is partly due to the prior assumptions on the source profile that are made in the estimation of FLUXDET.
Deviations from this assumed source profile result in errors in the derived flux density.

Second, the FLUXDET reliabilities appear to be higher at low SNR compared to the FLUX based reliability values.
At low values, the aperture flux-density based FLUX estimates tend to become increasingly affected by sky noise and even the background
estimation becomes more uncertain. FLUXDET values are derived assuming a fixed source shape and a flat background. As a result, for point sources,
the scatter in FLUXDET values at low SNR is smaller. When the error in the flux density estimate exceeds 30\%, the reliability decreases.
We note that the uncertainty on the reliability estimate is dominated by the Poisson statistics of the number of sources in the corresponding
SNR bins. That is, if the completeness in a particular SNR bin is low, the uncertainty on the reliability is high. The typical uncertainty
on the reliability estimate in a particular SNR bin is about 5\%. 

Also shown in Figure~\ref{fig:mcres} is the histogram of separations between the injected and extracted positions. The 1$\sigma$ positional offset
is approximately $230\,\textrm{arcsec}$ in low background regions, which is almost FWHM/10 at 30\,GHz.

The completeness plots for the Monte-Carlo point sources
are shown for illustrative purposes for the half of the sky with the lowest background RMS. Without factoring in the source
size distribution and flux density distribution of the real source populations, as well as the 
fraction of sky observed with a particular amount of exposure time,
the completeness plot cannot be used to directly infer the actual completeness of the ERCSC.

\begin{figure*}
\includegraphics[width=9cm]{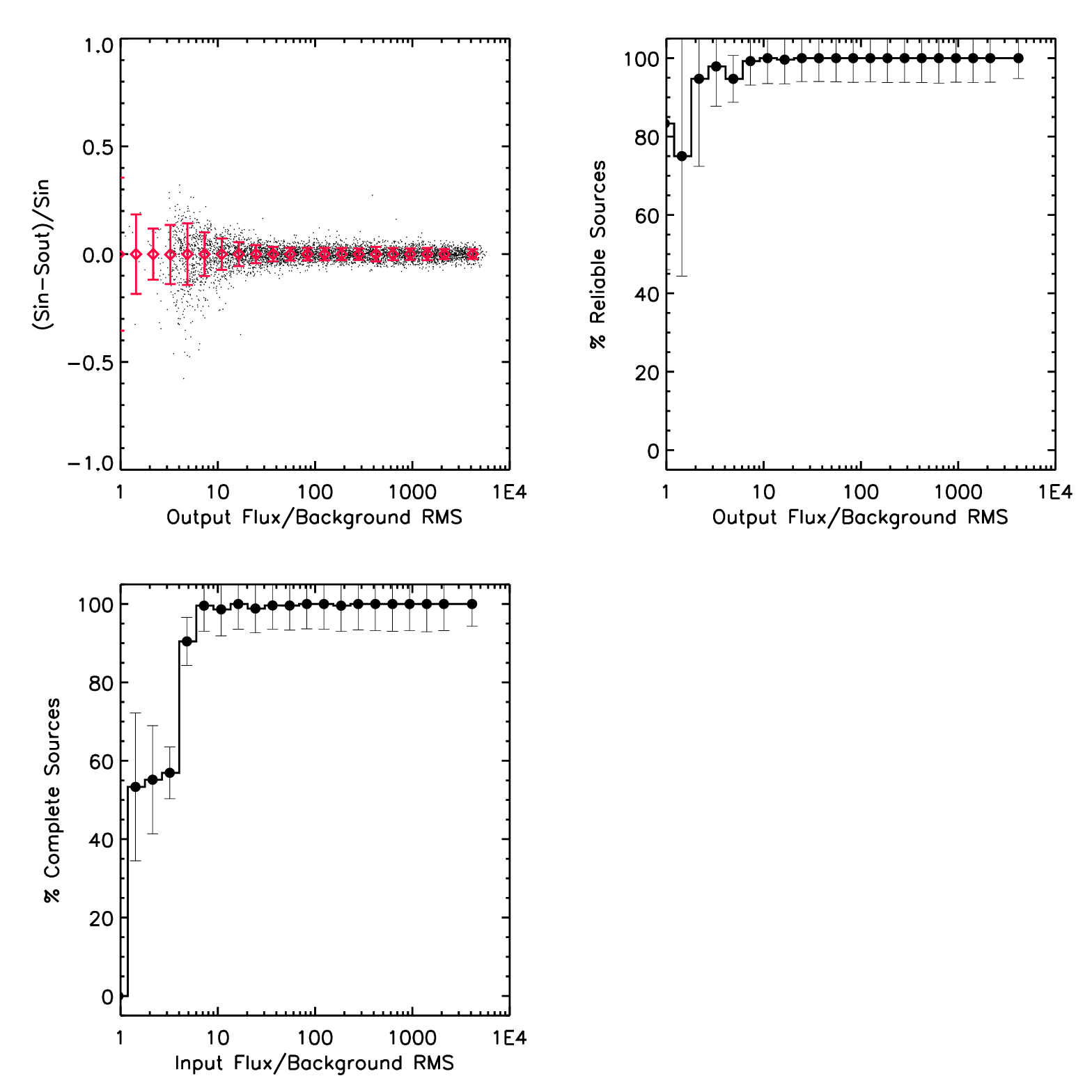}
\includegraphics[width=9cm]{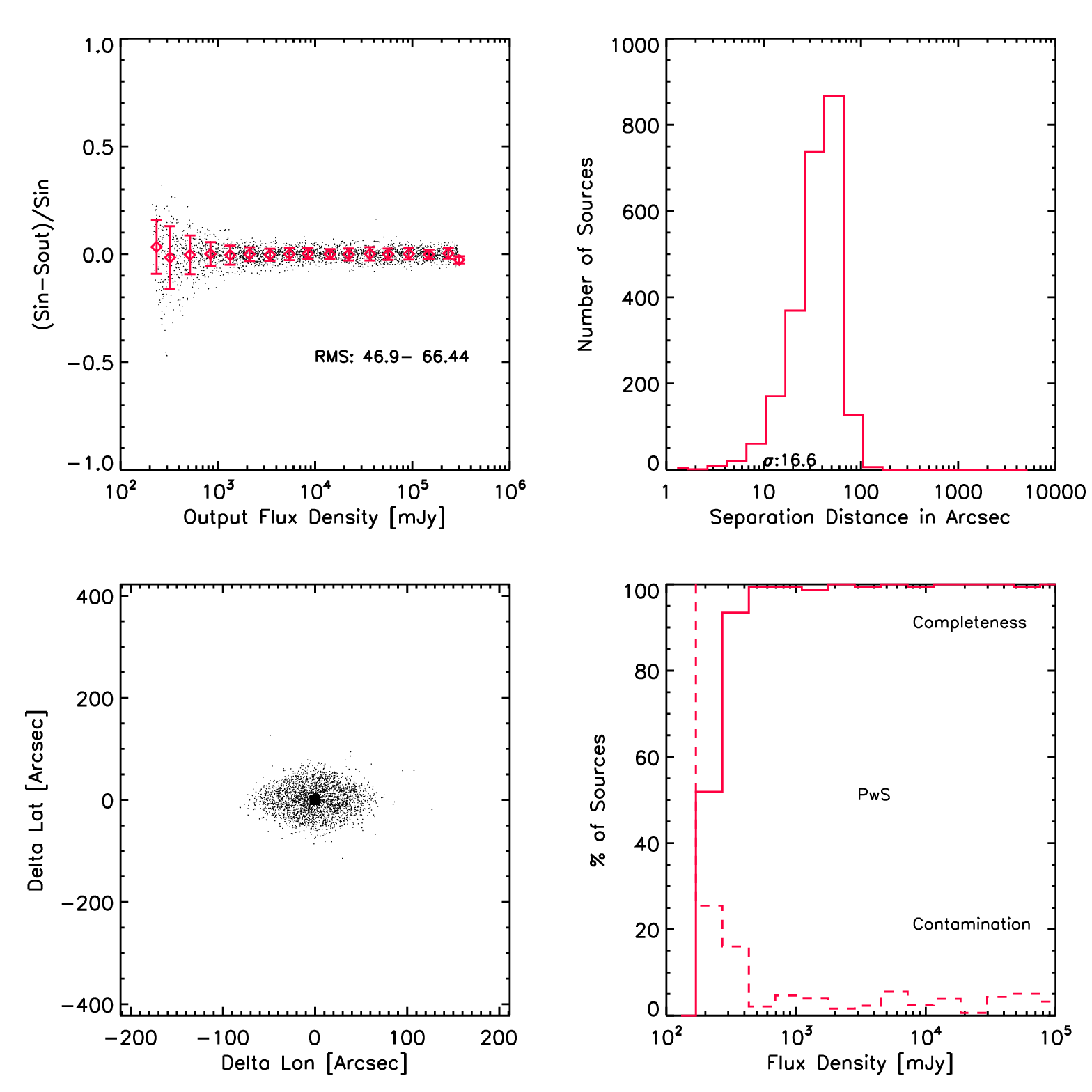}
\includegraphics[width=9cm]{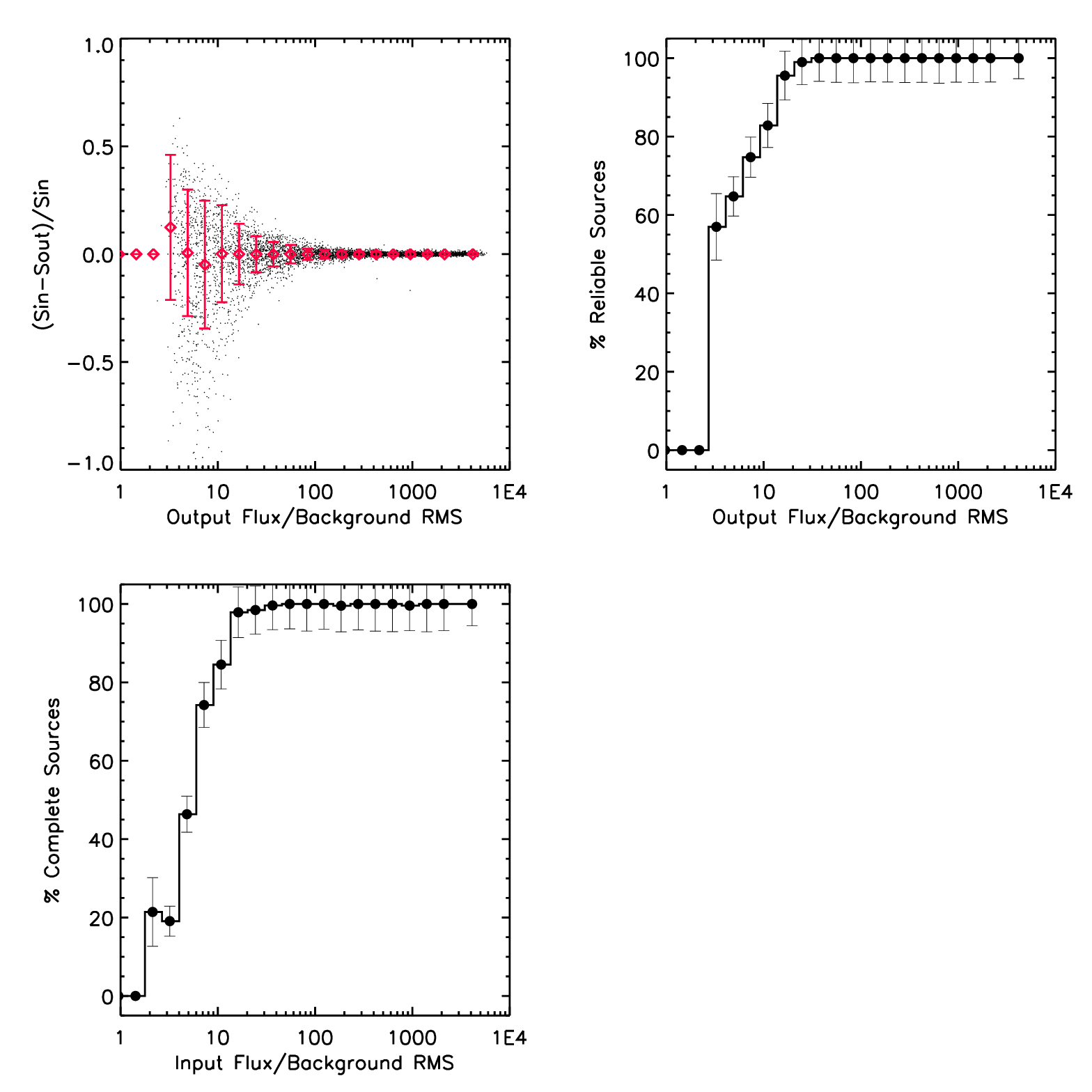}
\includegraphics[width=9cm]{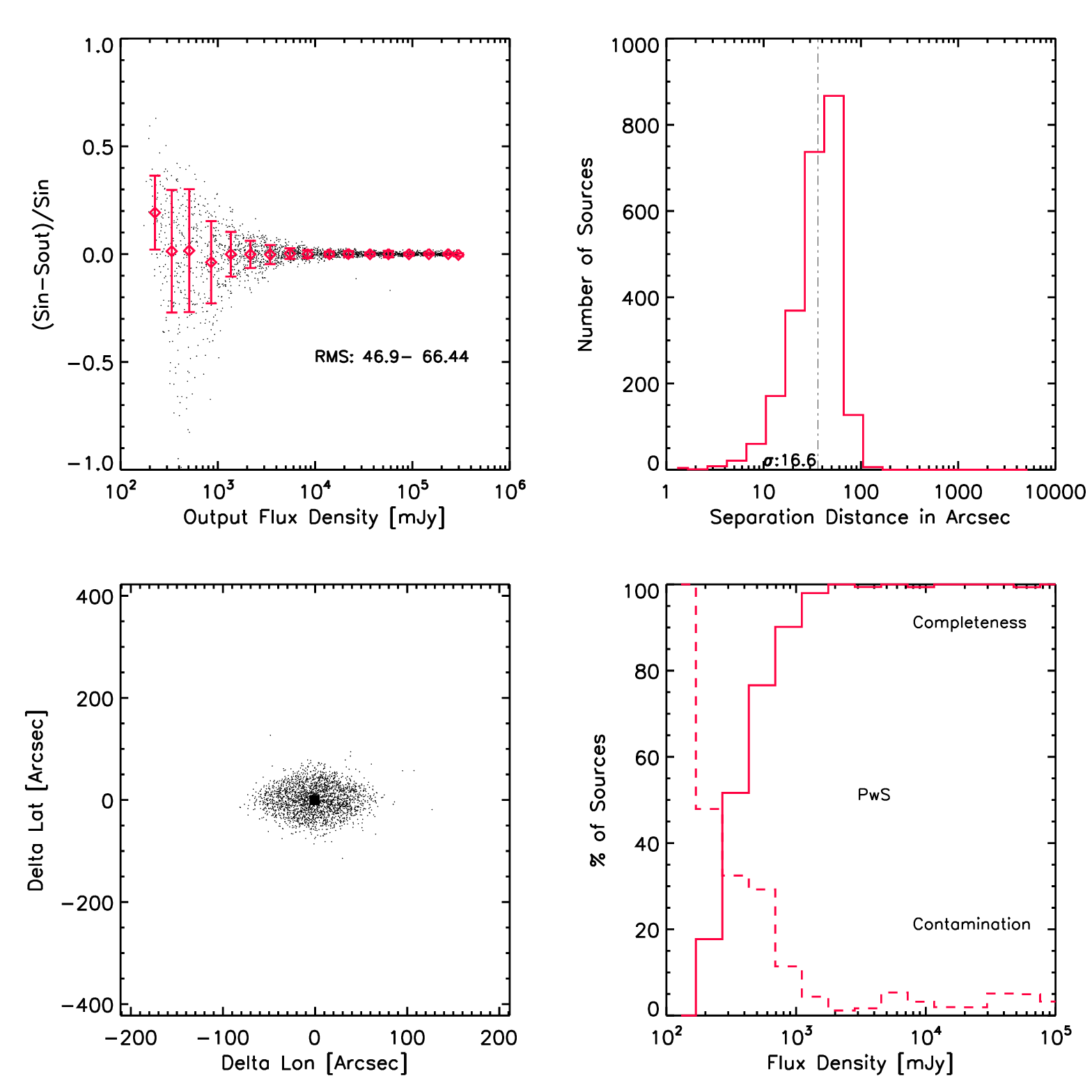}
\caption[Monte-Carlo analysis at 143\,GHz]{
As in Figure~\ref{fig:mcres} but at 143\,GHz with the PwS algorithm.
}
\label{fig:mcres143}
\end{figure*}

\begin{figure*}
\includegraphics[width=9cm]{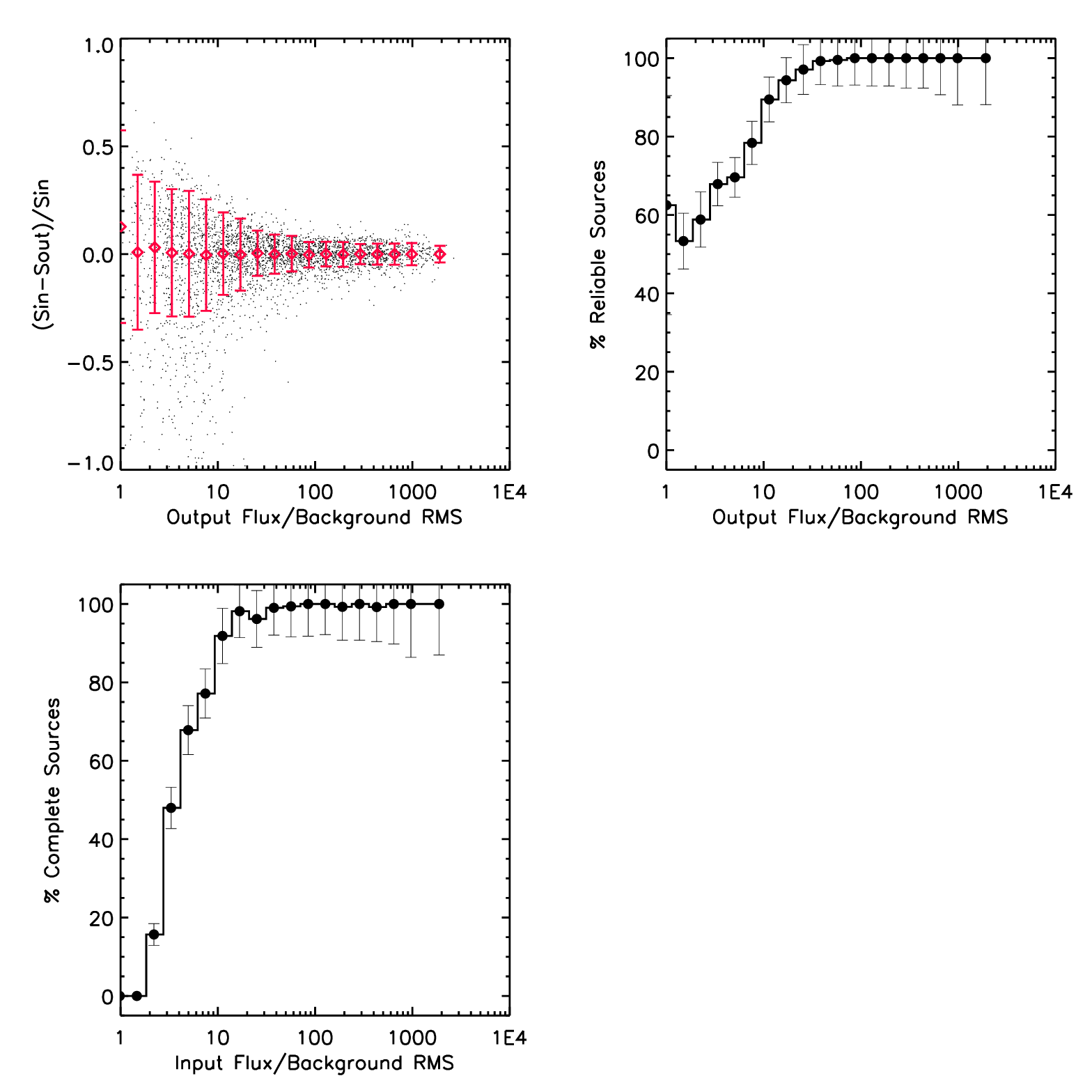}
\includegraphics[width=9cm]{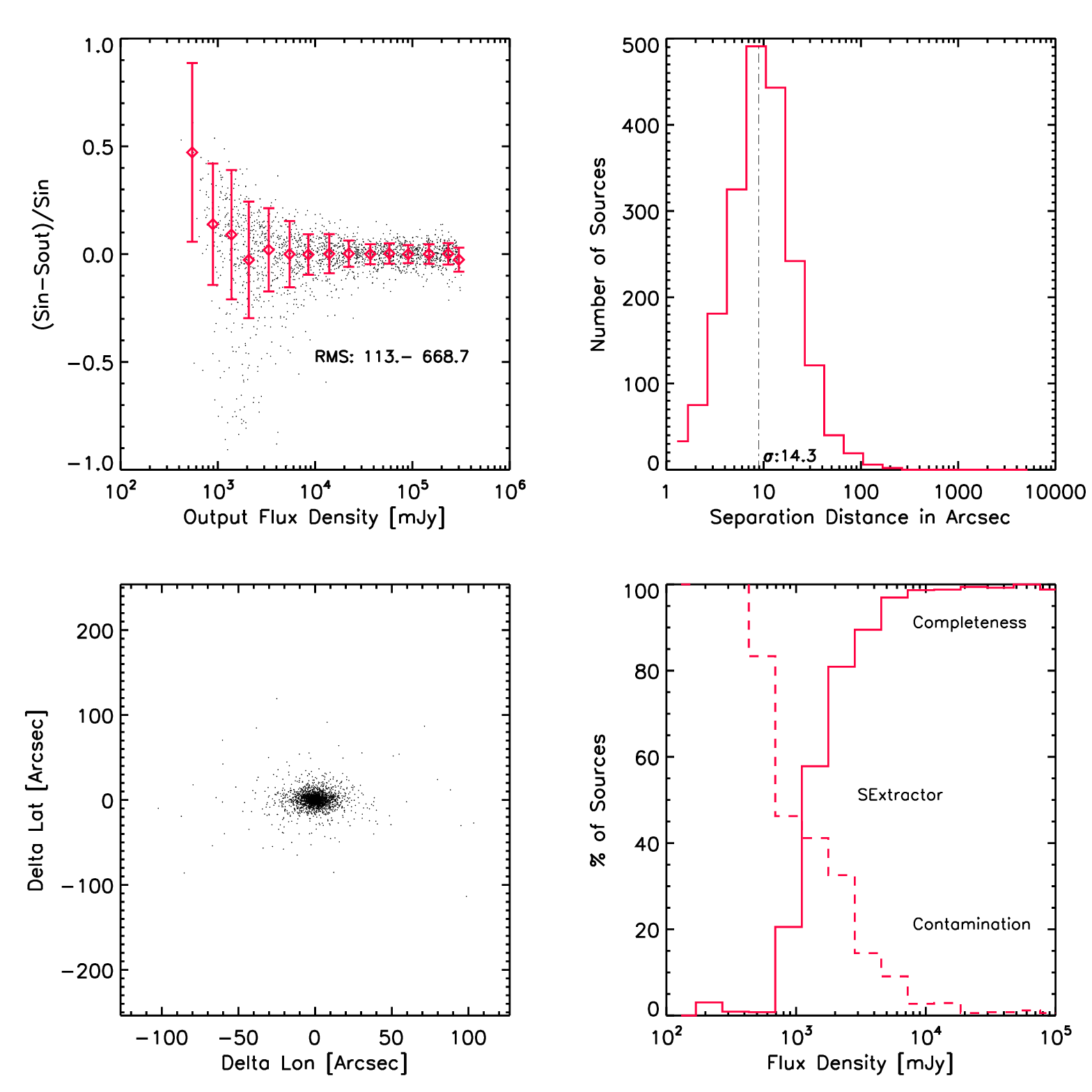}
\includegraphics[width=9cm]{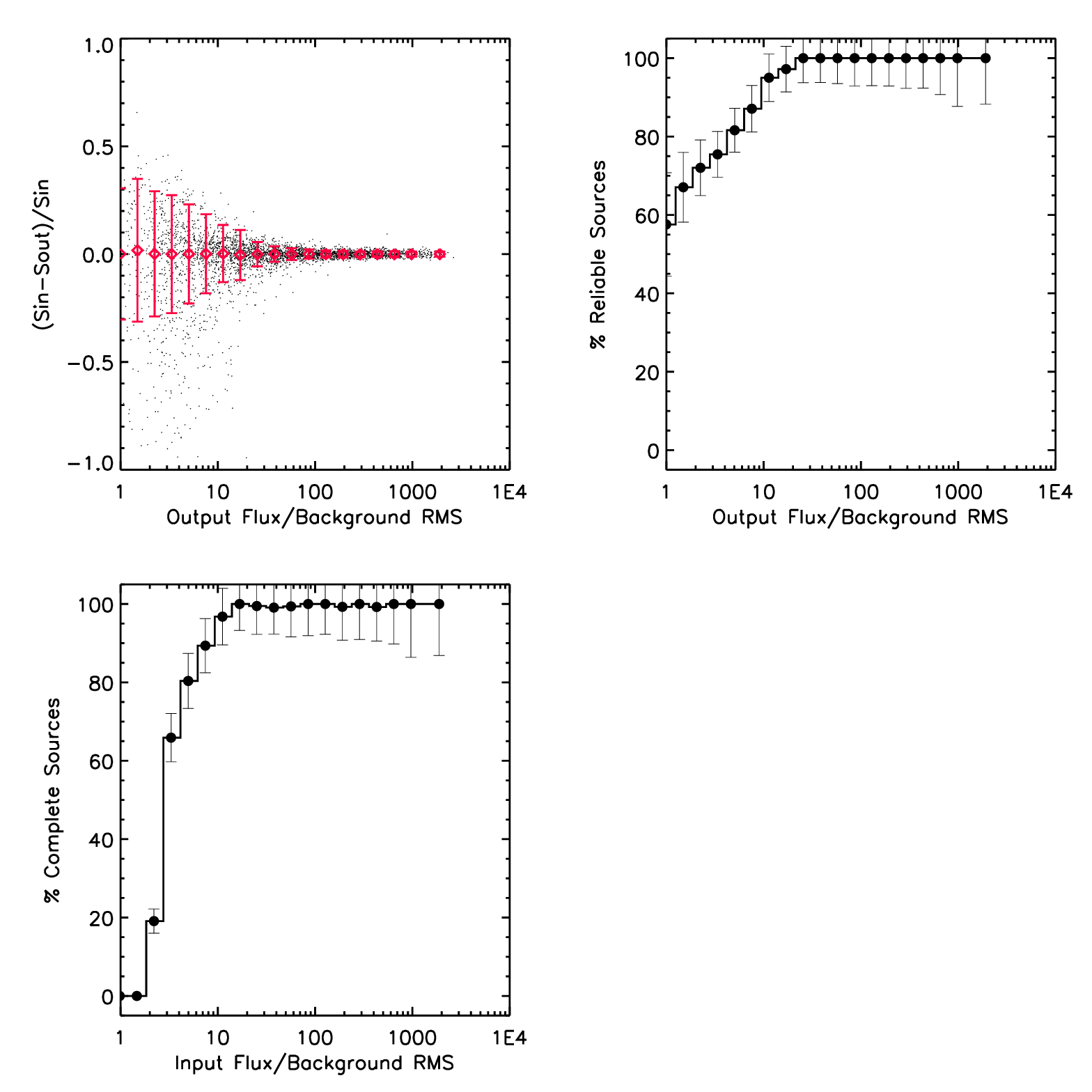}
\includegraphics[width=9cm]{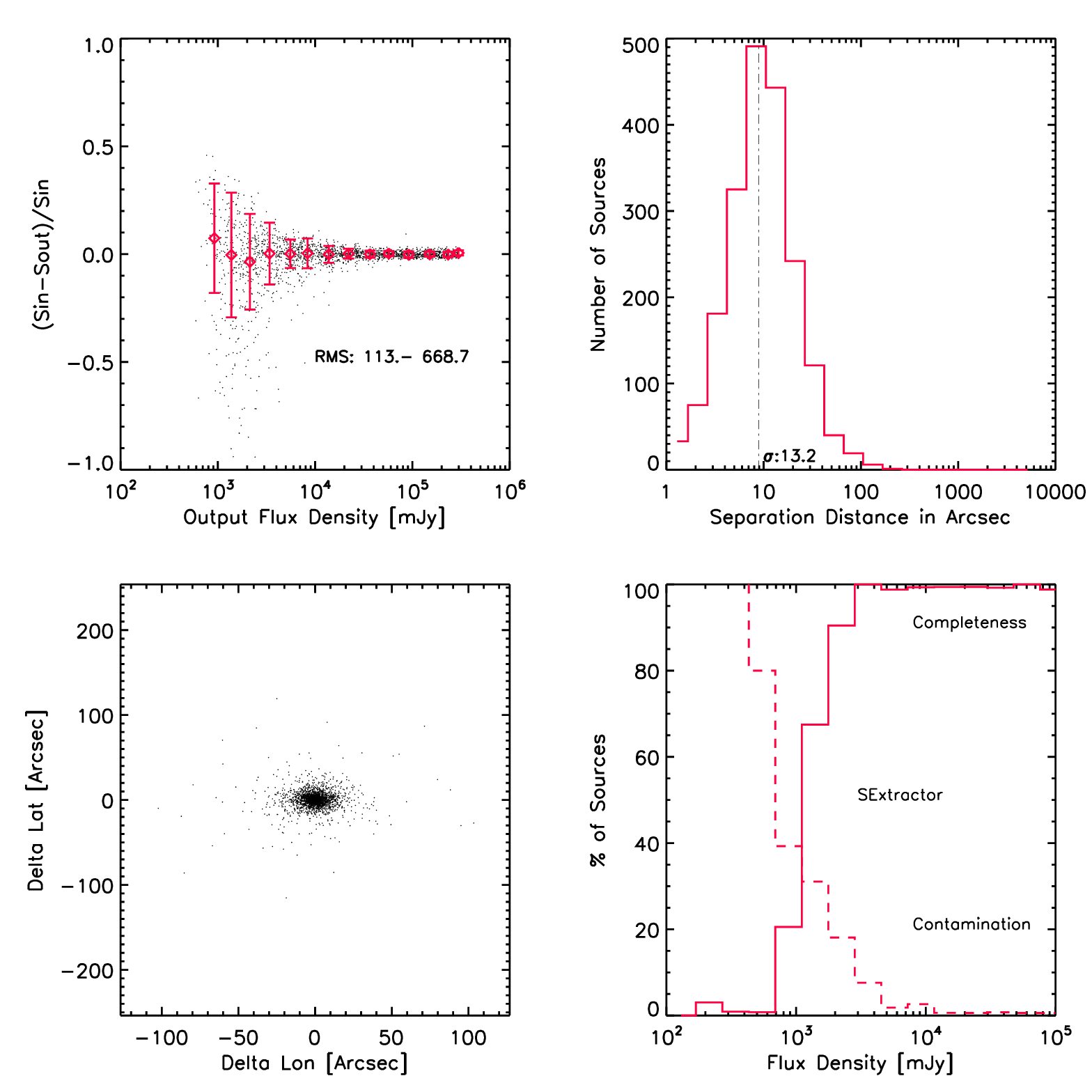}
\caption[Monte-Carlo analysis at 143\,GHz]{
As in Figure~\ref{fig:mcres} but at 857\,GHz with the SExtractor algorithm.
}
\label{fig:mcres857}
\end{figure*}

Figures~\ref{fig:mcres143} and \ref{fig:mcres857} show similar plots for the Monte-Carlo sources at 143 and 857\,GHz.
Interesting trends can be observed by comparing these plots. The obvious one is an improvement in positional accuracy 
with increasing
frequency due to the spatial resolution of \Planck\ improving with increasing frequency. Another interesting trend is the evolution in the range of background RMS
values for the cleanest half of the sky, which is provided in mJy in the numbers following ``RMS:''. The numbers indicate that the
background RMS is the largest at 857\,GHz due to the enhanced contribution of ISM emission. 

The SNR values of the real sources are then estimated from the ratio of FLUX/Background RMS or FLUXDET/Background RMS.
The reliability of the Monte-Carlo sources shown in these Figures is applied to the real sources using the SNR value as the comparison metric.
These reliability values are between 0 and 1 although the minimum over all frequencies
after the cumulative reliability cuts are applied is 0.74. If an 
arbitrary source has a reliability of 0.74, it implies that 74\% of the time, a source lying in a patch of sky with similar sky noise
will have an estimated flux density that is accurate to within 30\%.

Once the differential reliability of each source in the original map has been estimated, the sources are sorted
in decreasing order of SNR. The differential reliability is converted to a cumulative reliability by integrating the
differential reliability over increasing SNR values.
We imposed a cumulative reliability threshold of 90\% and a maximum standard deviation in the reliability of 10\%
for the ERCSC.  This is the primary criterion used to select high reliability sources. 
The reliability
cut is applied to both the FLUX/Background RMS as well as the FLUXDET/Background RMS, since these are two
distinct measures of flux density and the resultant catalogue is the union of the two reliability cuts.
The union is selected to maximise source counts since different measures of flux density tend to be more accurate in different regimes
as described in Sect.~\ref{sec:catalogue}.

The technique that is chosen at each frequency is the one that returns the maximum number
of $|b|>30\degr$ sources above a cumulative reliability of 90\%. These happen to be
our particular implementation of PwS between $30$ and $143$\,GHz and SExtractor between $217$ and $857$\,GHz.

\subsection{Secondary Cuts in Selection of Sources}

The primary selection criteria described above have been augmented by a set of secondary cuts 
which take into account known source artefacts in the maps.

First, the transit of bright sources (especially planets) across the beam results in a pattern of bright and dark patches
that is repeated every 36\,arcmin along the scan pattern,
in the upper HFI bands \citep{planck2011-1.7}. These are due to the imprecise removal of an instrumental 
artefact (the 4K cooler spectral line).
A subset of these patterns have been visually identified in the maps and masks have
been generated for those patches of sky. These masks are reflected in the incomplete sky coverage in Table \ref{tbl-ercsc}.
If more than 5\% of the pixels within one FWHM from the source fall on the mask, the source is rejected.

Second, there are known gaps in the maps associated with the masking of planets and asteroids\footnote{The following objects
have been masked in the map making. Asteroids: 10 Hygiea, 11 Parthenope, 128 Nemesis, 12 Victoria, 13 Egeria, 14 Irene, 
15 Eunomia, 16 Psyche, 
18 Melpomene, 19 Fortuna, 1 Ceres, 20 Massalia, 29 Amphitrite, 2 Pallas, 324 Bamberga, 3 Juno, 41 Daphne, 
45 Eugenia, 4 Vesta, 511 Davida, 52 Europa, 704 Interamnia, 7 Iris, 88 Thisbe, 8 Flora, 9 Metis.
Comets: Broughton, Cardinal, Christensen, d'Arrest, Encke, Garradd, Gunn, Hartley 2, Holmes, Howell, 
Kopff, Kushida, LINEAR, Lulin, McNaught, NEAT, Shoemaker-Levy 4, SidingSpring, Tempel 2, Wild 2.
}. If sources
have any of their pixels within one FWHM falling on such a gap, the source is rejected.
This prevents edge effects due to the side lobes of bright planets from being classified as sources and also prevents
the introduction of large errors in the photometry of sources.

Third, sources are also required to have either an aperture-photometry $\mathrm{SNR}\ge 5$ ($\mathrm{FLUX}/\mathrm{FLUX\_ERR}\ge 5$)
or a detection method photometry $\mathrm{SNR}\ge 5$ ($\mathrm{FLUX}/\mathrm{FLUXDET\_ERR}\ge 5$). The distinction
is important, due to the fact that the photometry from the PowellSnakes implementation
consistently underestimates the flux density for even marginally extended sources at
the lower frequencies.

Fourth, due to the requirements on the flux-density accuracy in the ERCSC, 
the standard deviation in the photometric error for the artificial sources with the same SNR as the real source is required to be less than 30\%. 

Fifth, in order to remove extended sources associated with substructure in the Galactic ISM, we 
eliminate non-circular sources (ELONGATION $\le$3) in the upper HFI bands. These are sources whose
ratio of major to minor axis is greater than three.

As a sixth criterion, we also insist that the aperture flux density is positive (APERFLUX $\ge$ 0), which alleviates
problems due to sources whose sky background estimate is biased high by the presence of bright sources in the sky annulus.
These sources will have uncertain photometry and are therefore rejected.

The final ERCSC compilation is the list of sources which have satisfied the primary Monte-Carlo based reliability criterion
as well as all the aforementioned secondary QA criteria.
These cuts imply that about half the sources in the uncut lower frequency
catalogues and about a third of the sources in the upper frequency catalogues are classified as
high reliability sources.

As mentioned earlier, each source has four different measures of flux density associated with it.
These flux density values have not been colour corrected. Users should 
identify appropriate colour corrections from \citet{planck2011-1.10sup} and apply them
to the flux densities. The absolute calibration uncertainty of the HFI and LFI instruments is better than
7\% at all frequencies \citep{planck2011-1.6,planck2011-1.7}.
However, the requirements on the ERCSC are a photometric accuracy of 30\%.

\section{Characteristics of the ERCSC}
\subsection{Sky Coverage and Sensitivity}

\begin{figure*}
\centering
\includegraphics{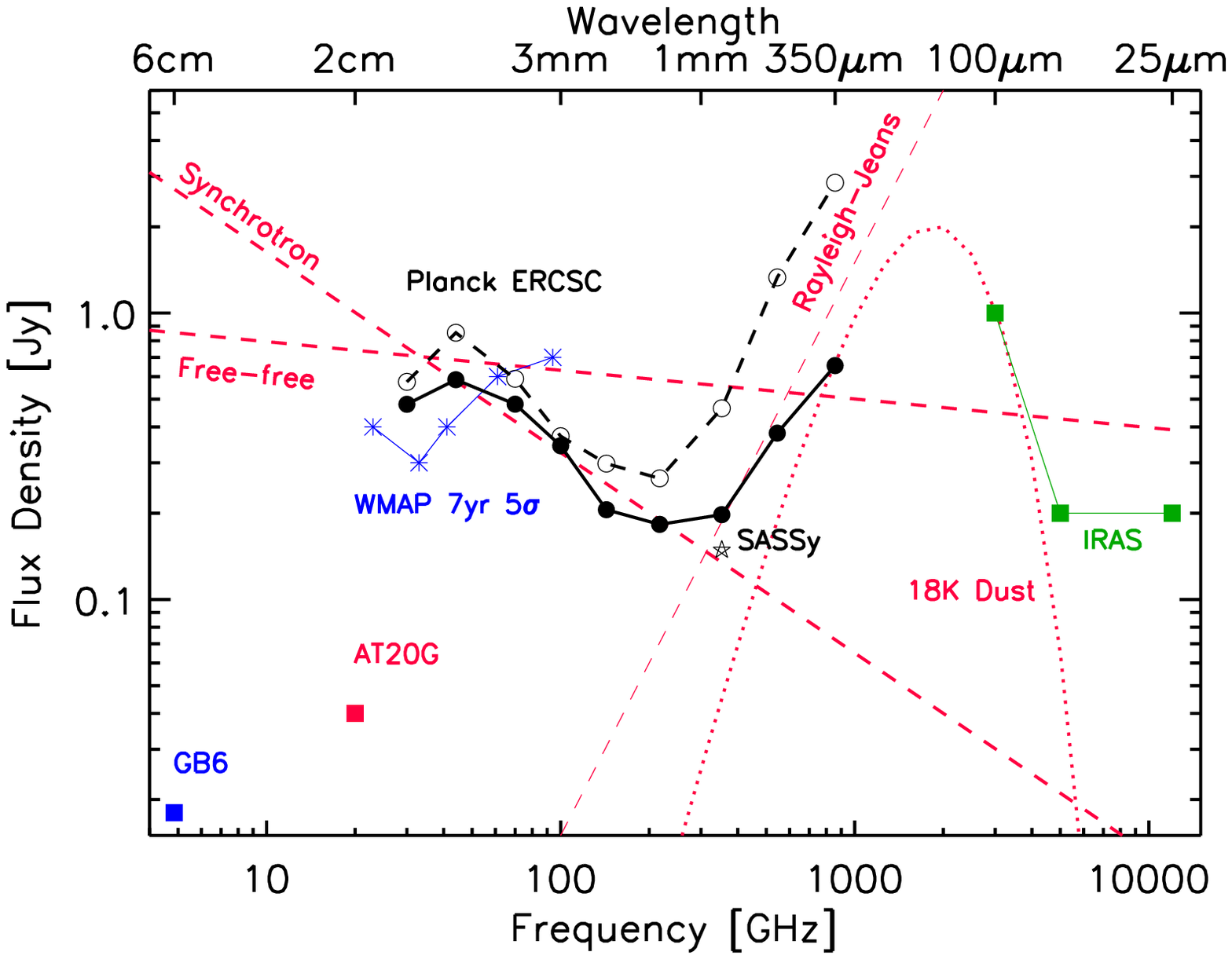}
\caption[Sensitivity of \Planck\ compared to other wide area surveys]{
The \Planck\ ERCSC flux density limit quantified as the faintest ERCSC source
at $|b|<10\degr$ (dashed black line) and at $|b|>30\degr$ (solid black line)
is shown relative to other wide area surveys. 
Also shown are the spectra of known sources of foreground emission as
red lines; these include a S$_{\nu}\sim\nu^{-0.7}$ synchrotron component, $\nu^{-0.1}$ free-free component, a 
Rayleigh-Jeans component and
a $\nu^{2}$ emissivity blackbody of temperature 18\,K. The ERCSC sensitivity is worse in the Galactic Plane due to the strong contribution of ISM emission especially
at submillimetre wavelengths. In the radio regime, the effect is smaller. The faintest 
{\it WMAP} 7 year 5$\sigma$ sources are derived from the 
catalogue of \citet{gold2010, wright2009}. Although the flux density limits of {\it WMAP} and \Planck\ appear to be comparable at the lowest
frequencies, the \Planck\ ERCSC is more complete as discussed in Section \ref{sec:valid}.
The GB6 sensitivity value is from \citet{greg}, AT20G flux limit from \citet{mur}, SCUBA-2 All Sky Survey (SASSy) 
limit from the Joint Astronomy Center website while the {\it IRAS} flux density limits are
from the {\it IRAS} explanatory supplement \citep{beichman}.
}
\label{fig:sens}
\end{figure*}

\begin{figure*}
\includegraphics[width=9cm]{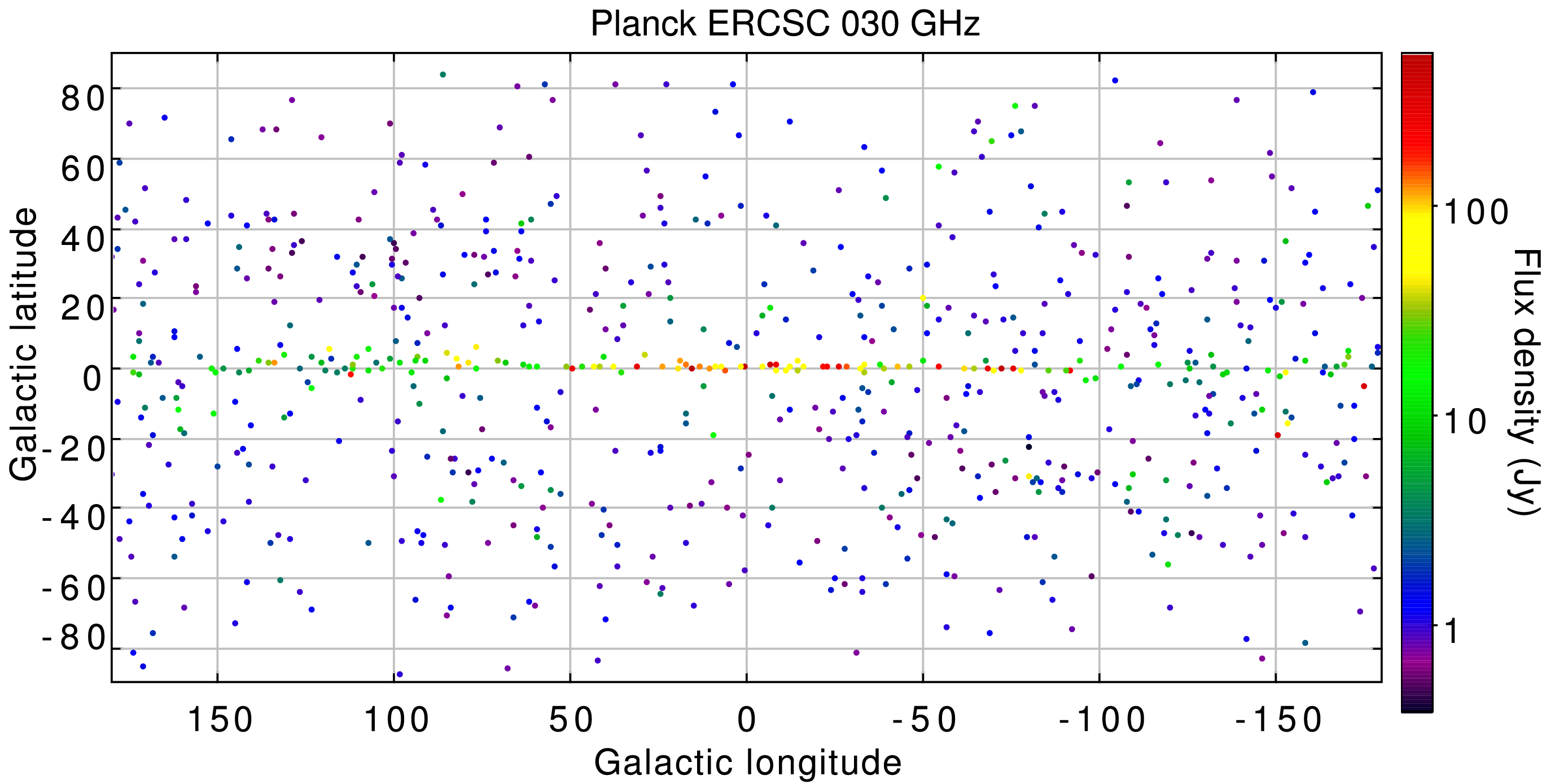}
\includegraphics[width=9cm]{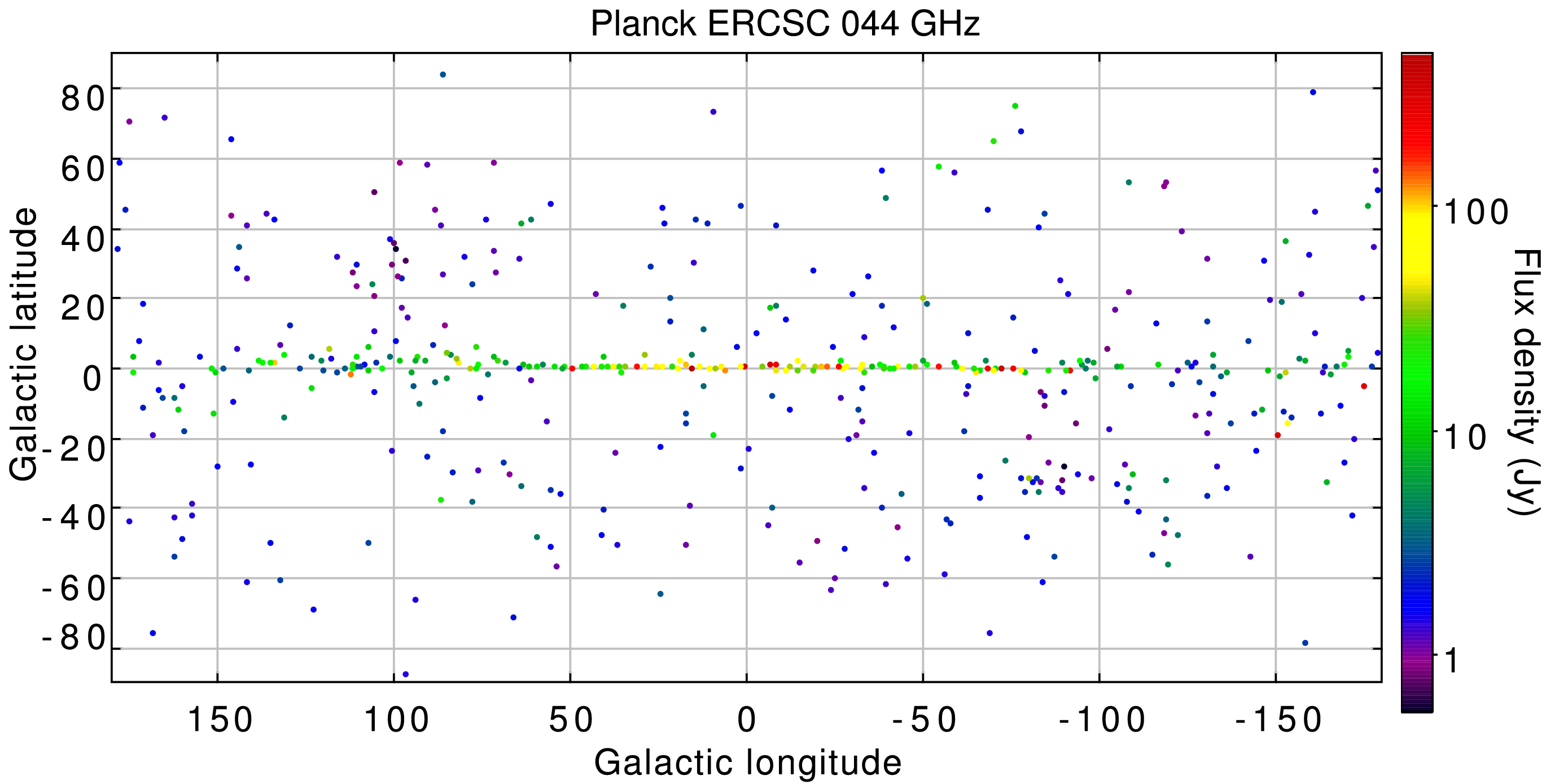}
\includegraphics[width=9cm]{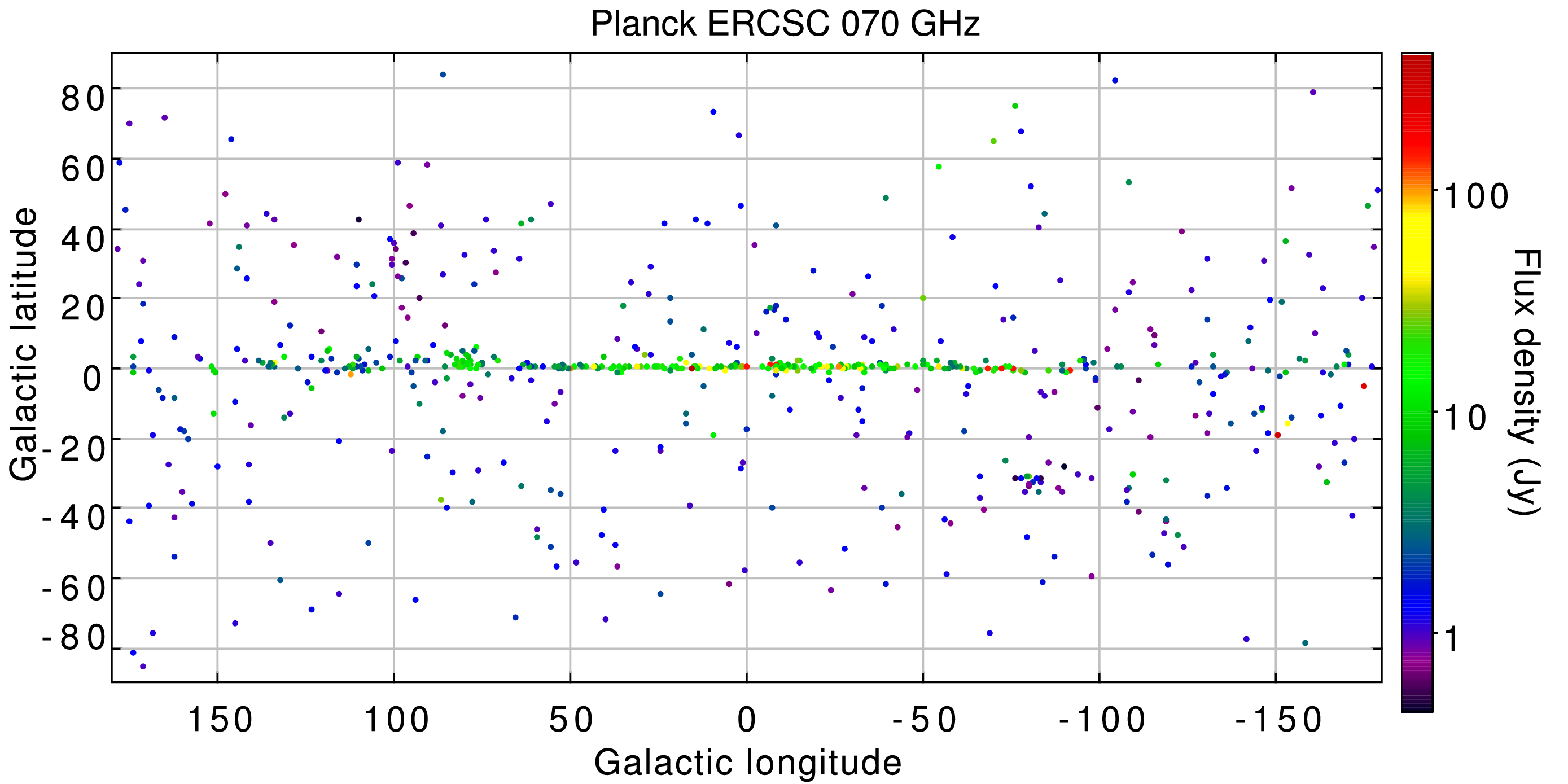}
\includegraphics[width=9cm]{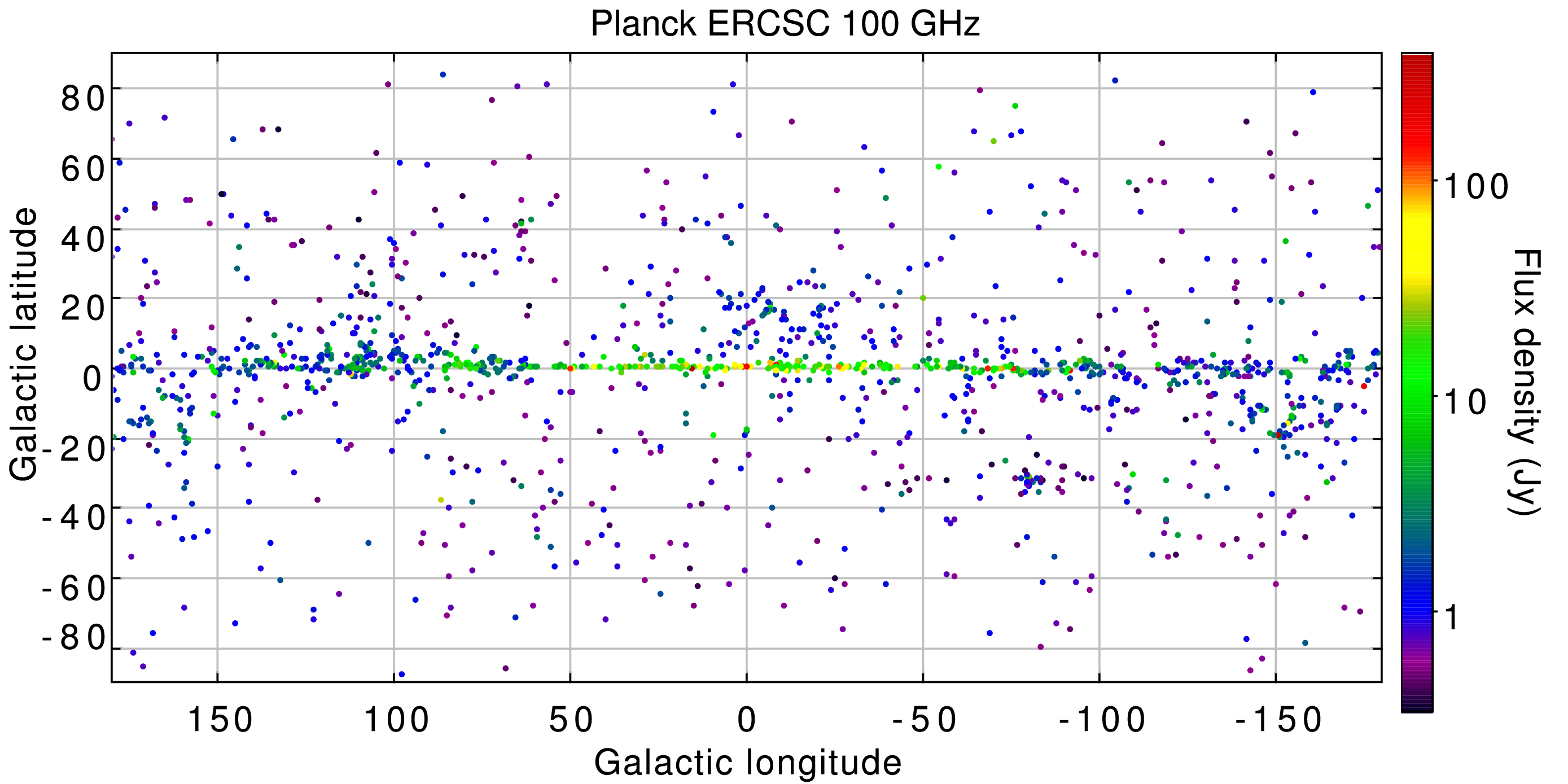}
\includegraphics[width=9cm]{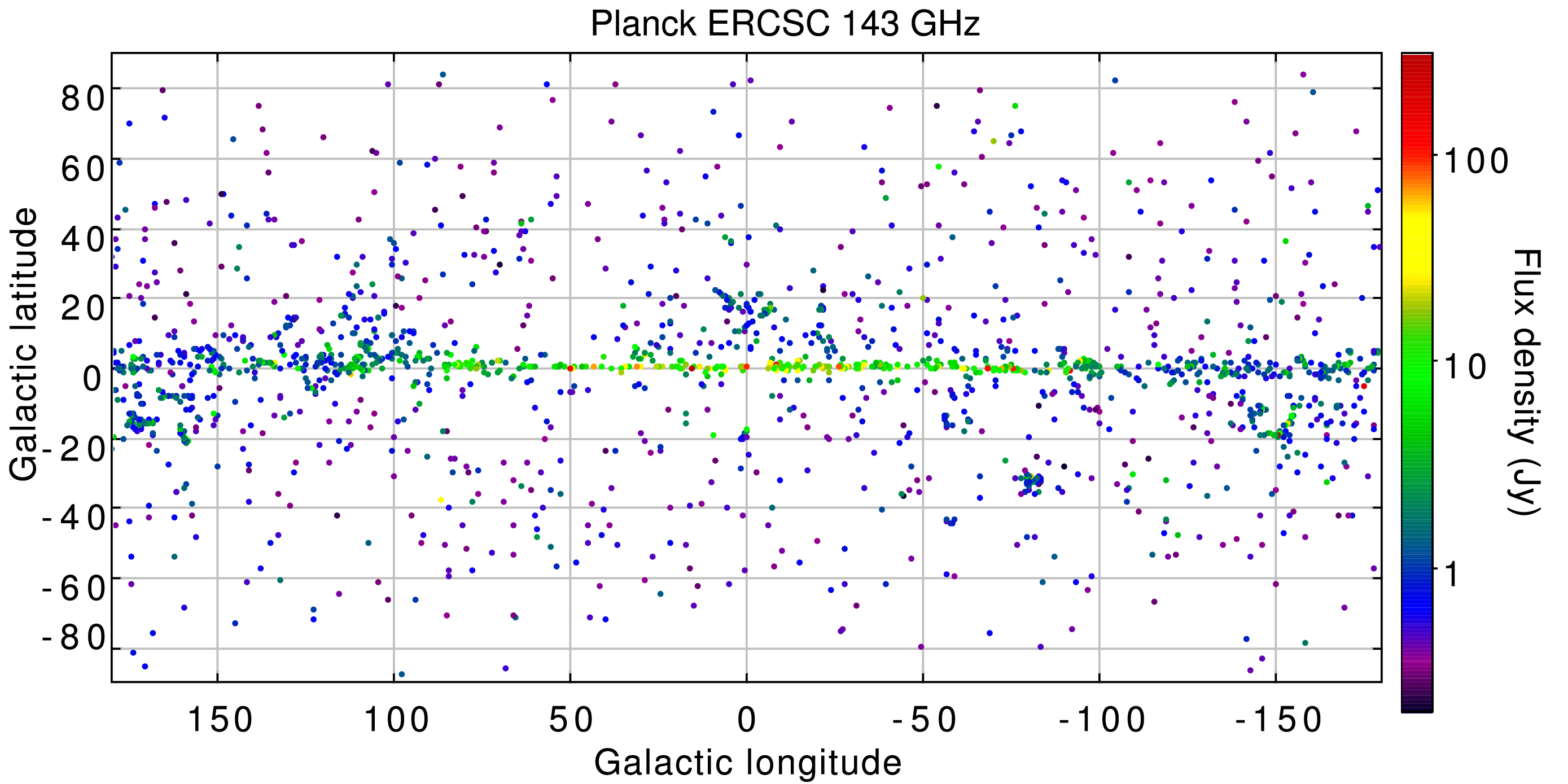}
\includegraphics[width=9cm]{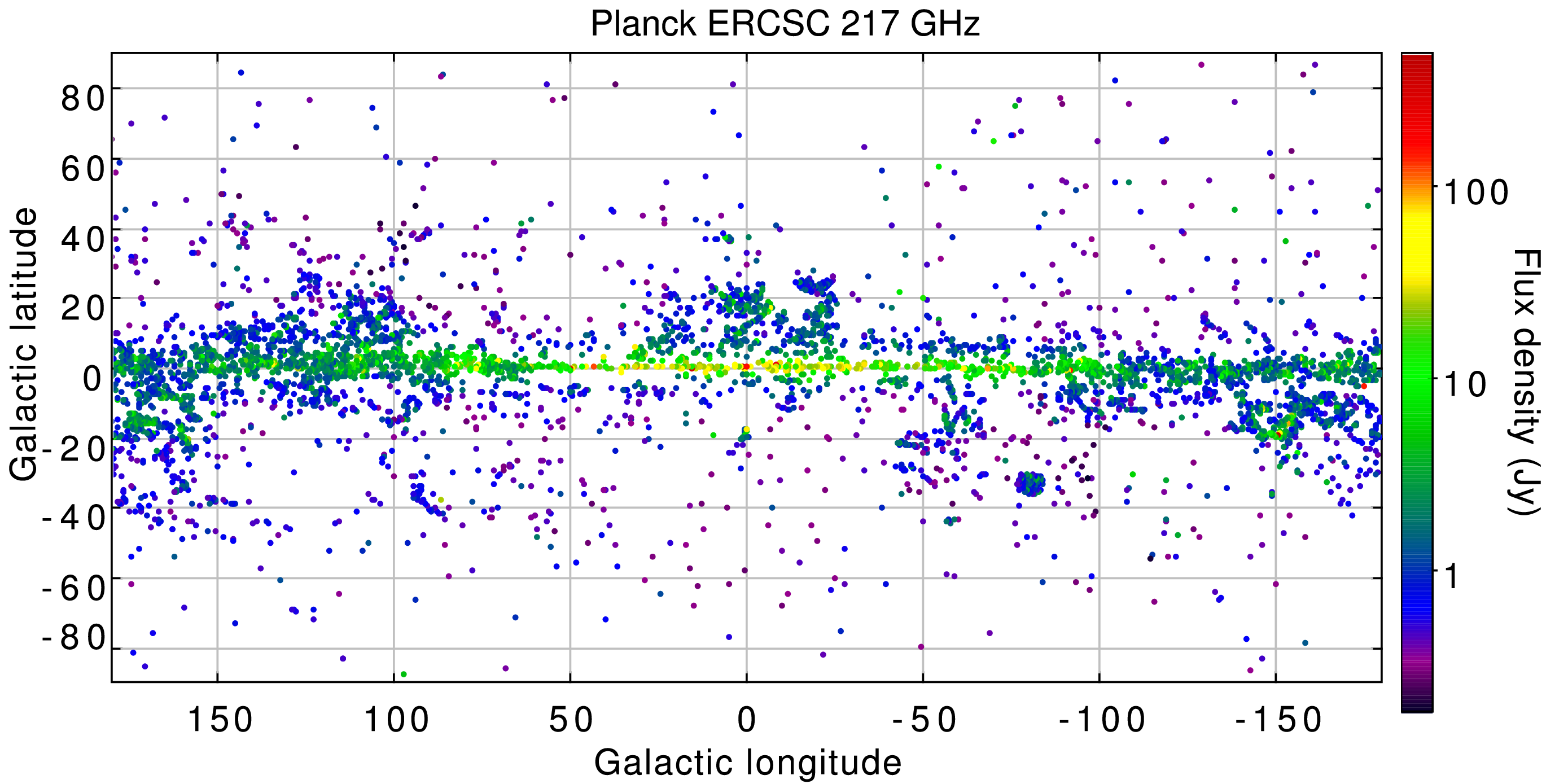}
\includegraphics[width=9cm]{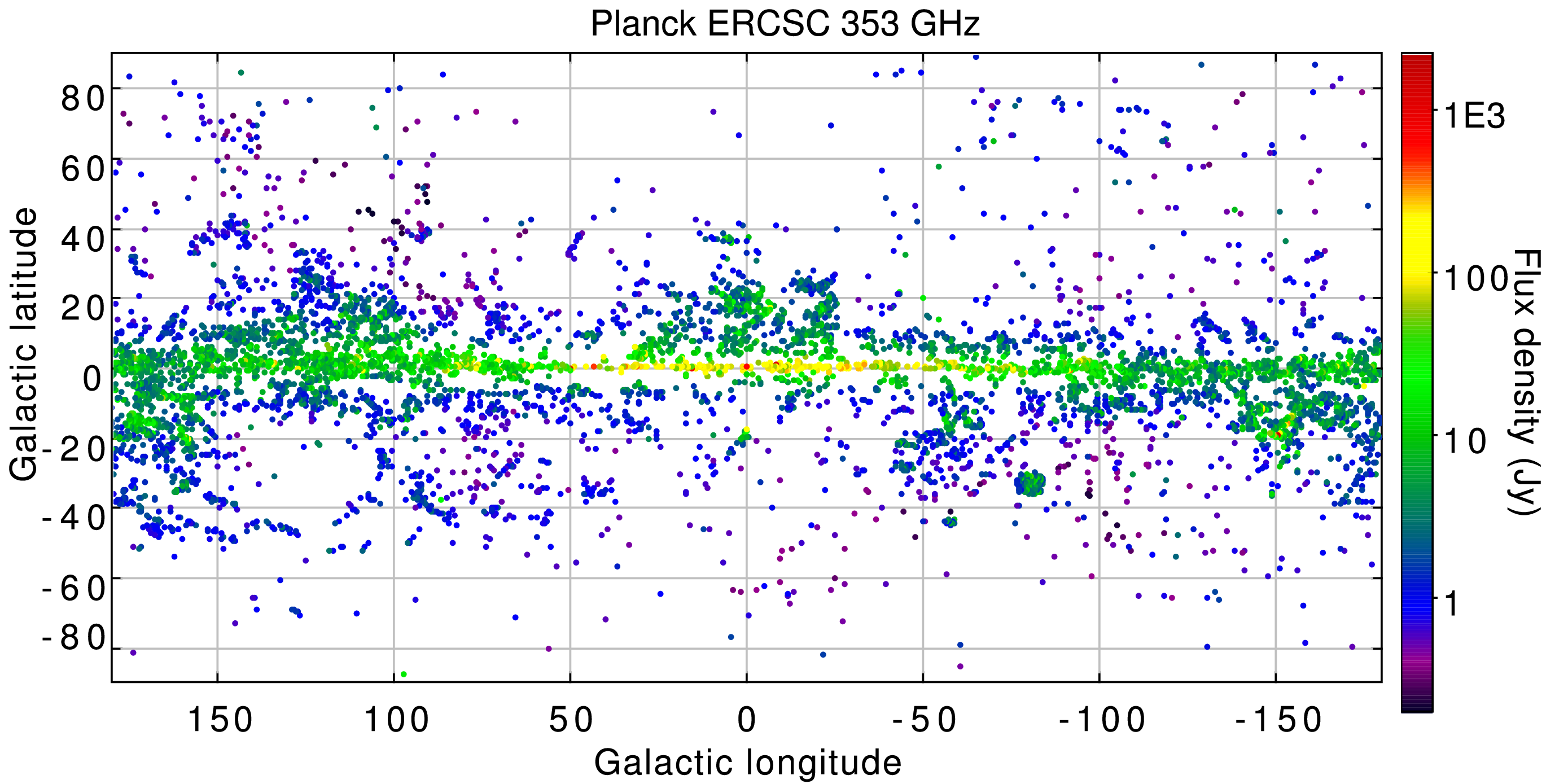}
\includegraphics[width=9cm]{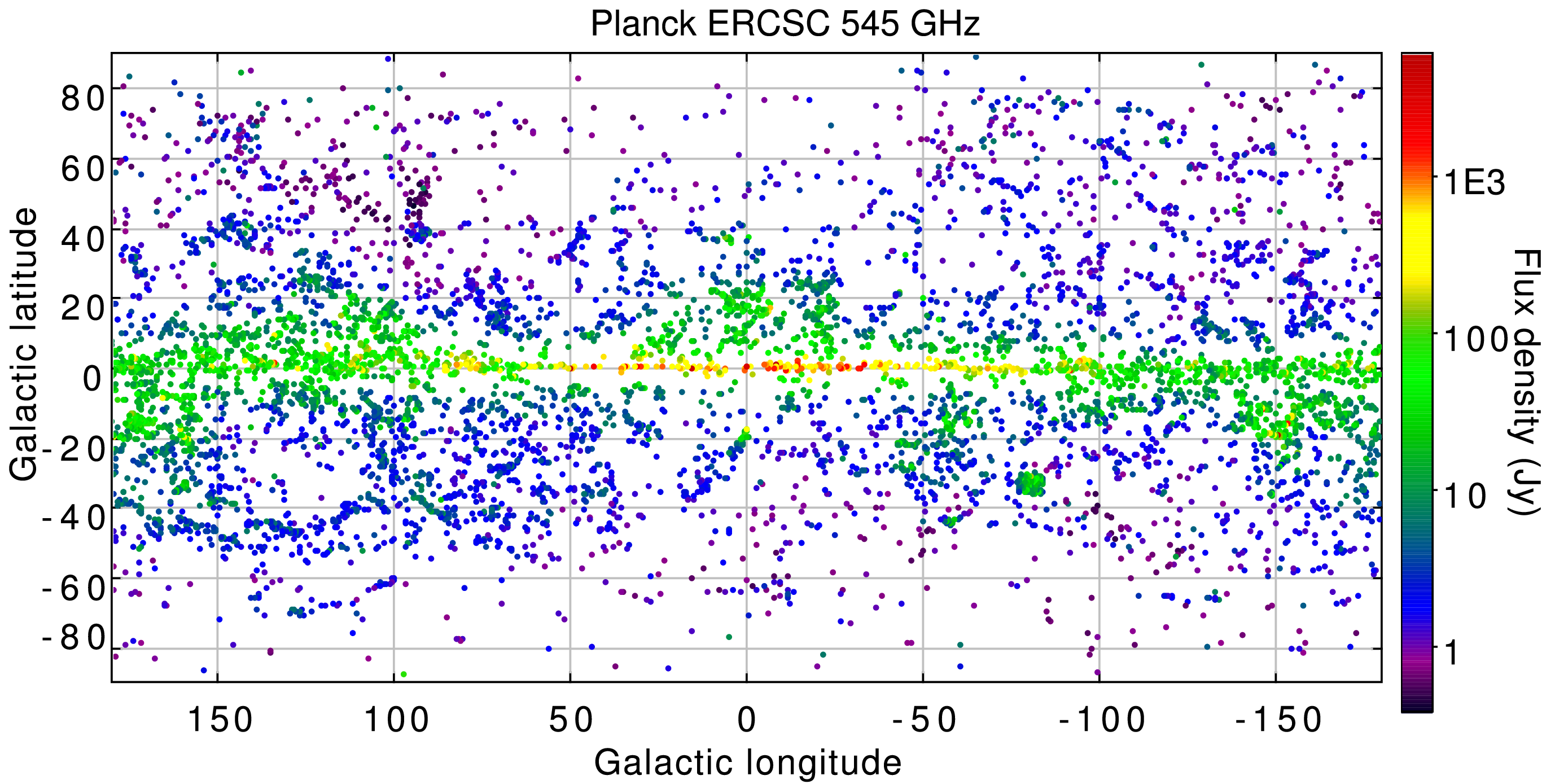}
\includegraphics[width=9cm]{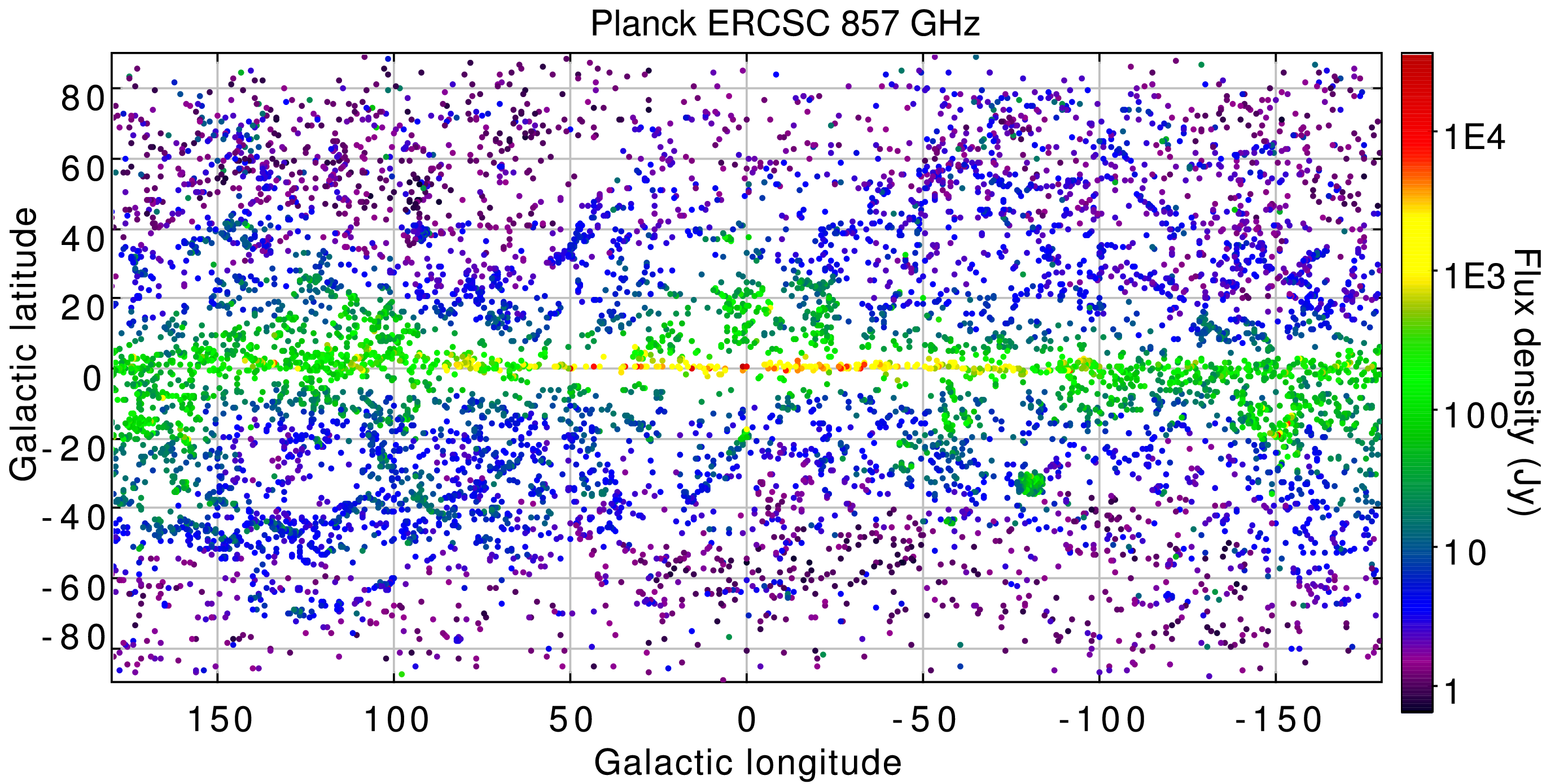}
\caption[Sky Distribution of ERCSC sources]{Sky distribution of sources in Galactic coordinates at all nine \Planck\
frequencies.
Sources are colour coded by flux density. In the Galactic Plane, due to strong emission from the ISM, there is a paucity 
of faint sources. The higher density of sources in the LMC region is also noticeable.}
\label{fig:skymap1}
\end{figure*}

Table~\ref{tbl-ercsc} shows the fraction of sky coverage, the beam FWHM and the sensitivity of 
the ERCSC after all cuts have been applied. Although the 10$\sigma$ values are quoted, sources
which are up to a factor of $\sim$2 fainter and located in regions of low sky background are included in the ERCSC
since they meet the high reliability criterion described in the previous section.
As an illustration, Figure~\ref{fig:sens} shows the flux density limit of \Planck\ both in the Galactic Plane ($|b|<10\degr$)
and at high Galactic latitude ($|b|>30\degr$) relative to other wide area surveys
at comparable frequencies. Also shown are the spectrum of typical sources of foreground emission. 
Figures \ref{fig:skymap1} show the all sky distribution of sources colour coded by flux density.

\begin{table*}
\centering
\caption{\Planck\ ERCSC Characteristics}
\label{tbl-ercsc}
\begin{tabular}{l l l l l l l l l l}
\hline\hline
Freq [GHz] & 30 & 44 & 70 & 100 & 143 & 217 & 353 & 545 & 857 \\
\hline
$\lambda$ [$\mu$m] &
10000 & 6818 & 4286 & 3000 & 2098 & 1382 & 850 & 550 & 350 \\
Sky Coverage in \% & 99.96 & 99.98 & 99.99 & 99.97 & 99.82 & 99.88 & 99.88 & 99.80 & 99.79 \\
Beam FWHM [arcmin]\tablefootmark{a} &
32.65 & 27.00 & 13.01 & 9.94 & 7.04 & 4.66 & 4.41 & 4.47 & 4.23 \\
\# of Sources & 705 & 452 & 599 & 1381 & 1764 & 5470 & 6984 & 7223 & 8988 \\
\# of $|b|>30\degr$ Sources & 307 & 143 & 157 & 332 & 420 & 691 & 1123 & 2535 & 4513 \\
\\
10$\sigma$\tablefootmark{b} [mJy] &
1173 & 2286 & 2250 & 1061 & 750 & 807 & 1613 & 2074 & 2961 \\
10$\sigma$\tablefootmark{c} [mJy] &
487 & 1023 & 673 & 500 & 328 & 280 & 249 & 471 & 813 \\
Flux Density Limit\tablefootmark{d} [mJy] &
480 & 585 & 481 & 344 & 206 & 183 & 198 & 381 & 655\\
\hline
\end{tabular}
\tablefoot{
\tablefoottext{a}{The precise beam values are presented in \citet{planck2011-1.6} and \citet{planck2011-1.7}. This table shows the values which were adopted for the ERCSC.}
\tablefoottext{b}{Flux density of the median $>$10$\sigma$ source at $|b|>30\degr$ in the ERCSC where $\sigma$ is the photometric uncertainty of the source.}
\tablefoottext{c}{Flux density of the faintest $>$10$\sigma$ source at $|b|>30\degr$ in the ERCSC.}
\tablefoottext{d}{Faintest source at $|b|>30\degr$ in the ERCSC.}}
\end{table*}

\subsection{Statistical Nature of Sources}

In this section, we characterise the sources detected by \Planck\ at each frequency.
A source, called source one, at frequency one is associated with a source, called source two, at frequency two, if it lies within $
\left( \mathrm{FWHM}_1 + \mathrm{FWHM}_2 \right)/2$, if source two is the closest source at frequency two to source one, and \emph{vice versa}. The results are summarised in Table~\ref{tbl-char}. 
Naturally, at the lowest frequency, 30\,GHz, it is impossible to
find associations at a lower frequency and hence columns B \& C are blank. Similarly, at the highest frequency, 857\,GHz,
it is impossible to find associations at a higher frequency and hence columns B \& D are blank.

We find that at 30\,GHz, where the radio spectrum might have a significant optically thin synchrotron
emission component (which implies decreasing flux density with increasing frequency), the
number of sources seen in the adjacent passband is 54\%. Similarly, at the highest frequency where
the thermal dust emission has a steep spectrum of the form $S_{\nu}\sim\nu^{3+\beta}$, the fraction of 857\,GHz 
sources seen at 545\,GHz is predictably low due to the relative sensitivities of the two bands.
However, at the intermediate frequencies, the fraction of sources which are associated with sources
in the adjacent bandpasses is high. Although the fraction of associations is not 100\%, we can
use the spectral information from these associations to characterise the nature of sources at each frequency.

\begin{table*}
\centering
\caption{ERCSC Source Characterisation}
\label{tbl-char}
\begin{tabular}{l l l l l l l}
\hline\hline
Frequency & A & B & C & D & E & F\\
\hline
   30 &  705&     ... &    ... &  379 &  379 & 0.54\\
    44 &  452&   334 &  379 &  388 &  433 & 0.96\\
    70 &  599&   363 &  389 &  520 &  546 & 0.91\\
   100 &  1381&  496 &  520 & 1104 & 1128 & 0.82\\
   143 &  1764&        929&       1106&       1357&       1534&    0.87\\
   217 &  5470&       1067&       1357&       4190&       4480&    0.82\\
   353 &  6984&       2848&       4189&       4244&       5585&    0.80\\
   545 &  7223&       3404&       4245&       5363&       6204&    0.86\\
   857 &  8988&          ...&       5365&          ...&       5365&    0.60\\
\hline
\end{tabular}
\tablefoot{
\tablefoottext{A}{ Total Number of sources detected}
\tablefoottext{B}{ Number of sources detected both at frequency just below and just above given frequency}
\tablefoottext{C}{ Number of sources detected at frequency just below given frequency}
\tablefoottext{D}{ Number of sources detected at frequency just above given frequency}
\tablefoottext{E}{ Number of sources detected either at frequency just below or just above given frequency}
\tablefoottext{F}{ Fraction of sources detected either at frequency just below or just above given frequency}
}
\end{table*}

The spectral index is calculated by fitting a single power law ($S_{\nu}\propto\nu^{\alpha}$)
to the flux density of sources in adjacent
bands. For 30\,GHz sources, only the 30 and 44\,GHz flux densities of sources are considered. Similarly,
at 857\,GHz, only the 545 and 857\,GHz flux densities of sources are fit. For all the intermediate bands,
the frequencies just below and just above are included in the fit, if the source is detected in the ERCSC.

\begin{figure*}
\centering
\includegraphics[width=12cm]{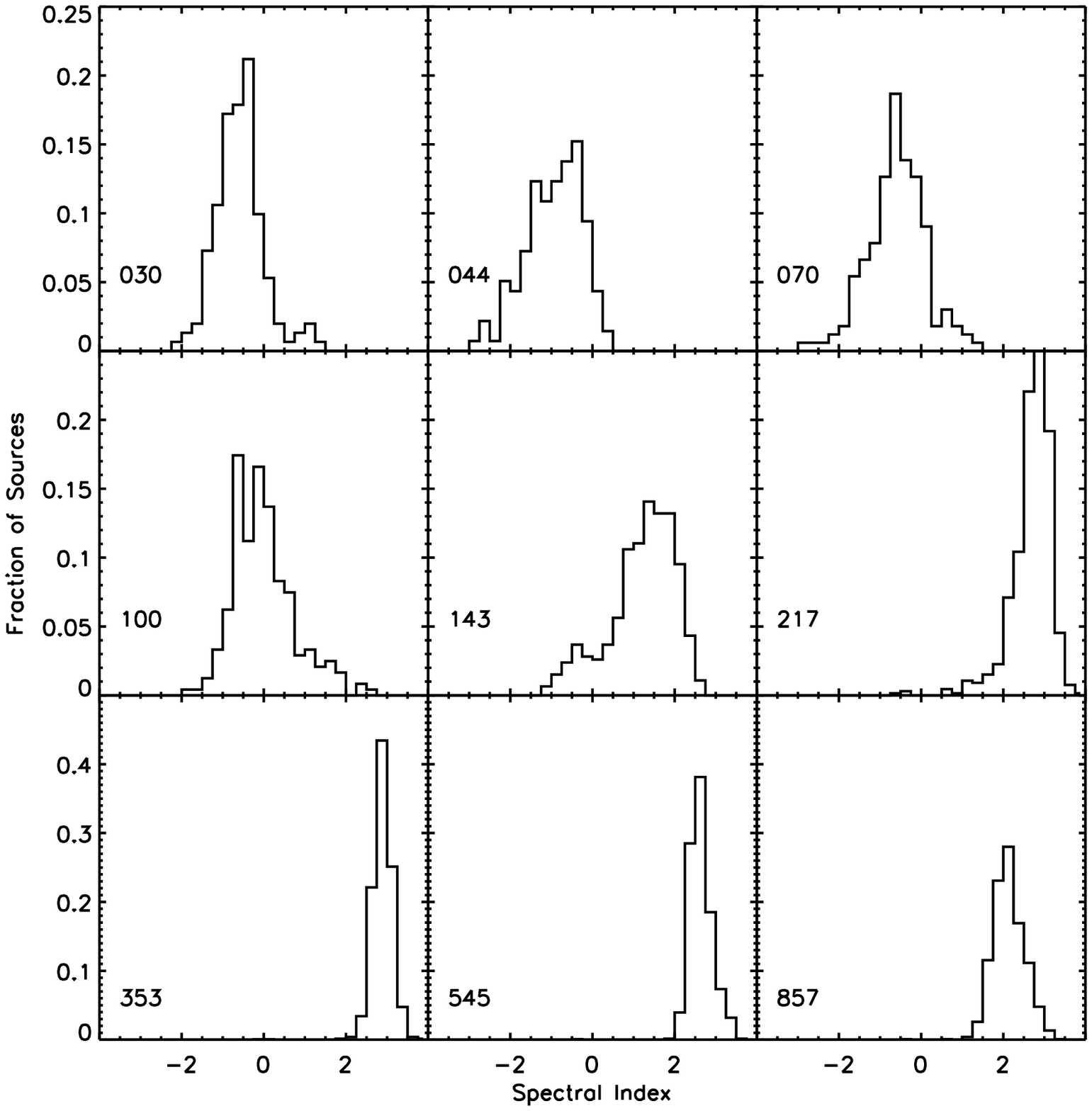}
\caption[Spectral index distribution of Galactic sources]{
The distribution of spectral indices ($\alpha$ where $S_{\nu}\propto\nu^{\alpha}$) for sources within 10$\degr$ of the Galactic Plane.
Each panel shows the spectral index distribution for ERCSC sources at the corresponding \Planck\ band. 
}
\label{fig:lowbsindx}
\end{figure*}

Figure~\ref{fig:lowbsindx} shows the distribution of spectral indices for the sources within $|b|<10\degr$
which are likely to be sources within our Galaxy. At low frequencies, the median SED of
sources in the Galactic Plane is an $S_{\nu}\propto\nu^{-0.5}$ spectrum. The distribution of $\alpha$
values significantly broaden between 30 and 100\,GHz, likely due to varying amounts of
free-free emission along different sightlines. 
At 100\,GHz, the spectrum becomes
noticeably flatter with a median $\alpha=-0.25$, partly due to the increasing contribution of
thermal dust emission and partly due to the large contribution
from the CO line to the 100\,GHz flux density. At 143\,GHz, the spectral index distribution shows
the presence of both radio sources as well as the dominant contribution from sources with thermal dust emission.
Expectedly, at higher frequencies, the distribution of spectral indices is narrow and
is centred between $\alpha=2$ and $\alpha=3$, tracing the Rayleigh-Jeans component of dust emission.
The reason the median $\alpha$ is almost 3 at 217 GHz but evolves to 2 at 857 GHz is a selection effect.
As can be seen in Figure \ref{fig:sens}, thermal dust emission with emissivity $>$0 would increase faster with
increasing frequency compared to a Rayleigh-Jeans spectrum, relative to the \Planck\ sensitivity.
As a result, sources at the intermediate frequencies can span a broader range of spectral indices than a faint source
at 857 GHz which would have an estimated spectral index only if it were detected at 545 GHz, and thereby preferentially
have a spectrum that is less steep.  

\begin{figure*}
\centering
\includegraphics[width=12cm]{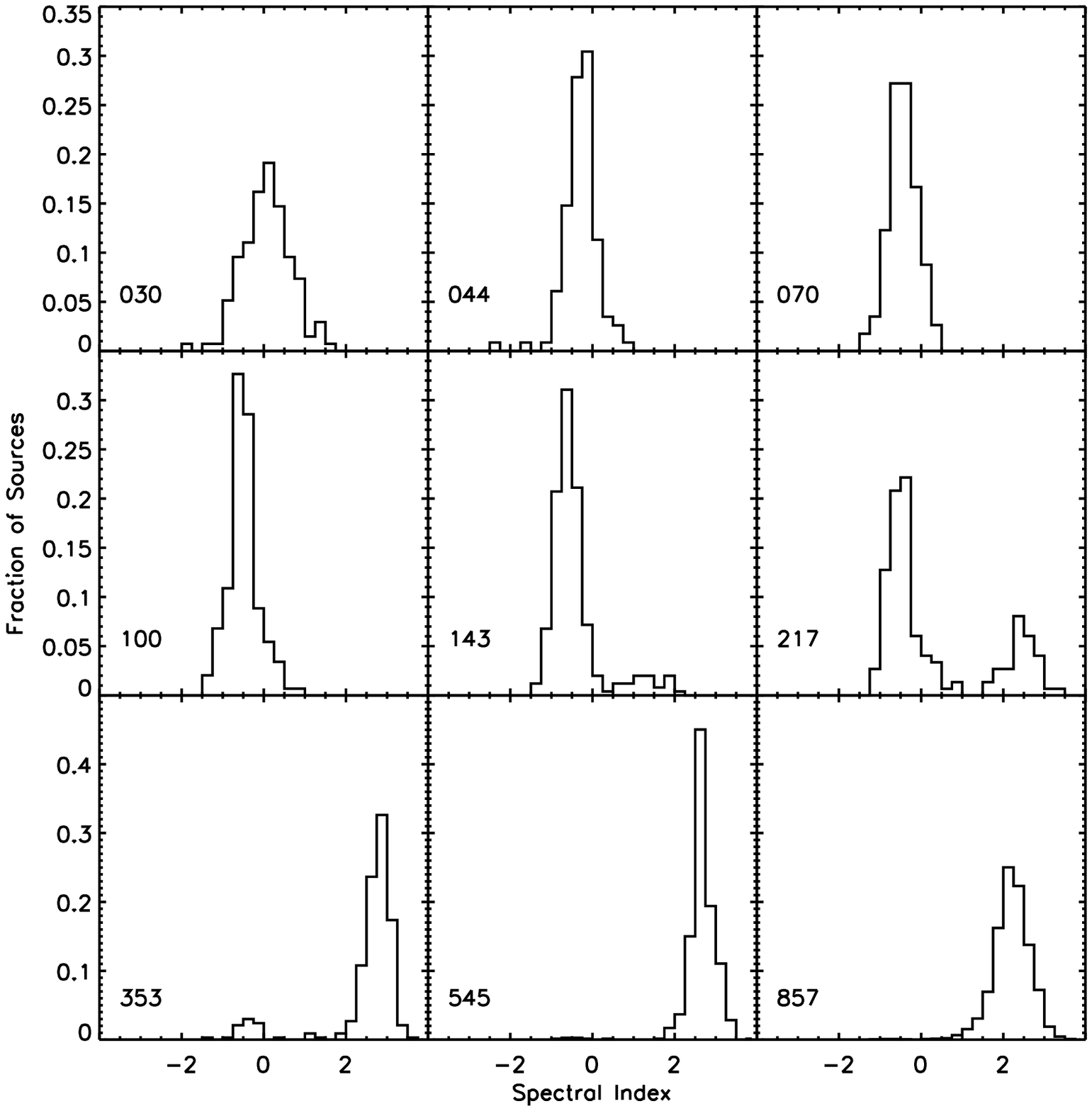}
\caption[Spectral index distribution of extragalactic sources]{
The distribution of spectral indices for sources above 30$\degr$ of the Galactic Plane, likely to be dominated by extragalactic sources.
}
\label{fig:hibsindx}
\end{figure*}

Figure~\ref{fig:hibsindx} shows the distribution of spectral indices for the sources at $|b|>30\degr$ which
are likely to be extragalactic. At the lower frequencies, the distribution of spectral indices is centred
at $\alpha=0$. However, unlike the Galactic sources where the distribution broadens with increasing frequency,
among the extragalactic sources, the spectral index distribution narrows between $30-100$\,GHz. 
There are two possible origins for this. One is that the CO contribution is generally negligible for the extragalactic
sources and that the larger distribution of spectral indices around 100 GHz for the Galactic sources is simply a tracer
of variation of the CO contribution to the broadband photometry. A second possibility is that the spectral index distribution 
of Galactic sources is intrinsically broader while the extragalactic sources at 100 GHz are dominated by a power-law distribution
of electrons produced in relativistic shocks, which tend to display a more uniform power-law index.
At 143 and 217\,GHz, 
the radio
source population continues to dominate although the dusty sources start to become significant.
This is in contrast to the Galactic population where the infrared luminous sources are the dominant contributor.
It is also striking that even at 353\,GHz, the radio source population continues to make a significant contribution.
At the highest frequencies, both the Galactic and extragalactic source populations show similar behaviour
expected from the Rayleigh-Jeans tail of dust emission.

A comparison with the statistical
properties of sources found in the South Pole Telescope 1.4mm and 2mm surveys \citep{Vieira2010} is warranted.
The SPT surveys found that
$\sim$30\% of the 1.4mm sources are dusty while the majority are synchrotron dominated.
This is similar to the results for the high Galactic latitude
ERCSC sources; $\sim$25\% of the ERCSC 217 GHz sources show an SED consistent with thermal dust emission. The difference
however is that the dusty sources observed by the SPT dominate at fainter flux densities ($<$15 mJy).
In contrast, the dusty population in the ERCSC appears
to be at brighter flux densities with a median FLUX of 2.4\,Jy, while the synchrotron sources
have a median FLUX of 0.8\,Jy. This difference is because the ERCSC 217 GHz dusty sources are associated with the Large
Magellanic Cloud and are thereby brighter than the typical dusty sources that the SPT has observed.

\subsection{Individual Case Studies}

\begin{figure}
\centering
\resizebox{\hsize}{!}{\includegraphics{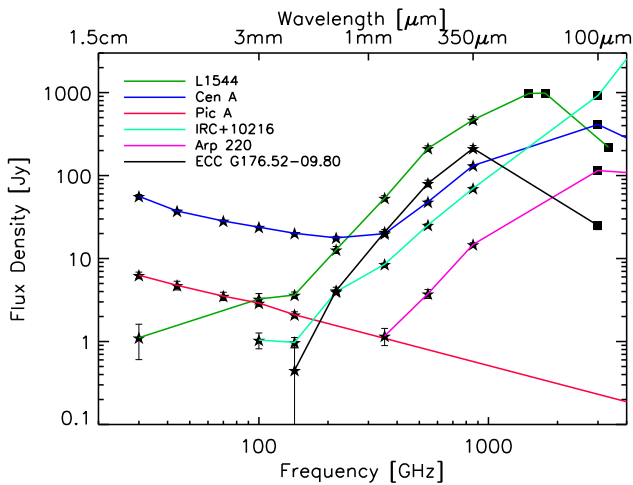}}
\caption[Representative Source Classes in ERCSC]{
The SED of representative source classes in the ERCSC. The plot shows
a pre-stellar core L1544, an extragalactic 
radio source Centaurus A, a synchrotron dominated radio galaxy Pictor A,
IRC+10216 which is the prototype of stars with dust shells, the starbursting ultraluminous infrared galaxy Arp 220 and the
cold stellar core ECC G176.52-09.80. IRAS or ISO flux densities are shown as solid squares while the \Planck\ flux densities
are shown as stars. The \Planck\ ERCSC enables a diverse class of sources to be studied over a broad range of frequencies
and flux densities.}
\label{fig:sedrange}
\end{figure}

The 
SED of representative sources of different classes that can be found in the ERCSC are presented in this sub-section. 
The selected sources are
a pre-stellar core L1544 \citep{WardThompson}, an extragalactic
radio source Centaurus A, a synchrotron dominated radio galaxy Pictor A,
IRC+10216 which is the prototype of stars with dust shells, the starbursting ultraluminous infrared galaxy Arp 220 and the
cold stellar core ECC G176.52-09.80. 
The sensitivity and wavelength coverage of \Planck\ enables
synchrotron emission, thermal bremsstrahlung emission, thermal dust emission as well as the transition frequencies between the emission
processes to be studied. Figure \ref{fig:sedrange} shows the SED of these representative sources. 

Since some of these sources are extended at the \Planck\ angular resolution, the ERCSC Gaussian fit flux densities (i.e. ``GAUFLUX")
are shown, except in the cases where
the fit failed in the low signal-to-noise regime. In those cases the aperture photometry values (i.e. ``FLUX'' in the ERCSC)
are plotted. Uncertainties 
include the Monte-Carlo estimate of flux density uncertainties. The plotted SED also show
IRAS and/or ISO flux densities at far-infrared wavelengths. \Planck\ can clearly reveal the contribution of cold dust
at wavelengths longward of IRAS/ISO and observe the transition from thermal dust emission to synchrotron/free-free radio emission.

\section{Validation of the ERCSC\label{sec:valid}}

At the three lowest frequencies of \Planck, it is possible to validate ERCSC source identifications, reliability, positional accuracy and flux density 
accuracy using external data sets, particularly large-area radio surveys. This external validation was undertaken using the 
following catalogues and surveys:  (1) full sky surveys and catalogues:  {\it WMAP} 5-year catalogue \citep{wright2009} and the NEWPS catalogue, based on 
earlier {\it WMAP} results \citep{massardi}; (2) in the southern hemisphere the AT20G survey at 20\,GHz \citep{mur}; (3) in the 
northern hemisphere, where no large-area, high-frequency survey like AT20G is available, we used CRATES \citep{hea}.

An ERCSC source was considered reliably identified if it falls within a circle of radius one half the \Planck\ beam FWHM 
which is centered on a source at the corresponding frequency in one of the above catalogs.
This means of identification was employed 
at $|b|>5\degr$ where confusion was less of a problem and the majority of the sources were extragalactic. Very few such sources were spatially
resolved by \Planck.
Table \ref{tbl:valid} shows the percentage of sources thus identified. For the three lowest \Planck\ frequencies and for $|b|>5\degr$, the ERCSC clearly 
meets its 90\% reliability specification as measured by this external validation.
Table \ref{tbl:valid} also displays results of an attempt to assess reliability of ERCSC sources in the Galactic Plane at $|b|<5\degr$. Here, an ERCSC source was considered reliably identified if it falls within $5\,\textrm{arcmin}$ from Galactic objects like planetary nebulae, supernova remnants, HII regions (or in a few cases, extragalactic sources that happen to be found at low Galactic latitude). The percentage of identifications in the Galactic Plane is lower, but 
still leaves the overall reliability figures at greater than 90\%, meeting the ERCSC specification for reliability.

\begin{table*}
\centering
\caption{ERCSC Source Validation}
\label{tbl:valid}
\begin{tabular}{l l l l l l l}
\hline\hline
Frequency & \# at $|b|>5\degr$ & \# Identified & \# at $|b|<5\degr$ & \# Identified & Total \# & \# Identified\\
\hline
30 &	563&	547 (97\%)&	142&	95 (67\%) &	705 &	642 (91\%)\\
44 &	278&	265 (95\%)&	176&	144 (82\%) &	454 &	409 (90\%)\\
70 &	320&	289 (90\%)&	280&	...        &    600 &  ...	\\
\hline
\end{tabular}
\end{table*}

\begin{figure*}
\includegraphics[width=8cm]{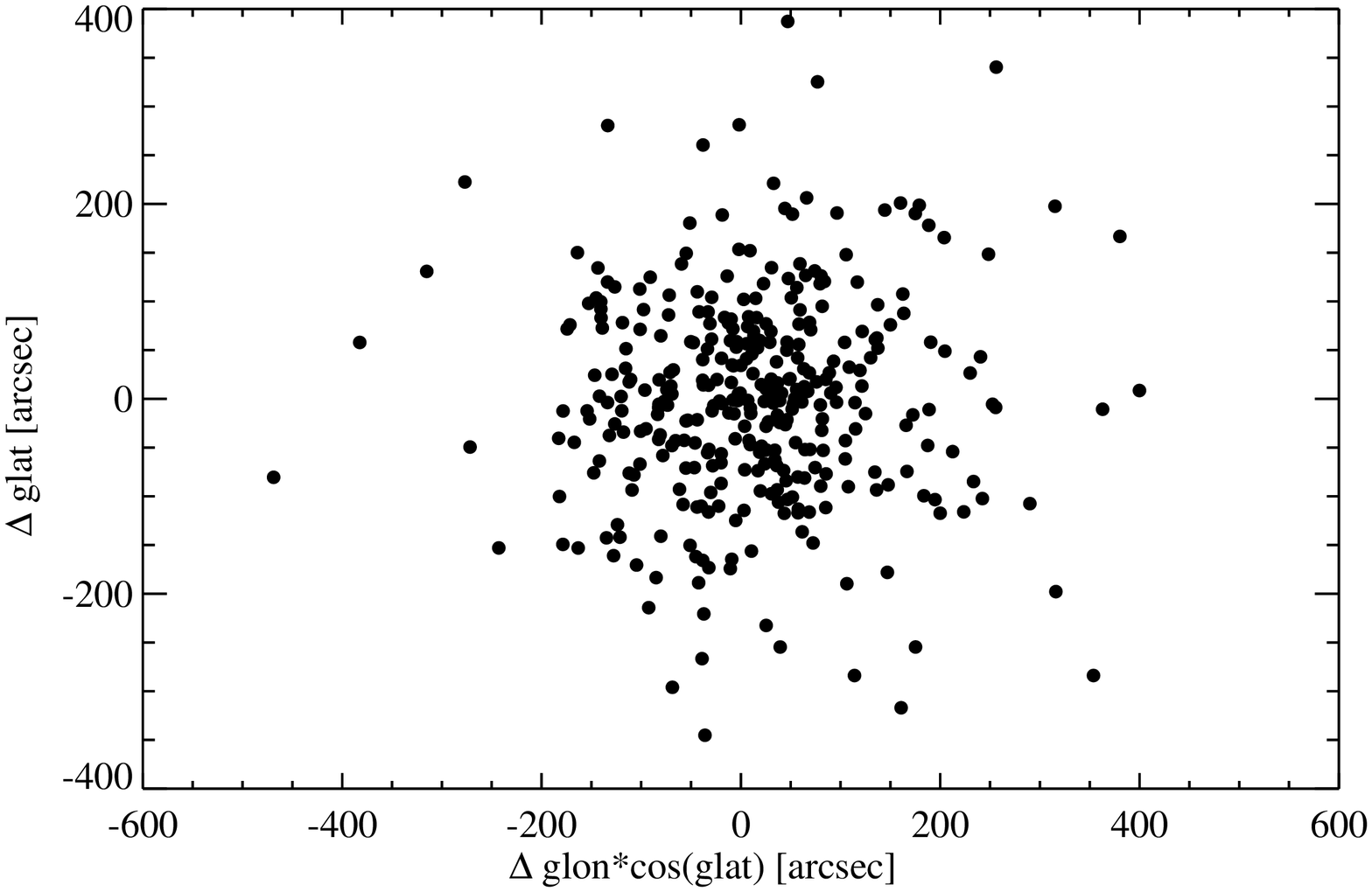}
\includegraphics[width=8cm]{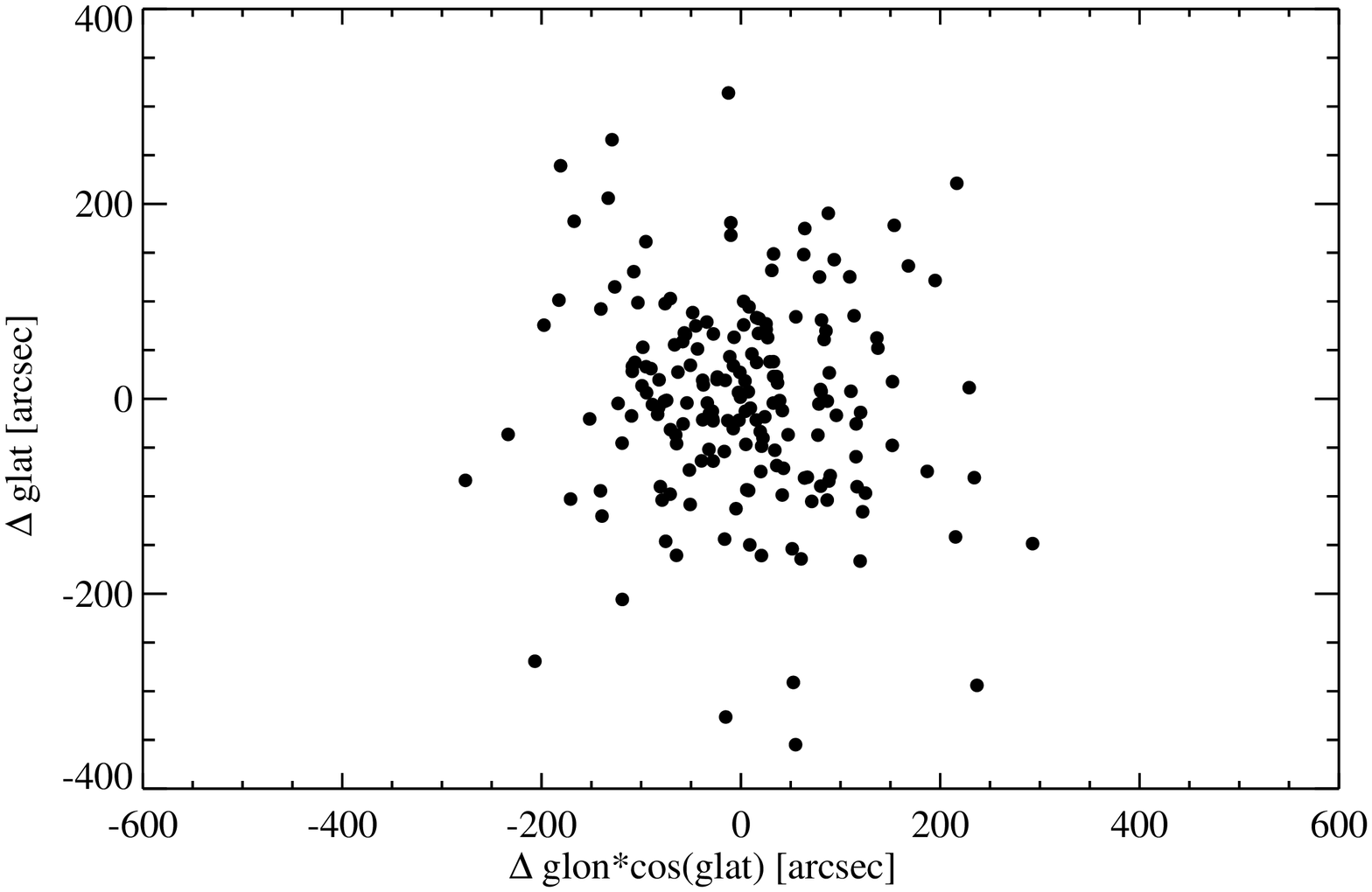}
\includegraphics[width=8cm]{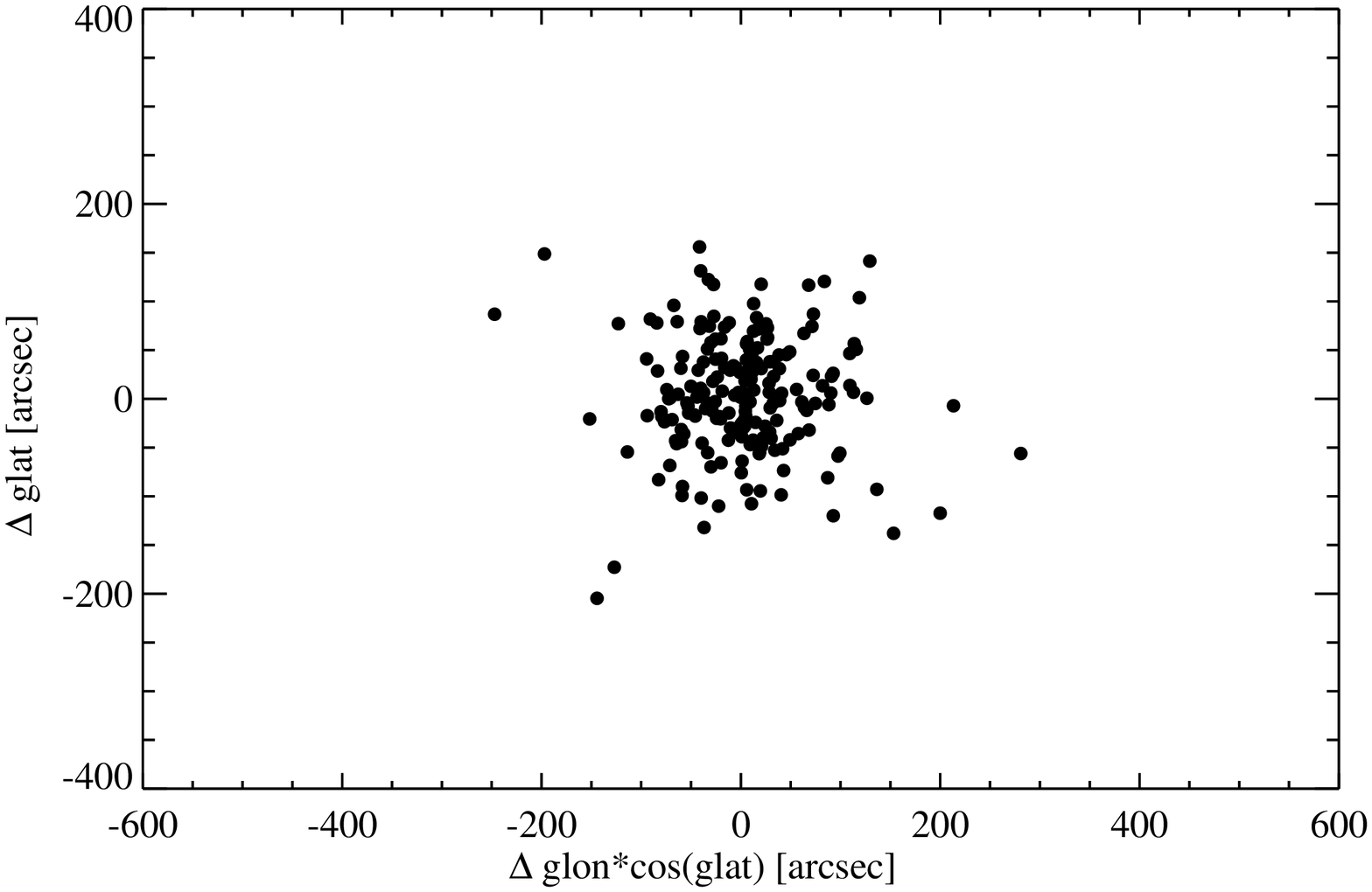}
\includegraphics[width=8cm]{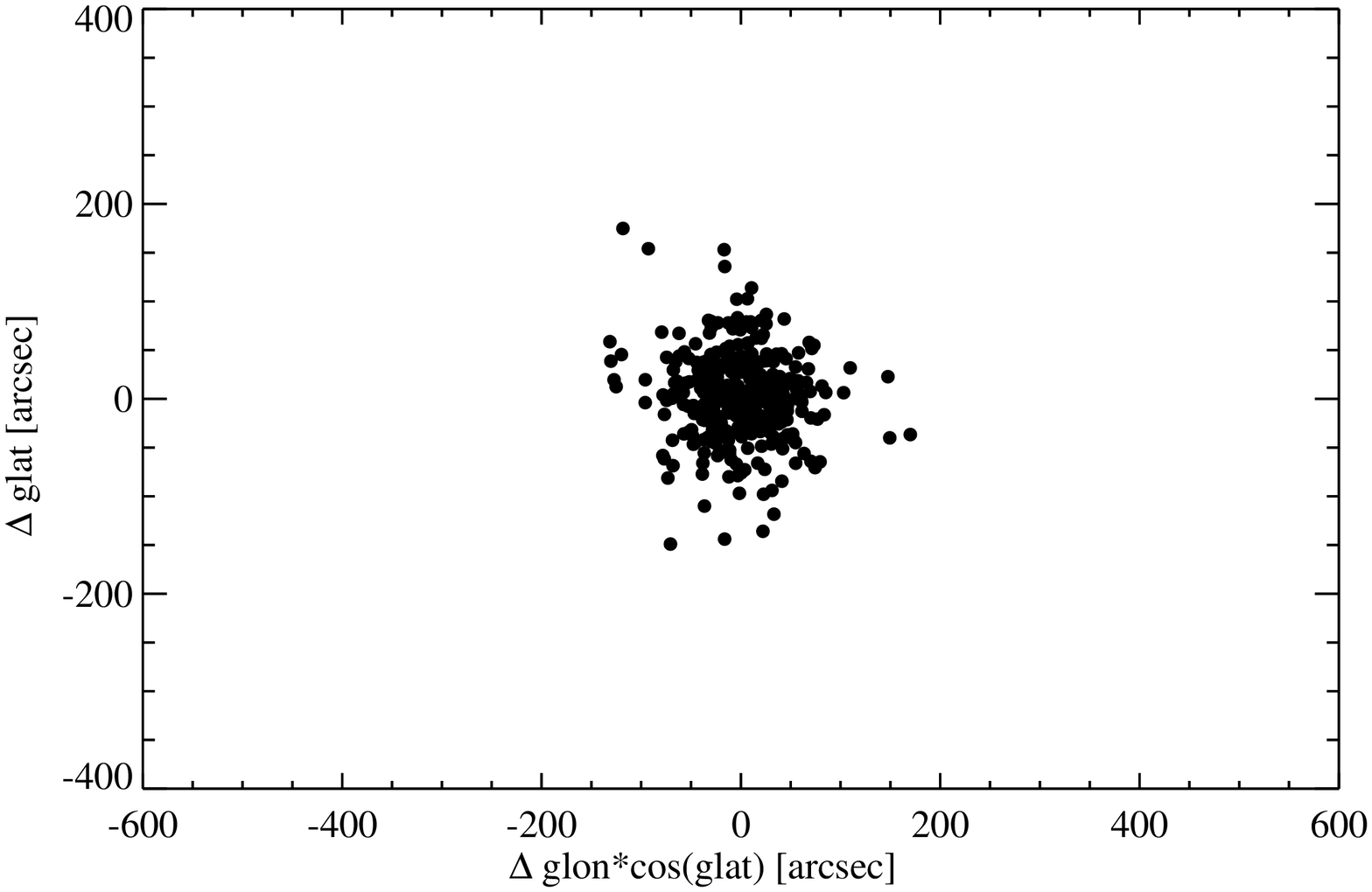}
\includegraphics[width=8cm]{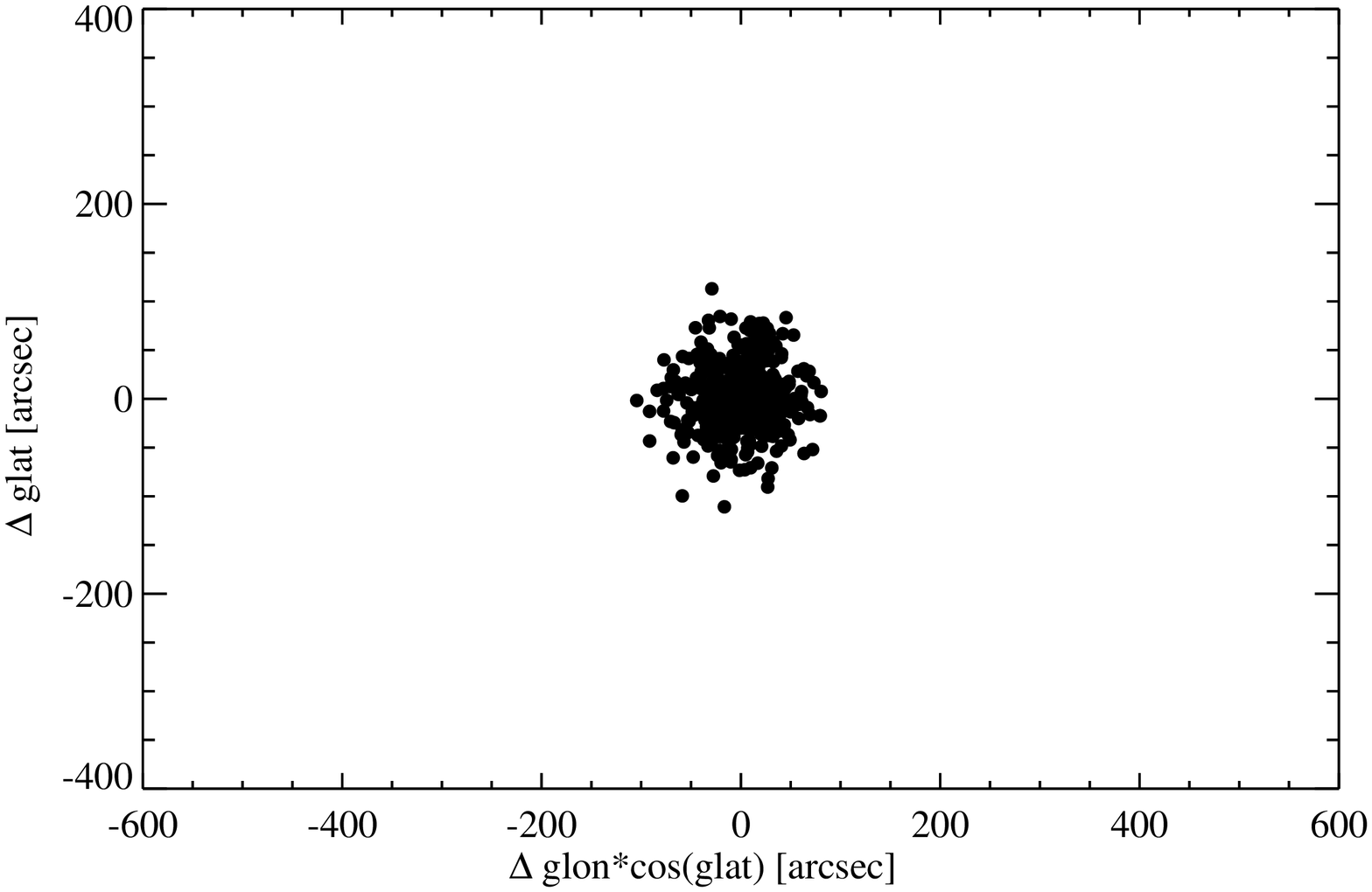}
\includegraphics[width=8cm]{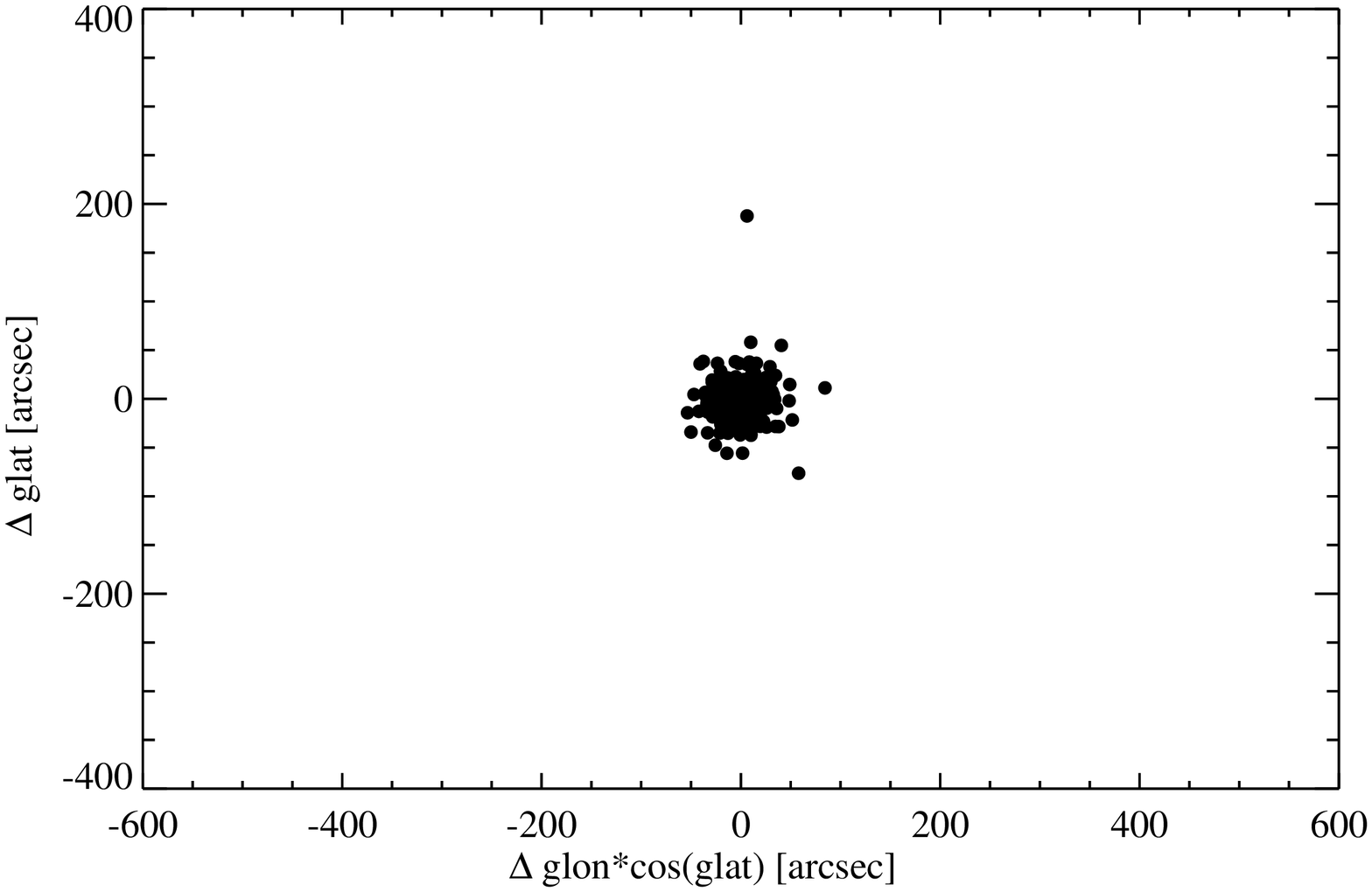}
\includegraphics[width=8cm]{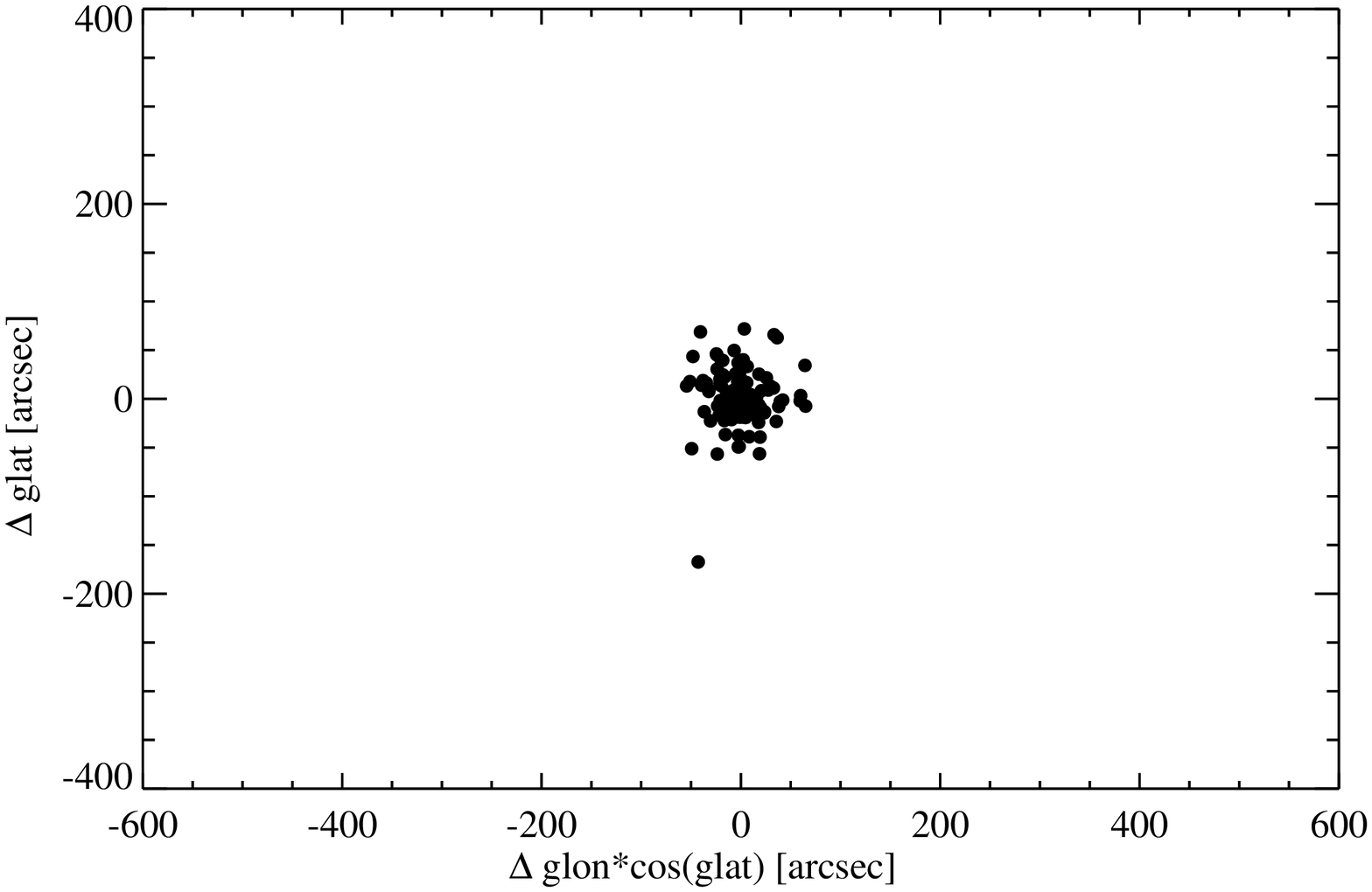}
\caption[Astrometric accuracy of ERCSC]{
Matches to quasars as a measure of positional offsets in the 
ERCSC 30 to 353\,GHz catalogues. The top row shows 30 and 44, the second row 70 and 100, the third row 143 and 217, and the 
final row 353\,GHz. There are insufficient numbers of detected quasars at the upper HFI frequencies.
}
\label{fig:posacc}
\end{figure*}

We also examine the positional accuracy of ERCSC sources by comparing positions taken from the ERCSC with those determined quasi-simultaneously using 
the Very Large Array  (VLA) of the US National Radio Astronomy Observatory. 
ERCSC positions were also compared to the positions of 
several hundred bright quasars (Figure~\ref{fig:posacc}) at frequencies of 353\,GHz and below where a significant fraction are detected. 
The median scatter in offset for frequencies 30--217\,GHz was 2.0, 1.7, 1.1, 0.8, 0.7, 0.3 and 0.35\,arcmin. The results of these 
two tests are consistent, and suggest that the ERCSC clearly meets its specification of RMS scatter in positions being less than FWHM/5.

\begin{figure*}
\centering
\includegraphics[width=12cm]{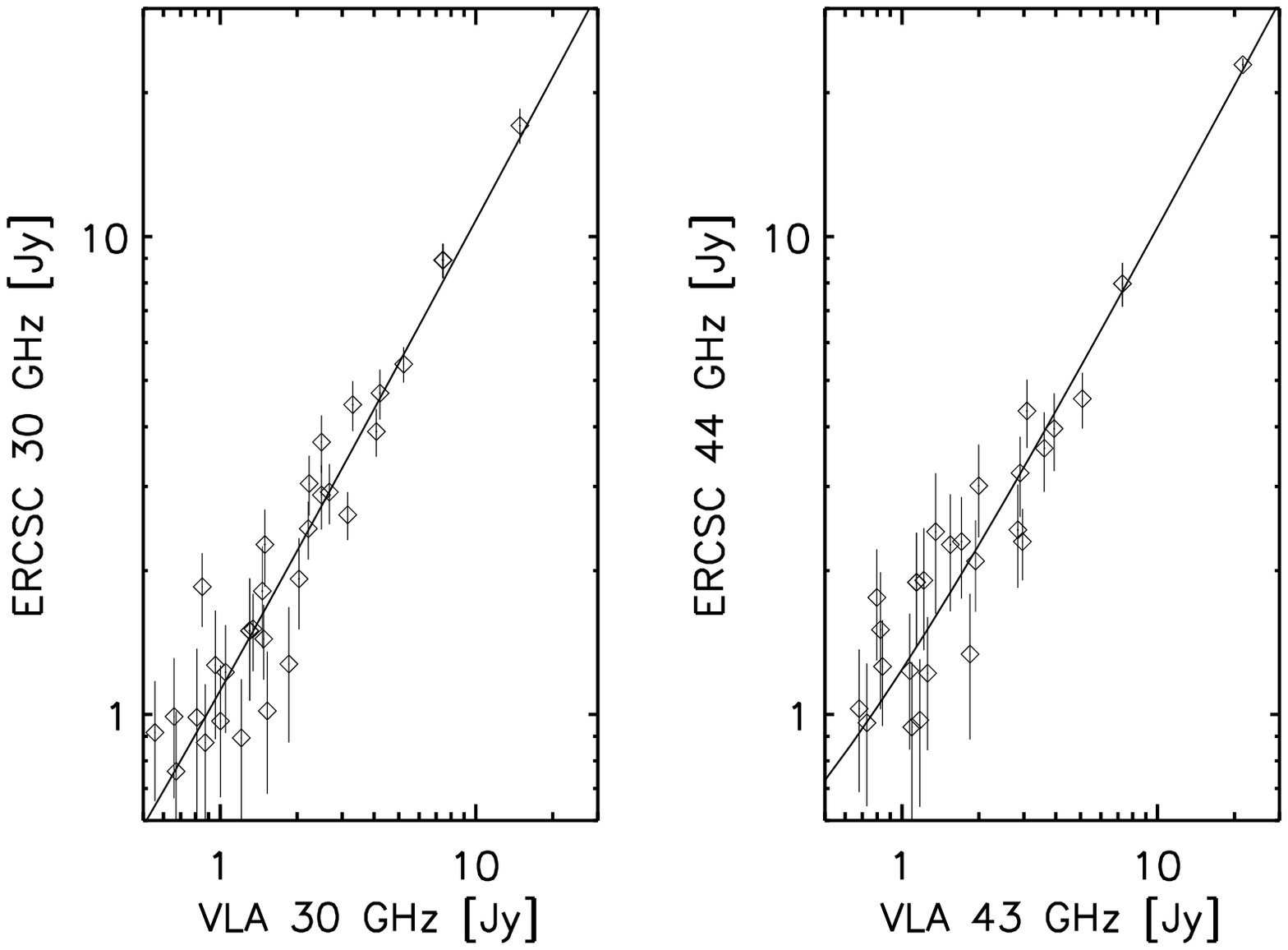}
\caption[Comparison between ERCSC and VLA]{The flux density of a subset of ERCSC sources at 30 and 44 GHz, with color corrections, compared to the
flux density obtained from VLA 22 and 43 GHz observations of the same sources translated to the \Planck\ effective frequency
\citep{planck2011-6.2}. 
The over-plotted lines are the first order polynomial resulting from
an uncertainty weighted fit to the VLA and ERCSC flux densities which partially takes into account Eddington bias. The 
slope of the fit is 1.08 at 30 GHz and 1.02 at 44 GHz indicating that both measurements are in good
agreement. The median ratio of the ERCSC flux density to the VLA flux density is 1.15 at both frequencies. The difference is most likely attributable to a combination of effects including cross-calibration uncertainties, contribution
from fainter sources within the \Planck\ beam, variability and
the fact that the smaller beam VLA measurements would be less sensitive to low surface brightness emission 
beyond the $2-4\arcsec$ primary VLA beam although care has been taken to use mainly unresolved sources. }
\label{fig:vlacomp}
\end{figure*}

A comparison between the ERCSC flux densities with VLA measurements of the same source has also been made (Figure \ref{fig:vlacomp}) 
and is discussed in \citet{planck2011-6.2}.
At both 30 and 44\,GHz the two flux density scales appear to be in good overall agreement
with any difference attributable partly to noise in the \Planck\ measurements and partly due to variability in the radio sources, 
since the \Planck\ and VLA measurements were not exactly simultaneous.
At 70\,GHz however, the comparison is challenging since the VLA measurements are made at 43\,GHz and an extrapolation needs to be made
assuming some spectral index for the source. If a simple extrapolation to a 70\,GHz flux density is made based on the VLA 22--43\,GHz spectral index,
the extrapolated VLA values are either too high or the flux density scale of \Planck\ is too low. The most likely interpretation of this
discrepancy is that the spectral index of radio sources detected
by \Planck\ steepens at frequencies above 44 or 70\,GHz. If, for instance, a spectral index change of
$\alpha = -0.5$ is allowed at frequencies above 43\,GHz, the agreement between the extrapolated and measured 70\,GHz fluxes would be entirely acceptable \citep{planck2011-6.2}.

\begin{figure*}
\centering
\includegraphics[width=8cm]{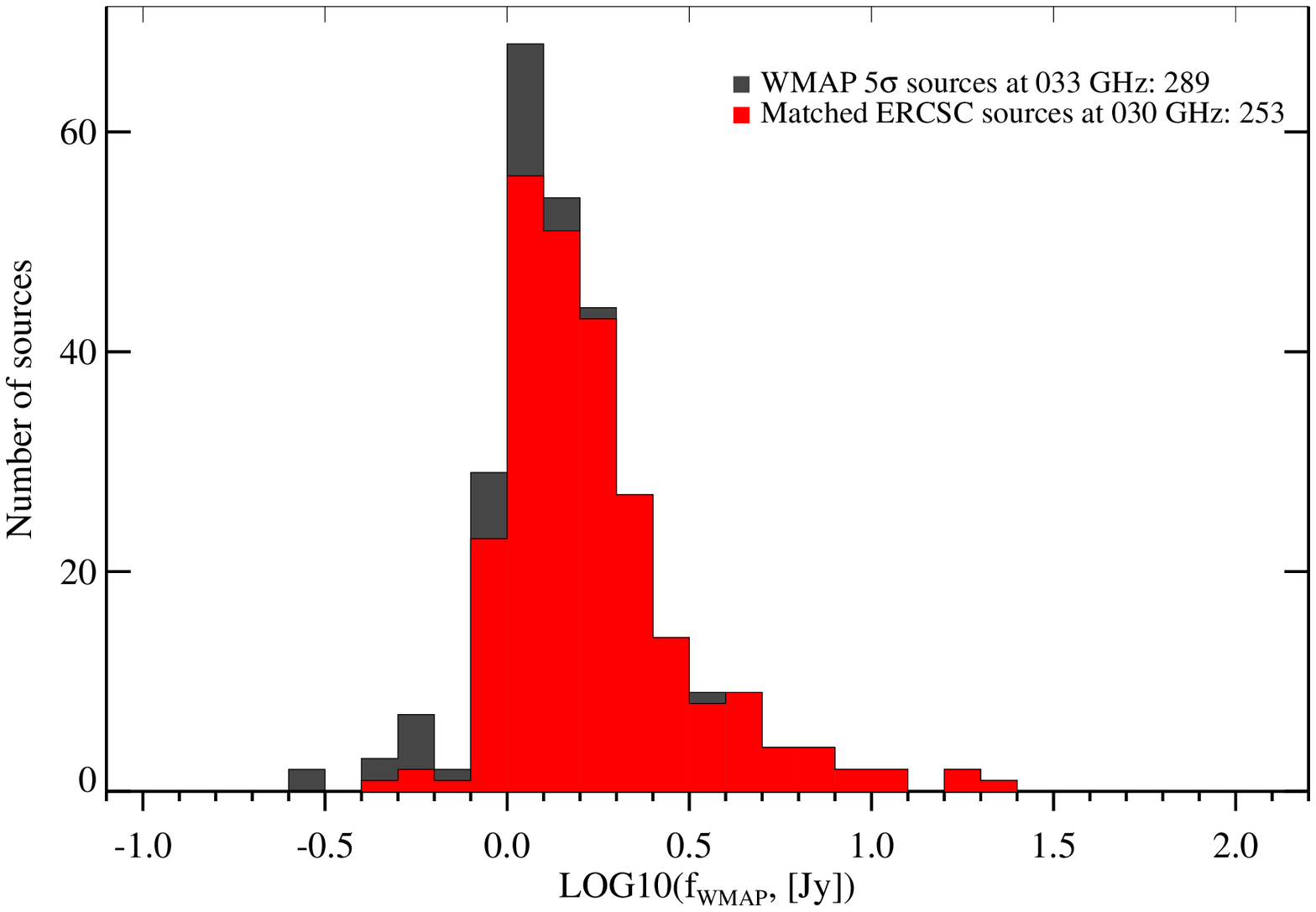}
\includegraphics[width=8cm]{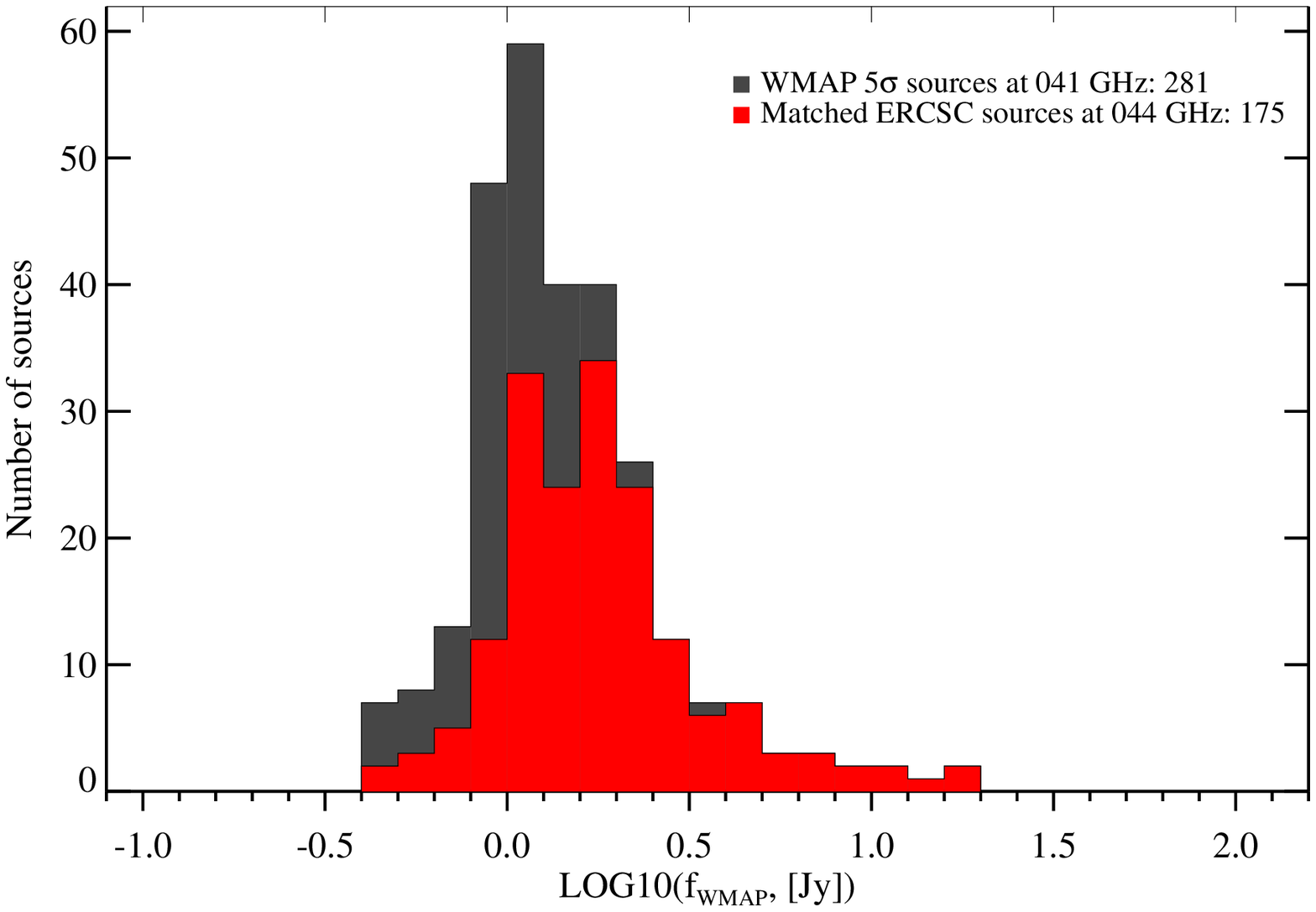}
\includegraphics[width=8cm]{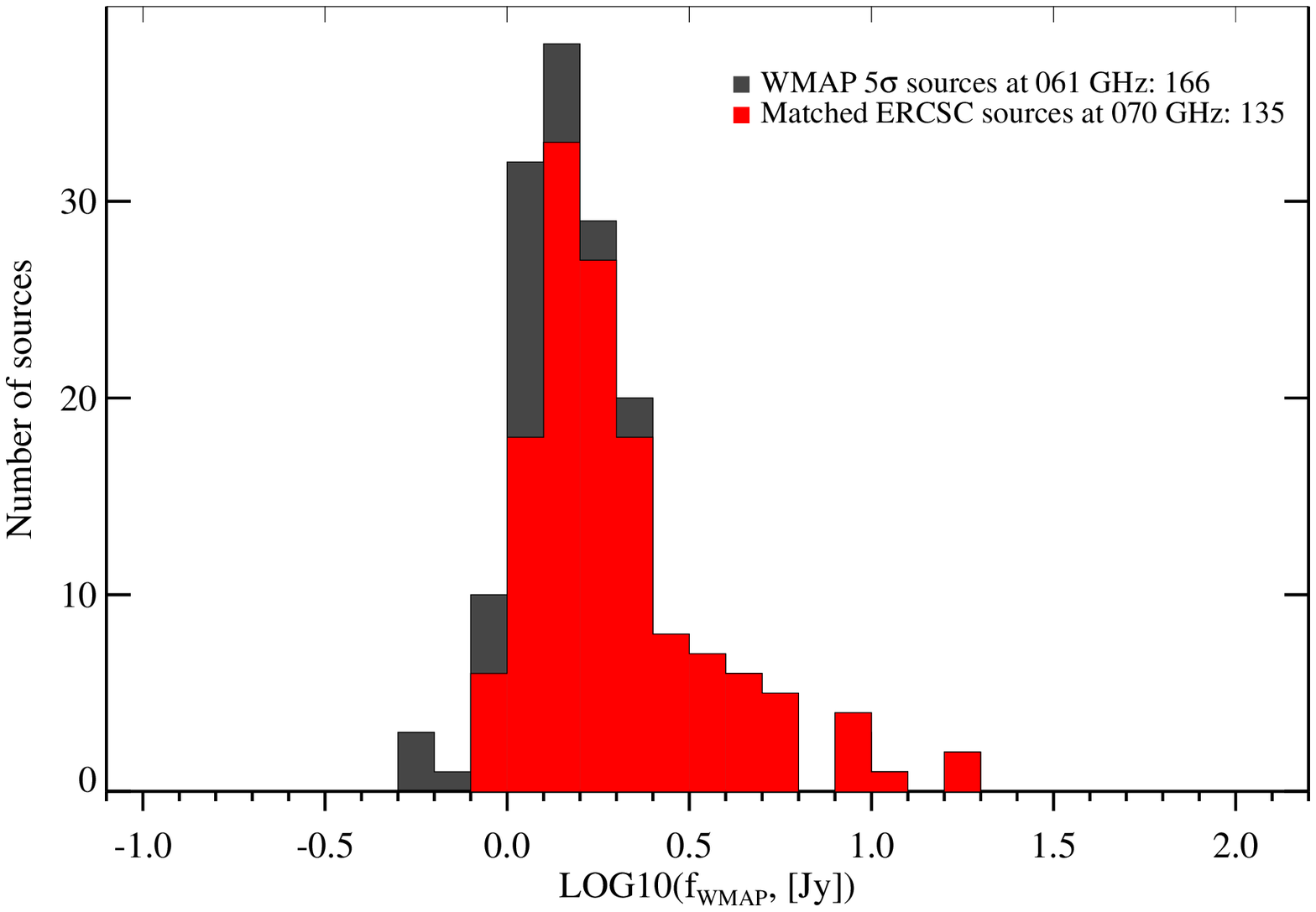}
\includegraphics[width=8cm]{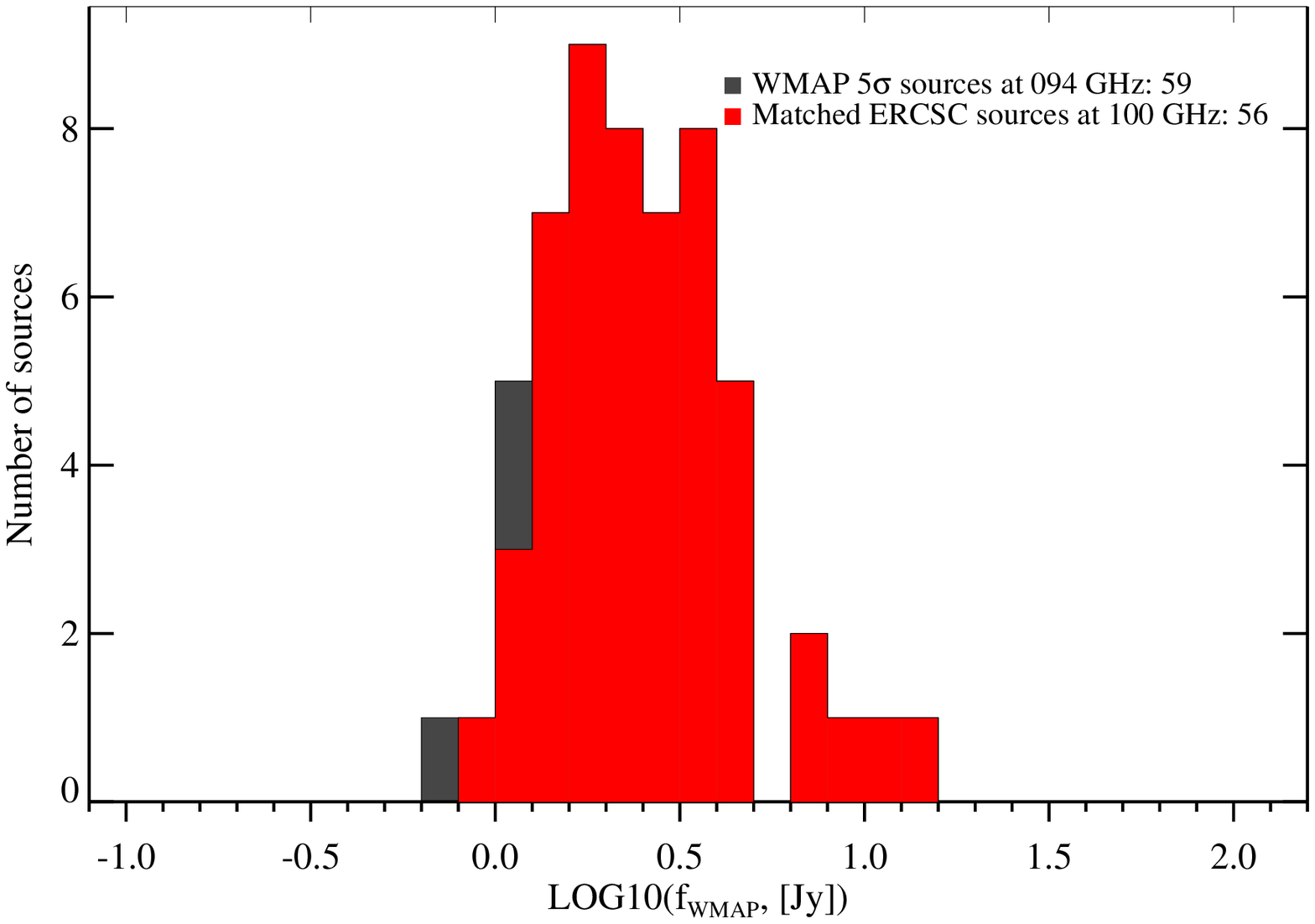}
\caption[Comparison between ERCSC and WMAP]{Histogram distribution of {\it WMAP} flux densities for all {\it WMAP}
5$\sigma$ sources in each band (gray region). The sources
that are detected in the ERCSC are shown as the red histogram. Some of the {\it WMAP} sources have been
missed because of source variability.}
\label{fig:hist_wmap}
\end{figure*}

\begin{figure*}
\centering
\includegraphics[width=8cm]{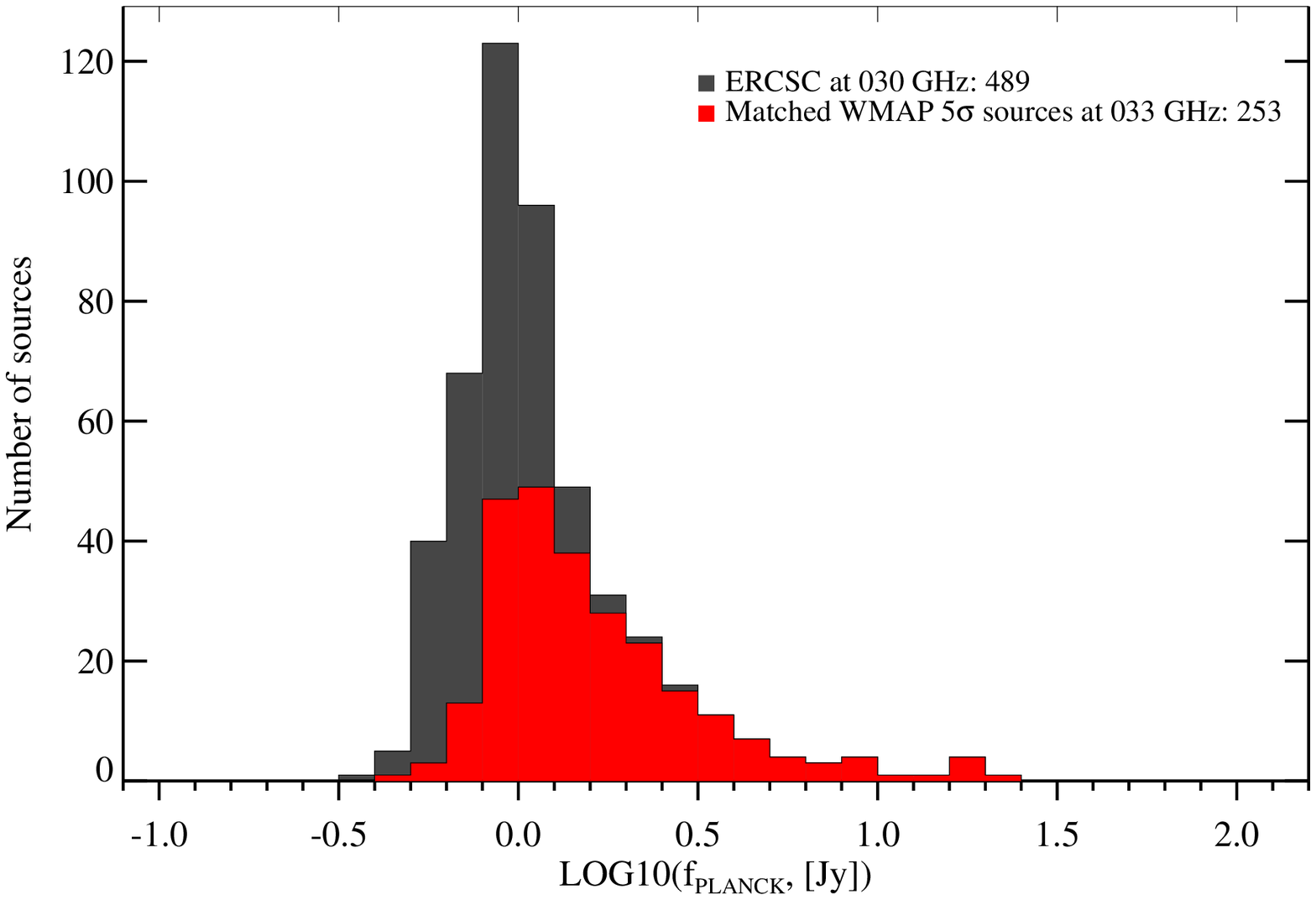}
\includegraphics[width=8cm]{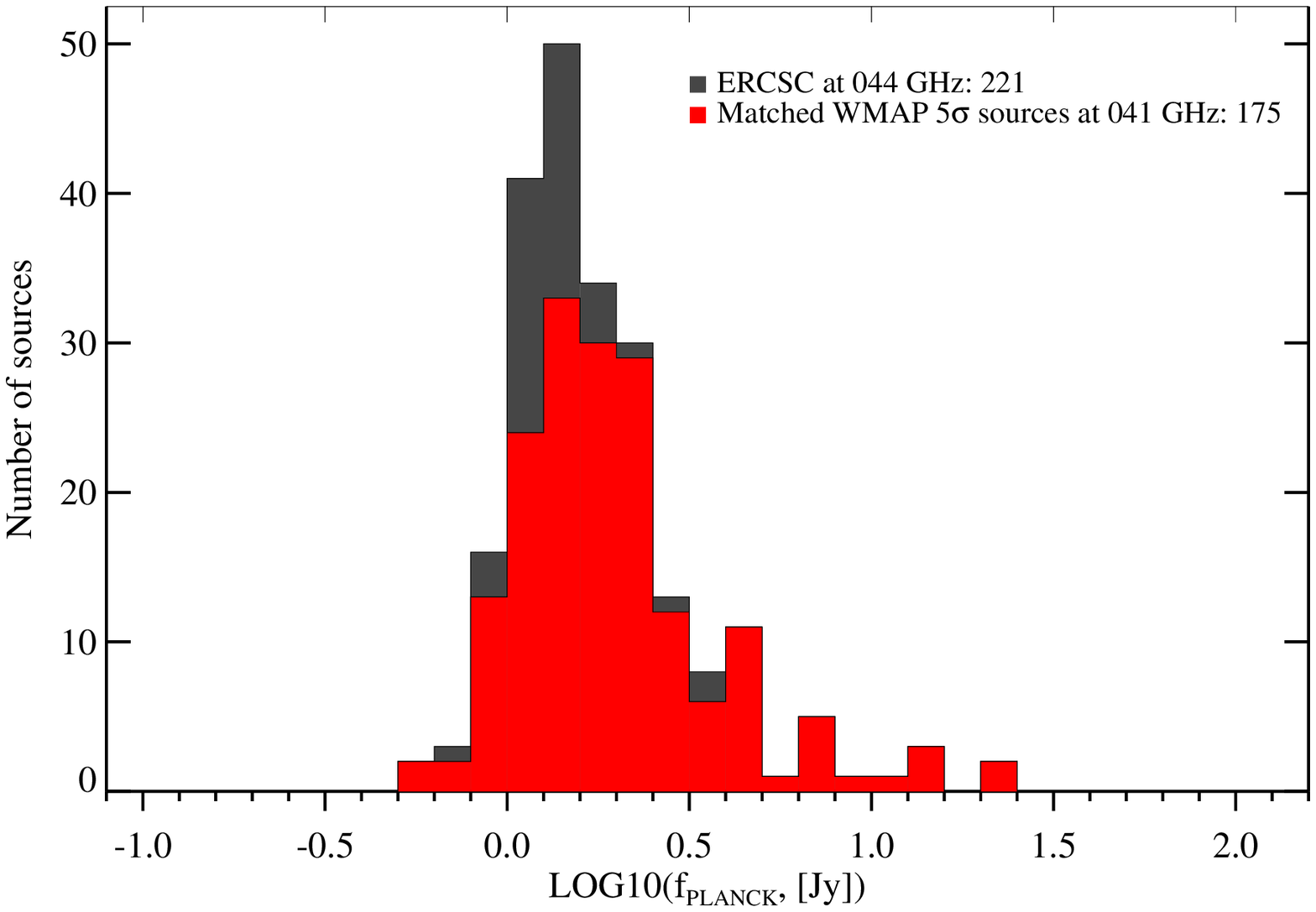}
\includegraphics[width=8cm]{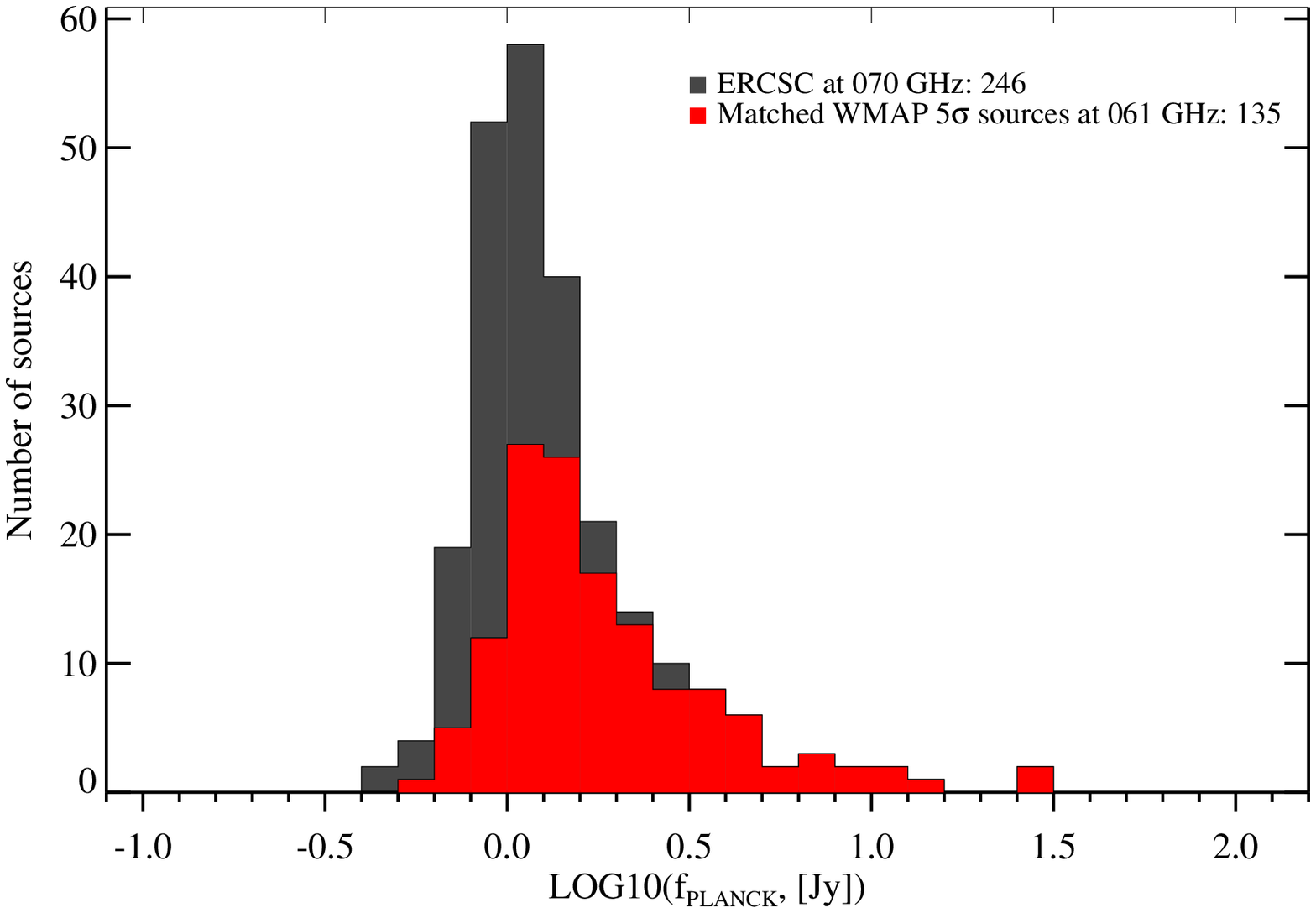}
\includegraphics[width=8cm]{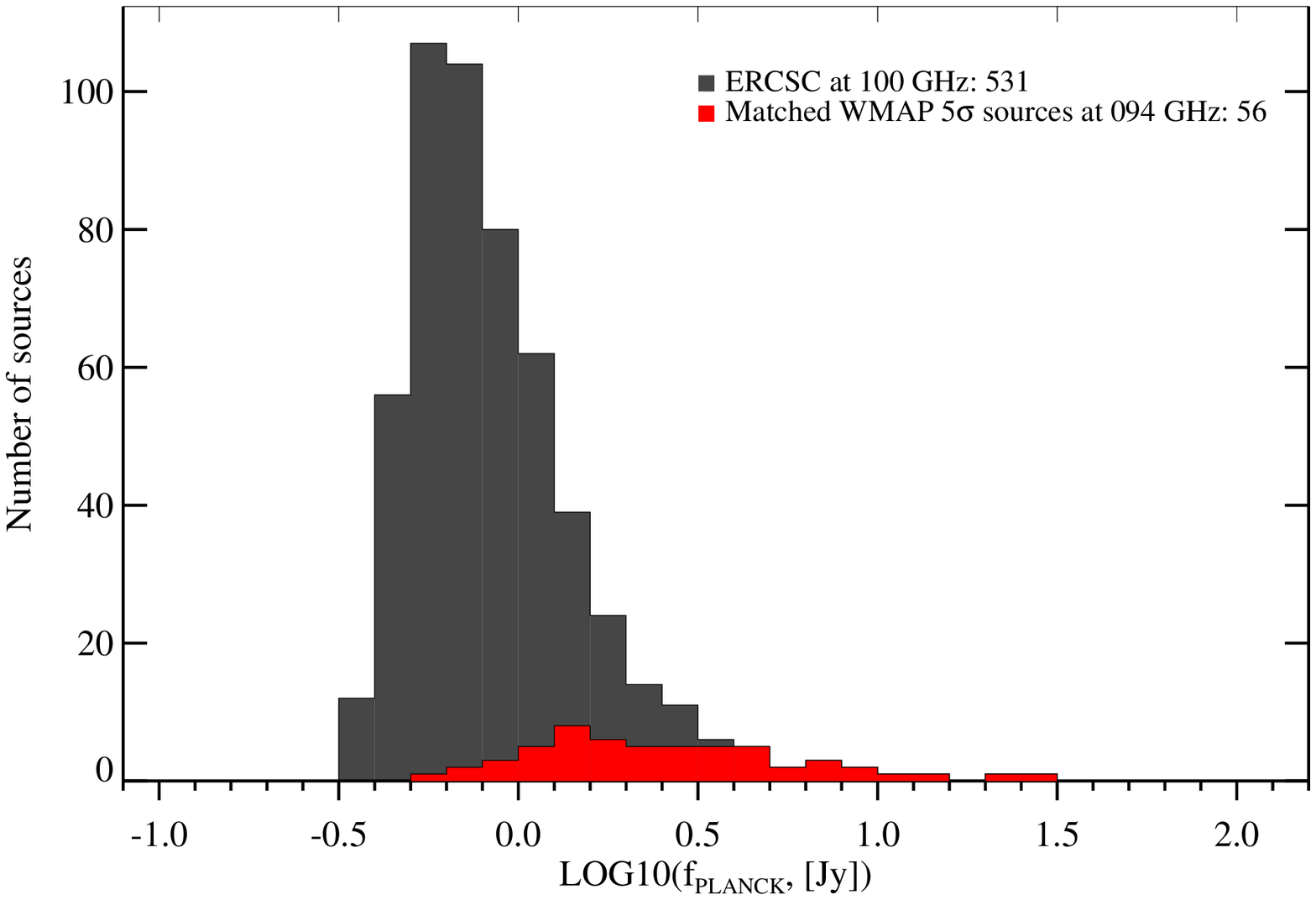}
\caption[Comparison between ERCSC and WMAP]{Histogram distribution of ERCSC flux densities
at each band in gray. ERCSC sources that are matched with
{\it WMAP} 5$\sigma$ sources in a similar band are shown as the red histogram. The {\it WMAP} 7 
year point source catalogue mask (see text) has been applied to the ERCSC to ensure the same sky coverage. }
\label{fig:hist_planck}
\end{figure*}

A comparison between the ERCSC sources and the {\it WMAP} point source catalogue was also undertaken.
The {\it WMAP} seven-year catalogue \citep{gold2010} contains a total of 471 sources in the five {\it WMAP} bands. We have 
compared the {\it WMAP} 5$\sigma$ sources at 33, 41, 61, 94 GHz with the sources in the ERCSC at 30, 44, 70 and 
100 GHz, respectively. A search radius corresponding to the FWHM of 
the {\it WMAP} beam at each frequency (0.66$\degr$, 0.51$\degr$, 
0.35$\degr$, 0.22$\degr$ at 33 to 94 GHz channels) is used to find a match of {\it WMAP} sources in the ERCSC. 
Figure \ref{fig:hist_wmap} shows the histogram distribution of {\it WMAP}
flux densities; the {\it WMAP} 5$\sigma$~sources are shown in gray, and the ones with an 
ERCSC match are in red. The ERCSC include 88$\%$, 62$\%$, 81$\%$ and 95$\%$~of the {\it WMAP} 5$\sigma$ sources 
at the four bands, individually. Figure \ref{fig:hist_planck} is a similar plot, but shows the 
histogram distribution of the ERCSC flux densities: the ERCSC 
sources are shown in gray, and the ones with a {\it WMAP} match are in red. The {\it WMAP} seven-year 
point source catalogue mask which excludes the Galactic Plane and the LMC/SMC
region has been applied to the ERCSC beforehand to ensure the same sky coverage. It 
is evident that the ERCSC is a much deeper and more complete catalogue than the {\it WMAP} 7 year catalog, 
especially at the 100 GHz channel. 

The {\it WMAP} 5$\sigma$ detections that are missed in the ERCSC at 30, 70 and 100 GHz are further 
investigated. The 44 GHz channel is skipped since it is known to have lower sensitivity compared to the {\it WMAP} 7-year data. It is found that 
at 100 GHz, all the missed {\it WMAP} sources can be explained by either the {\it WMAP} source not having a 5 GHz counterpart or only being weakly associated
with a 5 GHz source suggesting that the {\it WMAP} source might be spurious. At 70 GHz,  $\sim41\%$ of the unmatched sources 
are variable (this is a lower limit as the variability info was obtained from the {\it WMAP} five-year catalog, which is a subset of the {\it WMAP} 
seven-year catalog), $\sim13\%$ of the unmatched sources have no 5 GHz ID or are only loosely associated with a 5 GHz source, $\sim38\%$ are recovered after the CMB subtraction. At 30 GHz, $\sim17\%$ of the unmatched sources are 
variable (again, this is only a lower limit), $\sim34\%$ of the unmatched sources have 
no solid identification, $\sim54\%$ are recovered after the CMB subtraction. This analysis
suggests that the reason these sources are not detected in \Planck\ is a combination of source variability, map 
sensitivity (different scanning strategy of {\it WMAP} and \Planck\ result in a difference in the local background noise; also the
ERCSC is based on 1.6 sky surveys whereas the {\it WMAP} catalogue is based on 14 sky surveys), and incompleteness of the ERCSC.

\begin{figure*}
\centering
\includegraphics[width=8cm]{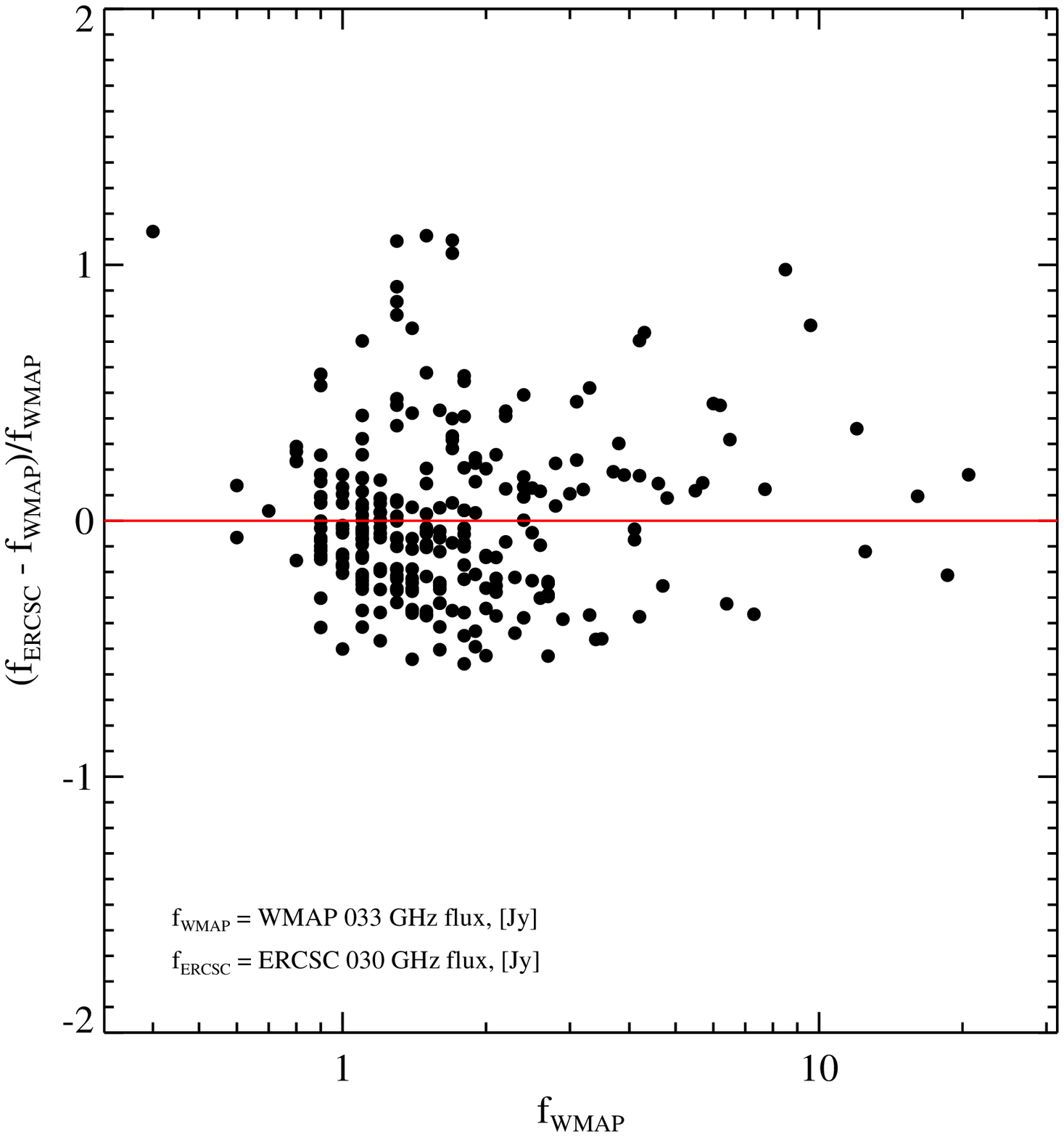}
\includegraphics[width=8cm]{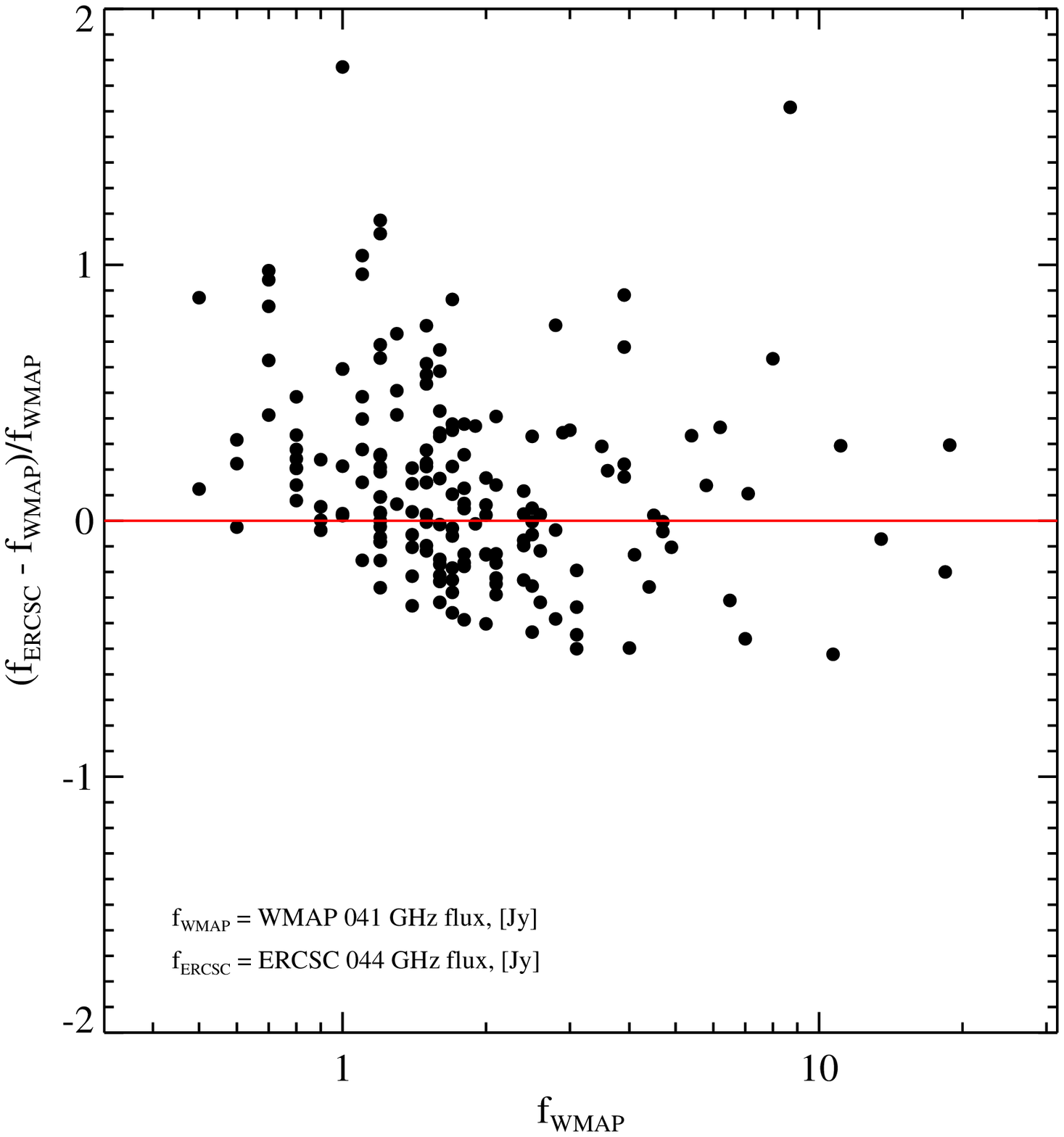}
\includegraphics[width=8cm]{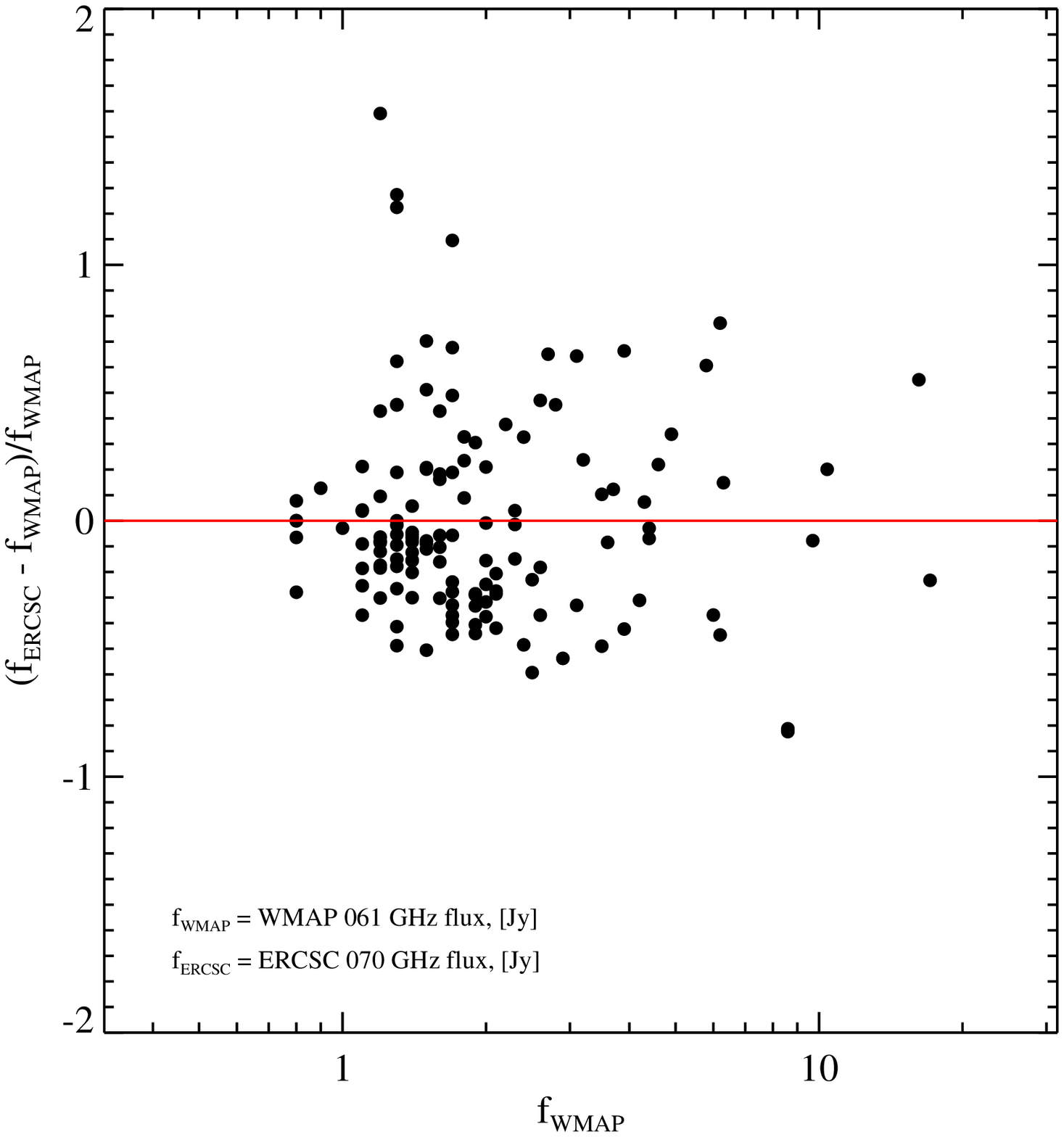}
\includegraphics[width=8cm]{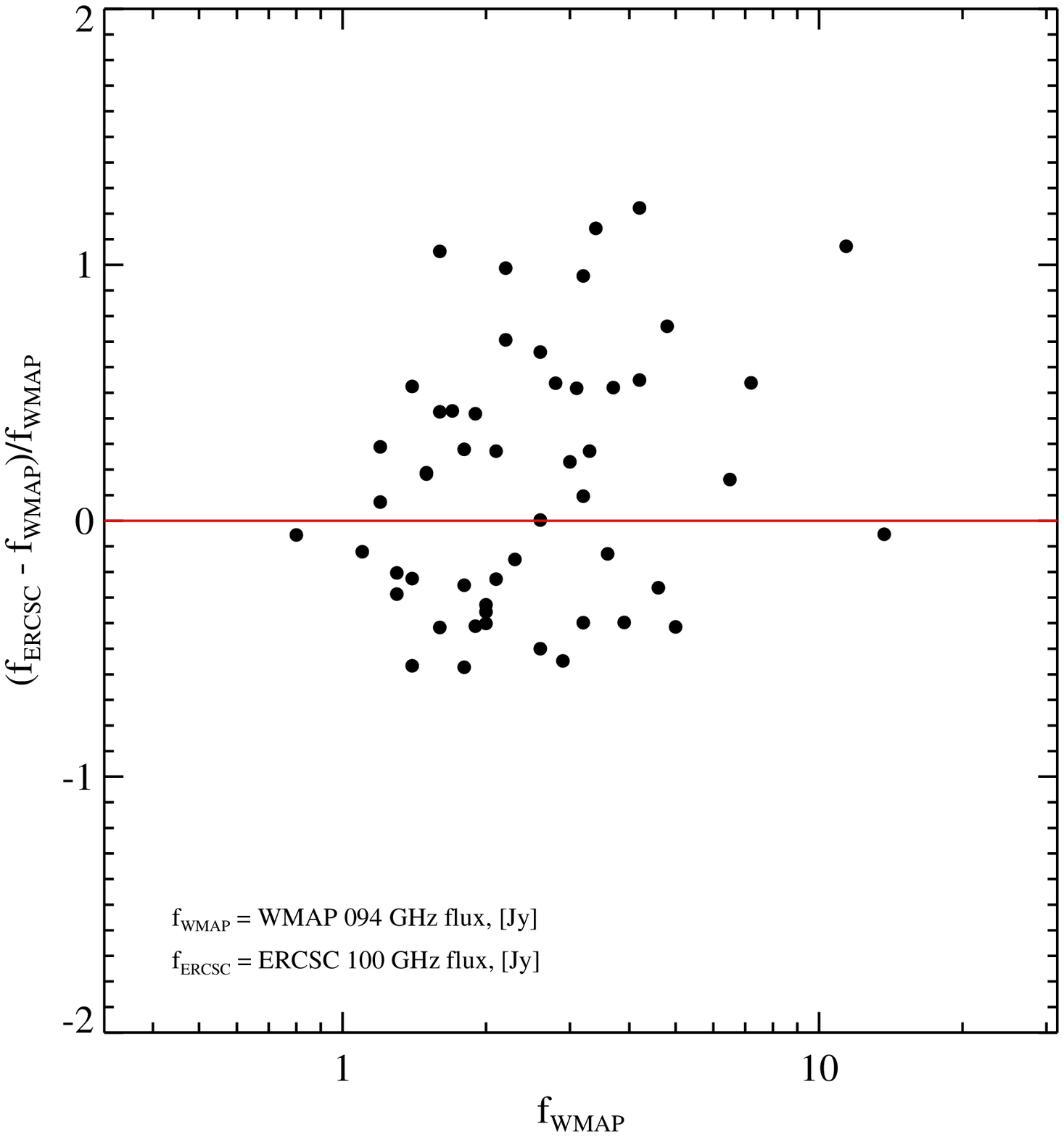}
\caption[Comparison between ERCSC and WMAP flux densities]{Fractional difference between the ERCSC flux densities and {\it WMAP}
flux densities at 30, 44, 70 and 100 GHz. The unit of the abcissa is Jy while the ordinate shows the fractional difference.
No correction has been applied to the {\it WMAP} flux densities to account for the difference
in bandpass compared to \Planck. The agreement is good but the significant scatter that can be seen is most likely
due to source variability.}
\label{fig:wmapfnu}
\end{figure*}

The similarity between the {\it WMAP} frequencies and \Planck\ bands also motivates a comparison between their flux densities which is shown in Figure \ref{fig:wmapfnu}. Overall we find
there is no systematic difference between the
{\it WMAP} and ERCSC flux densities at the corresponding bands. The significant scatter in Figure \ref{fig:wmapfnu}
again indicates that variability is an issue. There is no variability analysis of the {\it WMAP} seven-year 
point sources, but an analysis of the variability on the five-year {\it WMAP} point sources mentioned above,
shows that a high fraction of the 
sources are variable at greater than 99$\%$ confidence, and these are in general the brighter sources \citep{wright2009}.

\section{The ERCSC: Access, Contents and Usage}\label{sec:ERCSC_access}

The ERCSC is available from both the ESA \Planck\ Legacy 
Archive\footnote{\url{http://www.sciops.esa.int/index.php?project=planck&page=Planck_Legacy_Archive}} and the NASA
Infrared Science Archive\footnote{\url{http://irsa.ipac.caltech.edu/Missions/planck.html}} (IRSA).

The source lists contain 35 columns per source at the LFI bands and 36 columns at the HFI bands. The 857\,GHz source list has six additional columns 
which consist of the band-filled flux densities and flux uncertainties
at the three adjacent lower frequencies, 217, 353 and 545\,GHz for
each source detected at 857\,GHz. The locations of sources are provided in Galactic coordinates. In addition, we 
also provide
for each detected source, a postage-stamp cutout of the source
from the all sky map of the corresponding frequency after the CMB has been subtracted. The size of the cutout is a square of side 4$\times$FWHM
at the corresponding frequency. The primary purpose of these cutouts is to aid in the visual
validation of sources. We also provide notes in a text file, one per frequency, for particular sources in the catalogues which
state associations of the ERCSC source with sources in ancillary catalogues (e.g. IRAS, GB6, WMAP)
as well as potential variability information.

Including the ECC and ESZ,
the entire data release thus consists of 11 source list files, 11 all-sky source distribution maps, 11 notes files and
postage-stamp cutouts in JPEG format of all the sources detected at the nine individual frequencies as well as in the ECC list.
No postage stamp cutouts are provided for the ESZ. 

\subsection{Catalogue Contents and Usage}\label{sec:catalogue}

The key columns in the catalogues are:
\begin{enumerate}
\item source identification: NAME (string)
\item position: GLON, GLAT, POS\_ERR which gives the Galactic coordinates in degrees and the estimated
1$\sigma$ positional uncertainty in arcminutes.
\item flux density: FLUX, FLUX\_ERR in mJy measured in a  circular aperture with radius equal to the nominal FWHM of the beam.
\end{enumerate}
The one additional column for the HFI bands compared to the LFI bands is due to the inclusion of a cirrus estimate, described below.

Individual sources can be searched for in the list either by
Galactic coordinates (GLON, GLAT), or by the equivalent J2000 equatorial
coordinates (RA, DEC). The 1$\sigma$ positional uncertainty for a
source, given by POS\_ERR in arcminutes, depends on the local background RMS
and SNR. This uncertainty is only a measure of the
uncertainty for fitting the location of the source in the maps and
does not take into account any astrometric offset in the maps. 
Furthermore, POS\_ERR is measured from the positional
uncertainty of artificial point sources injected into the maps. As a result,
sources might have larger positional uncertainties which are
not reflected in this value (See Section 2).

When a source is classified as extended, we set EXTENDED=1.
This implies that the square root of the product of the major
and minor axis of the source is 1.5 times larger than the square root
of the major and minor axis of the estimated \Planck\ point spread
function at the location of the source, i.e., 
\begin{eqnarray}
\sqrt{\mathrm{GAU\_FWHMMAJ} \times \mathrm{GAU\_FWHMMIN}} &&\nonumber \\ > 1.5\times
\sqrt{\mathrm{BEAM\_FWHMMAJ} \times \mathrm{BEAM\_FWHMMIN}}
\end{eqnarray}
In the
upper HFI bands, sources which are extended tend to be associated with
structure in the Galactic interstellar medium although individual 
nearby galaxies are also extended sources as seen by \Planck\ [see \citet{planck2011-6.4a}].
The choice of the threshold being set at 1.5 times the beam is motivated by the accuracy with which
source profiles can be measured from maps where the Point Spread Function is critically sampled (1.7\arcm\ pixel scale
for a $\sim$4\arcm\ FWHM).
Naturally, faint sources for which the Gaussian profile fit might have failed do not have the EXTENDED tag set.

As described in Sec.~\ref{sec:flux_density_estimation}, four measures of flux density are provided in mJy. 
For extended sources, both FLUX and PSFFLUX will likely be 
significant underestimates of the true source flux density. Furthermore, at faint flux densities
corresponding to low signal-to-noise ratios (less than 20), the PSF fit might
have failed. This would be represented either by a negative flux
density or by a significant difference between the PSFFLUX and FLUX
values.  In general, for bright extended sources, we recommend using
the GAUFLUX and GAUFLUX\_ERR values although even these might be
biased high if the source is located in a region of complex, diffuse
foreground emission.

Uncertainties in the flux density measured by each technique are
reflected in the corresponding ``\_ERR'' column.  The flux
uncertainties derived from the artificial point sources injected into the
maps are available in MCQA\_FLUX\_ERR.  MCQA\_FLUX\_ERR is the
standard deviation of the dimensionless $(S_\textrm{input}-S_\textrm{output})/S_\textrm{input}$ for input and output flux densities
$F$ based on the aperture
flux density (i.e., FLUX) at the signal-to-noise ratio of the source.
We believe that the most conservative
flux uncertainty is the quadrature sum of the Monte-Carlo flux
uncertainty and the ``\_ERR'' value relevant for the appropriate flux density
(FLUX, PSFFLUX, GAUFLUX or FLUXDET).

MCQA\_FLUX\_BIAS provides the median in the difference between the injected flux
and extracted aperture flux of the artificial point sources. In principle, the
bias should be close to zero if the aperture corrections are precisely known, the
aperture is perfectly centred on each source,
and the background can be precisely estimated. In practice, there is an offset of a 
few percent,
which can become large at the lowest signal-to-noise ratios or in high-background
regions. This is a median offset estimated as a function of SNR
from the artificial point sources,
and has already been applied to the FLUX value of all sources. The bias correction
has been applied such that the FLUX in the catalogue is the measured flux density
divided by (1-MCQA\_FLUX\_BIAS) and increases the flux density values by about 5\%.

The 1$\sigma$ point source flux uncertainty due to structure in the
background is given in BACKGROUND\_RMS in units of mJy.  At the lowest
frequencies this is a combination of CMB noise and instrumental noise,
with the latter dominating.  At 143\,GHz, the noise is dominated by the
CMB. At higher frequencies, it is dominated by Galactic ISM. The
ratio of source flux density to BACKGROUND\_RMS is the primary
parameter which is used to calibrate the RELIABILITY of sources.

The dates on which the source was observed are included in DATESOBS
(UTC) in the yyyymmdd format. This will be useful in the analysis of
time-variable sources. The flux-density value in the ERCSC
is an average over all the dates of observations.

Sources in the HFI bands each have a CIRRUS number which is based on the
number of sources (both low and high reliability) within a 2$\degr$ radius of the source,
in raw 857\,GHz catalogues derived from the maps. The number has been normalised
to a peak value of one. The normalisation factor is in practice,
derived from the number density of sources in the Large Magellanic Cloud region
where the maximum number of 857\,GHz sources is located.

Finally, each source has a CMBSUBTRACT flag. This flag has values of
0, 1 or 2. The value is 0 if the source is detected in the CMB-subtracted maps and has an aperture flux difference
$|S_\textrm{intensity}-S_\textrm{nocmb}|/S_\textrm{intensity}<0.30$. CMBSUBTRACT=1 if the
source is detected in the CMB subtracted maps but has a flux
difference of greater than 30\%.  CMBSUBTRACT=2 if the source is not detected
in the CMB subtracted maps. CMB subtraction results in artefacts in
the maps which might remove real sources. It is recommended that a
conservative user who wants a guarantee of source detection in follow-up observations neglect sources with CMBSUBTRACT=2.

\begin{figure}
\centering
\resizebox{\hsize}{!}{\includegraphics{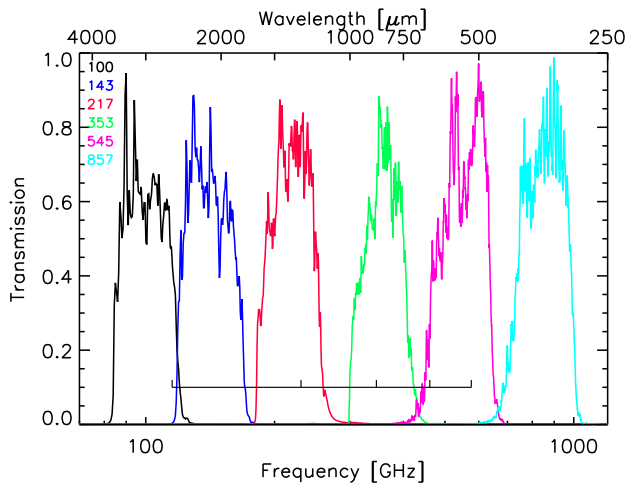}}
\caption[HFI Bandpasses]
{Bandpasses of HFI with the location of bright CO rotational lines ($J=1\rightarrow0$, $2\rightarrow1$, $3\rightarrow2$, $4\rightarrow3$, and $5\rightarrow4$ from left to right)
shown as a horizontal black line with tick marks.
The CO lines can introduce a significant positive bias in the flux density of the sources, particularly those associated
with Galactic star-forming regions. The effect is the most significant at 100\,GHz where the flux density may be
boosted by more than 50\%. See \citet{planck2011-1.7} for details.
\label{fig:bpco}}
\end{figure}

\subsection{Cautionary Notes in Usage of Catalogues}\label{sec:caveats}

In this section, we list some cautionary notes associated with usage of the ERCSC list.

\begin{itemize}
\item {\it Statistical Character:}
The ERCSC list is an early list of highly reliable sources from the first \Planck\ all sky survey.
It is not a flux density limited sample or even a complete sample
of sources and therefore care should be taken before undertaking statistical studies
such as source counts. 
This is partly due to the fact that the scan strategy results in
significant variation in instrumental sensitivity as a function of position on the sky. In addition, the
relative contribution of astrophysical
sources of ``noise" such as the CMB and the emission from the Galactic interstellar medium (ISM)
vary across the \Planck\ frequencies. The CMB contribution peaks between 100 and 143 GHz while the ISM contribution
peaks above 857 GHz. In conjunction with the varying spatial resolution, this results in varying limits to the 
sensitivity of sources that can be detected both as a function of position on the sky and as a function of frequency.
The Monte-Carlo analysis presented
later, does quantify this variation in sensitivity for the overall catalogue. However, the estimates for the
fraction of sky area above a particular completeness limit, have not been factored into the catalogue.

\item {\it Variability:}
At radio frequencies, many of the extragalactic sources are highly variable. A small fraction of them
vary even on time scales of a few hours based on the brightness of the same source as it passes through the different \Planck\ horns. 
Follow-up observations of these sources might show significant differences
in flux density compared to the values in the data products. 
Although the maps used for the ERCSC are based on 1.6 sky coverages, 
the ERCSC provides only a single average flux density estimate over all \Planck\ data samples that were included in the all sky maps
and does not contain any measure of the variability of the sources. 
The \Planck\ Quick Detection System \citep[QDS;][]{Aatrokoski2010} attempts to quantify the variability of sources seen by \Planck. 
The information from the QDS has been included in the notes for certain sources.

\item {\it Contamination from CO:}
At infrared/submillimetre frequencies (100\,GHz and above), the \Planck\
bandpasses straddle energetically significant CO lines (Figure~\ref{fig:bpco}). The effect is the most significant at 100\,GHz, where
the line might contribute more than 50\% of the measured flux density. Follow-up observations of these sources, especially
those associated with Galactic star-forming regions, at a similar frequency but different bandpass, should correct for
the potential contribution of line emission to the continuum flux density of the source. 
See \citet{planck2011-1.7} for details.

\item {\it Photometry:}
Each source has multiple measures of photometry FLUX, GAUFLUX, PSFFLUX and FLUXDET as defined above. The appropriate photometry
to be used depends on the nature of the source. For sources which are unresolved at the spatial resolution of \Planck,
FLUX and PSFFLUX are most appropriate. Even in this regime, PSF fits of faint sources fail and consequently these have
a PSFFLUX value of ``NaN'' (``Not a Number''). For bright resolved sources, GAUFLUX might be most appropriate although GAUFLUX appears to
overestimate the flux of sources close to the Galactic plane due to an inability to fit for the
contribution of the Galactic background at the spatial resolution of the data. For faint resolved sources
in the upper HFI bands, FLUXDET, which is the flux density in an elliptical Kron aperture provided by SExtractor,
might give the most accurate numbers. The user should also note that the absolute calibration of flux-density values
are required to be accurate to within about 30\% although the signal-to-noise of the sources are much higher.

\item {\it Cirrus/ISM:}
A significant fraction of the sources detected in the upper HFI bands could be associated with Galactic
interstellar medium features or cirrus. The {\it IRAS} 100\micron\ surface brightness in MJy\,sr$^{-1}$ for each of the sources,
which is commonly used as a proxy for cirrus,
is available through a search of the ERCSC with IRSA.
Candidate ISM features can also be selected by choosing objects with EXTENDED=1 although nearby Galactic and extragalactic sources which
are extended at \Planck\ spatial resolution will meet this criterion. Alternately, the value of CIRRUS in the catalogue can be utilised
to flag sources which might be clustered together and thereby associated with ISM structure.

\end{itemize}

\section{Astrophysical Source Classes Identified by their Multifrequency Signature}

In addition to the single frequency catalogs described in the previous sections, there are two 
other source catalogs that are provided as part of the ERCSC.
These two additional 
catalogs leverage the spectral signature of two specific classes of astrophysical sources through the \Planck\ bands
and are generated using specialized multifrequency algorithms which have been developed within the \Planck\ Collaboration.
These are a list of galaxy clusters detected through the Sunyaev-Zeldovich effect and cold pre-stellar cores
identified by the derived far-infrared color temperature in fits to the \Planck\ photometry.

\subsection{The Early Sunyaev-Zeldovich Cluster Catalogue}

The \Planck\ Early Release Sunyaev-Zeldovich (ESZ) cluster sample [described in more detail in \citet{planck2011-5.1a}] is a list of 189 SZ
cluster candidates which are detected by their multi-frequency signature in
the \Planck\ bands. The thermal SZ effect is the result
of energetic electrons in the hot intra-cluster medium inverse-Compton scattering
off the CMB photons. The net result is a distortion in the shape of the CMB
spectrum, which results in a deficit of flux density below $\sim$220\,GHz and an
increment in flux density at higher frequencies \citep{sz72, carl02}. By
utilising a matched multi-frequency filter, the spectral signature of this
distortion can be detected and measured in the \Planck\ all-sky maps, which
enables cluster candidates to be detected.

The ESZ sample generated as part of the \Planck\ early data release is
the result of a blind multi-frequency search in the all sky maps, i.e.,
no prior positional information on clusters detected in any
existing catalogues was used as input to the detection algorithm. 
The ESZ sample is produced using one of the four matched
multi-frequency filter (MMF) algorithms available within the
\Planck\ collaboration (hereafter MMF3; see \citet{melin:2010} for
details of the comparison of the cluster extraction algorithms
available within the collaboration).  
MMF3 is an all-sky extension of
the algorithm described in \citet{melin2006} and is run blindly over
the six HFI frequency maps. The technique first divides the all-sky maps into a
set of overlapping square patches.  The matched multi-frequency filter
then optimally combines the six frequencies of each patch, assuming the
SZ frequency spectrum and using the \citet{arnaud:2010} pressure profile
as the cluster profile.  Auto- and cross- power
spectra used by the MMF are directly estimated from the data. They are
thus adapted to the local instrumental noise and astrophysical
contamination such as ISM emission.  For each patch, the scale radius of the cluster
profile is varied to maximise the signal-to-noise ratio of each
detection. The algorithm thus assigns to each detected source an
estimated size and an integrated flux.  The detected sources extracted
from individual patches are finally merged into an all-sky cluster
list. Non-SZ sources captured by the MMF algorithm can contaminate the
list and an additional step of validation of the detection is needed.

Unlike the individual frequency source list or the ECC list, which are validated
through a Monte-Carlo technique, the reliability of the ESZ list has
been estimated through a validation process based on internal checks
and on cross-checks with ancillary optical/near-infrared and X-ray
cluster catalogues or images. Cross-matches with the 
Meta-Catalogue of X-ray detected
Clusters of galaxies \citep[MCXC hereafter;][]{Piff2010}, Abell and Zwicky
catalogues, SDSS-based catalogues, MAXBCG and Wen et al. (2009) and a compilation of SZ
observed clusters were undertaken.  
For each known X-ray cluster, several entries are available
among which the identifiers, redshift, coordinates, total mass
$M_{\mathrm 500}$, and radius $R_{\mathrm 500}$ were 
used during the external validation process.
R$_{\mathrm 500}$ is the radius 
that encompasses a mean matter density which is 500 times the critical density 
at the corresponding redshift. R$_{\mathrm 500}$ is less than the virial radius of the 
cluster. $M_{\mathrm 500}$ is the mass within $R_{\mathrm 500}$.
Further searches in Virtual Observatory (VO) and in logs of
observatories were performed. The goal of this search was to identify cluster candidates which might already
have ancillary data available for community access.
Of the 189 cluster candidates, 169 are associated
with known X-ray or optical clusters and the \Planck\ data provides the first
measure of the SZ signature for the majority of them.
In addition to the cross-check with ancillary data, follow-up observations with XMM
confirmed an additional 11 new clusters which are described in \citet{planck2011-5.1b}.
9 other new clusters have not been confirmed in the X-ray as yet.

\begin{figure}
\centering
\resizebox{\hsize}{!}{\includegraphics[width=15cm]{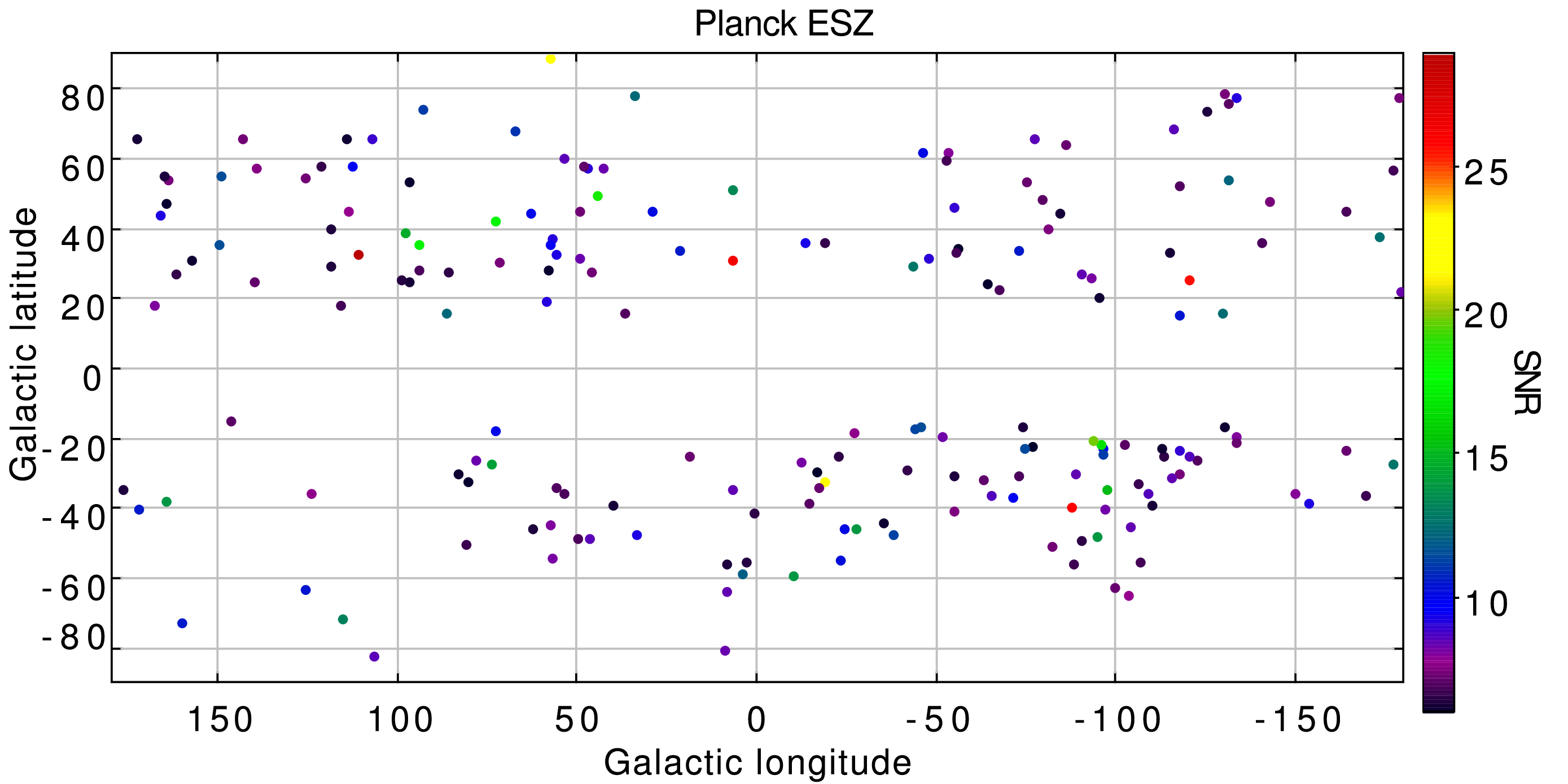}}
\caption[All-sky distribution of ESZ cluster candidates]{
Plot showing the all sky distribution of the ESZ cluster
candidates colour coded by signal-to-noise ratio. Sources close to the Galactic 
Plane have been excluded since the spurious fraction is
high.
}
\label{fig:esz1}
\end{figure}

\begin{figure}
\centering
\resizebox{\hsize}{!}{\includegraphics{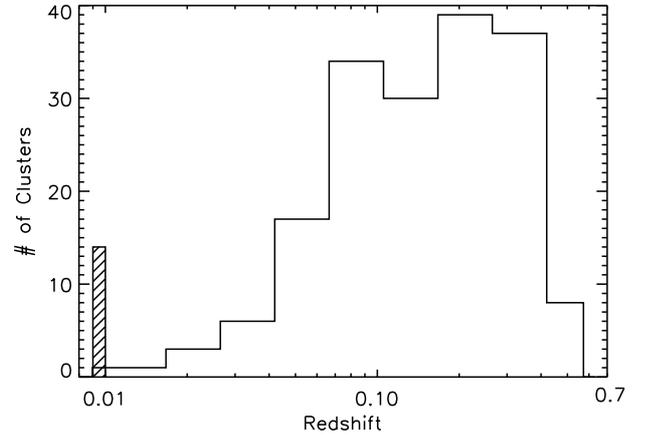}}
\caption[Redshift distribution of ESZ cluster candidates]{
Redshift distribution of ESZ cluster candidates.
ESZ clusters which do not have a redshift are shown 
as the hatched region at $z=0.01$. 
}
\label{fig:esz2}
\end{figure}

A full description of the validation effort is in \citep{planck2011-5.1a}.
Figure~\ref{fig:esz1}
shows the all sky distribution of the clusters and cluster candidates while
Figure~\ref{fig:esz2} shows their redshift distribution.
Table \ref{tab:eszcol} gives the list of columns in the ESZ catalogue.

All clusters have a \Planck\ name which is given in
the column NAME. This name is constructed from 
GLON and GLAT, the best estimated Galactic coordinates of
the SZ signal. SNR gives the
detection's signal-to-noise ratio as defined by the matched 
multi-filter method MMF3.

When a \Planck\ SZ cluster candidate is identified as an X-ray cluster
in the MCXC the coordinates of the X-ray 
counterpart (i.e., the X-ray centroid) is given. 
The same positional information is given for the \Planck\ cluster candidates
confirmed by XMM-Newton observation (apart from one candidate
identified with a double cluster, see notes below). For those
clusters with an X-ray counterpart, the Compton-$Y$ parameter, which is the integral
of the Compton-$y$ over the cluster area, is re-extracted from the \Planck\ maps using the X-ray centroid coordinates
and X-ray size THETA\_X as priors, yielding the value 
Y\_PSX and its error Y\_PSX\_ERR. The Compton-$Y$ parameter measured
using the X-ray position and size priors is known to be more robust
than the blind value estimated without priors \citep{planck2011-5.1a}.

For cluster candidates without available estimates of X-ray
position or size, the derived SZ parameters THETA, Y, and
the associated errors THETA\_ERR and Y\_ERR are the values
returned directly by the matched filter. These are likely to be more
uncertain than cases where the cluster has been confirmed in the X-ray
data. THETA and THETA\_X are the estimated angular size of the cluster
at 5 times R$_{\mathrm 500}$. 

Notes on individual clusters
can be found in \citet{planck2011-1.10sup}.
These notes include cross-matches with ERCSC sources as well as the origin
of the redshift.

\begin{table*}
\centering
\caption{ESZ Catalogue Columns}
\label{tab:eszcol}
\begin{tabular}{l l}
\hline\hline
Keyword & Type \\
\hline
INDEX &  Index of clusters i.e., 1, 2, 3...\\
NAME  & \Planck\ name of cluster candidate \\
GLON  & Galactic Longitude from \Planck\ (deg)\\
GLAT  & Galactic Latitude from \Planck\ (deg)\\
RA    & Right Ascension (deg) from \Planck\ (J2000) \\
DEC   & Declination (deg) from \Planck\ (J2000) \\
SNR   & Signal-to-noise ratio returned by the matched multi-filter (MMF3) \\
ID    & External identifier of cluster e.g., Coma, Abell etc. \\
REDSHIFT & Redshift of cluster from the MCXC X-ray cluster
compilation (Piffaretti et al. 2010) unless stated otherwise in the notes\\
GLON\_X  & Galactic Longitude of the associated X-ray cluster (deg) \\
GLAT\_X  & Galactic Latitude of the associated X-ray cluster (deg) \\
RA\_X    & Right Ascension (deg) of the associated X-ray cluster (J2000) \\
DEC\_X   & Declination (deg) of the associated X-ray cluster (J2000) \\
THETA\_X & Angular size (arcmin) at 5R500 from X-ray data. \\
Y\_PSX   & Integrated Compton-Y (arcmin$^{2}$) at X-ray position and within 5R500 (THETA\_X) \\
Y\_PSX\_ERR & Uncertainty in Y\_PSX \\
THETA &      Estimated angular size (arcmin) from matched multi-filter (MMF3) \\
THETA\_ERR & Uncertainty in THETA \\
Y         & Integrated Compton-Y (arcmin$^{2}$) at \Planck\ position and within THETA from matched multi-filter (MMF3) \\
Y\_ERR    & Uncertainty in Y \\
\hline
\end{tabular}
\end{table*}

\subsection{The Early Cold Cores Catalogue}

Pre-stellar cloud cores represent the transition from turbulence dominated
large scales to the gravitation dominated protostellar scales and are
therefore a crucial step in the process of star formation. Imprinted in their
structure and statistics is information of the properties of the parental
clouds and the core formation processes where interstellar turbulence,
magnetic fields, self-gravity, and external triggering all play a role.

The \Planck\ {\em all-sky} submillimetre/millimetre survey has both the
very high sensitivity and spatial resolution required for the detection of
compact cores.
The highest frequency channels at 857, 545 and 353\,GHz
cover the frequencies around and longwards of the intensity maximum of
the cold dust emission: $B_{\nu}(T=10{\rm K}) \nu^2$ peaks at a wavelength close to
300\micron while, with a temperature of $T\sim 6$\,K, the coldest dust inside
the cores has its maximum close to 500\micron. When \Planck\ data are combined
with far-infrared data like the {\it IRAS} survey, the observations enable accurate
determination of both the dust temperatures and its spectral index. 
For historical reasons, we use ``Cold Cores'' to designate the entries in
the ECC, since pre-stellar cores were a major scientific goal of this product.
However, as two  companion papers \citep{planck2011-7.7a, planck2011-7.7b} demonstrate,
most of these entries are more correctly described as ``cold clumps", intermediate
in their structure and physical scale between a true pre-stellar core
and a molecular cloud.  This is of course to be expected as the \Planck\
effective beam dictates a preferred angular scale for ECC detection, and the selection process
places their emission peak in the submm range.

In order to detect the cold cores, a warm background determined by the scaled
{\it IRAS} 100\micron\ emission is subtracted from the \Planck\ maps at 217, 353 and 545\,GHz \citep{Montier2010, planck2011-7.7b}.
The scaling factor is determined by measuring the sky background in a
disk of 15\,arcmin outer radius.
We search for the presence of a source in the residual emission, and perform photometry at the location of detected sources.
The band-merging process positionally matches objects in the 353\,GHz detection list,
which contains the least number of entries, 
against both the 545 and 857\,GHz catalogues using a 5$\arcmin$ matching radius.
Sources only detected in one or two bands are discarded. The SNR and position 
of the detection having the greatest SNR are assigned to the band-merged entry.

Aperture photometry is performed on the {\it IRAS} 100\micron\ and 353, 545, and
857\,GHz maps using a source radius of five arcminute and a background annulus
spanning radii from five to ten arcminutes. An unconstrained three-parameter (T, $\beta$, and 
S$_{857}$) greybody is fit to the four-band aperture photometry with
the fitted temperatures used in the source selection process.

\begin{figure}
\centering
\resizebox{\hsize}{!}{\includegraphics[width=15cm]{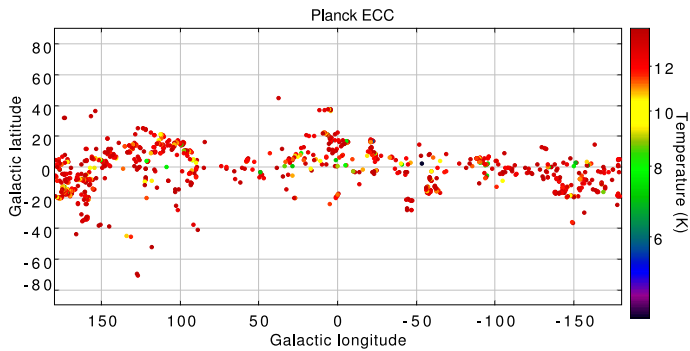}}
\caption[All-sky distribution of ECC sources]
{Sky distribution of ECC detections having SNR $\ge$ 15 and T $\le$ 14\,K is shown.
The symbols are colour-coded by temperature using the scale shown on the right.
\label{fig:ecc-spatial}}
\end{figure}

As for the ERCSC, a Monte-Carlo process is used to define signal-to-noise thresholds where
the derived temperatures are consistent with being $<$14\,K. A full description of the process
can be found both in \citet{planck2011-1.10sup}. 
The delivered ECC catalogue consists of 915 objects (Figure~\ref{fig:ecc-spatial}) meeting the
ECC selection criteria of SNR $\ge$ 15 and T $\le$ 14\,K., after
removal of selected sources having obviously discrepant SEDs or are
closely positionally matched to bright AGN, e.g., 3C~273. The columns in the ECC catalogue are shown in
Table \ref{tab:ecccol}.

It should be noted that the derived
temperatures of the Cold Cores are degenerate with the derived emissivity due to the absence of more
than one flux-density estimate at wavelengths shortward of the peak in the blackbody spectrum.
This issue is discussed in detail in \citet{planck2011-7.7b}.

Further information on the ECC is given in \citet{planck2011-1.10sup, planck2011-7.7b}.
Additional remarks on individual sources, including cross-matches with ERCSC sources are provided in the notes file that accompany
the individual catalogs (see Sect.~\ref{sec:ERCSC_access}). 

\begin{table*}
\centering
\caption{ECC Catalogue Columns}
\label{tab:ecccol}
\begin{tabular}{l l}
\hline\hline
Keyword & Type \\
\hline
NAME & Source name \\
SNR & Signal to Noise ratio of detection \\
\\
GLON &        Galactic longitude (deg) based on bandmerge algorithm \\
GLAT   &      Galactic latitude (deg) based on bandmerge algorithm \\
RA & Right Ascension in degrees (J2000) \\
DEC & Declination in degrees (J2000) \\
\\
APFLUX353 &        Aperture flux density at 353\,GHz (mJy) \\
APFLUX545 &        Aperture flux density at 545\,GHz (mJy)\\
APFLUX857 &        Aperture flux density at 857\,GHz (mJy )\\
APFLUX3000 &        Aperture flux density at 3000\,GHz (mJy)\\
APFLUX353\_ERR &   Uncertainty (1 sigma) in APFLUX353 \\
APFLUX545\_ERR &    Uncertainty (1 sigma) in APFLUX545  \\
APFLUX857\_ERR &    Uncertainty (1 sigma) in APFLUX857  \\
APFLUX3000\_ERR &    Uncertainty (1 sigma) in APFLUX3000 \\
\\
TEMPERATURE       &        Temperature from greybody fit (K) \\
BETA    &        Emissivity index from greybody fit\\
S857 &     Flux density at 857\,GHz from greybody fit (mJy) \\
TEMPERATURE\_ERR   &       Uncertainty (1 sigma) in TEMPERATURE (K)\\
BETA\_ERR &      Uncertainty (1 sigma) in BETA \\
S857\_ERR & Uncertainty (1 sigma) in S857 \\
BESTNORM & Summed squared residuals for best fit (mJy$^2$) \\
\\
TEMPERATURE\_CORE       &   Core Temperature from greybody fit to cold residual emission (K) \\
BETA\_CORE    &        Emissivity index from greybody fit to cold residual emission \\
MAJ\_AXIS\_FWHM\_CORE &  Ellipse major axis of cold residual emission (arcmin) \\
MIN\_AXIS\_FWHM\_CORE &  Ellipse minor axis of cold residual emission (arcmin) \\
TEMPERATURE\_CORE\_ERR       &    Uncertainty (1 sigma) TEMPERATURE\_CORE (K) \\
BETA\_CORE\_ERR    &       Uncertainty (1 sigma) BETA\_CORE  \\
MAJ\_AXIS\_FWHM\_CORE\_ERR &  Uncertainty (1 sigma) MAJ\_AXIS\_FWHM\_CORE \\
MIN\_AXIS\_FWHM\_CORE\_ERR  & Uncertainty (1 sigma) MIN\_AXIS\_FWHM\_CORE \\
\hline
\end{tabular}
\end{table*}

\section{Concluding Remarks}

\Planck\ is the third generation space based CMB experiment with more than an order of magnitude higher spatial resolution 
than {\it COBE} and with a broader range of frequency coverage than {\it WMAP}. The completion of the first sky survey in April 2010
yields a unique opportunity to study the classes of astrophysical sources that are foreground contributors to the CMB.
The ERCSC is a catalogue with $>90$\% reliability and is based on 1.6 sky coverages by \Planck. It has been produced with a very
rapid turnaround time to facilitate follow up observations with existing and future telescope facilities. 
The \Planck\ Collaboration expects that the diversity 
of sources present in the ERCSC, ranging from protostellar cores to SZ selected clusters, radio galaxies and
luminous star-forming galaxies, will provide a rich opportunity for follow-up studies of interesting astrophysical phenomena.

\begin{acknowledgements}

The production of the Planck Early Release Compact Source Catalogue was funded by NASA and carried out at the U.S. Planck Data Center at the Infrared Processing and Analysis Center (IPAC), California Institute of Technology, on behalf of and in collaboration with 
the LFI and HFI Data Processing Centers and with many contributions by members of the \Planck\ Collaboration.
The \Planck\ Collaboration acknowledges the support of: ESA; CNES and CNRS/INSU-IN2P3-INP (France); ASI, CNR, and INAF (Italy); NASA and DoE (USA); STFC and UKSA (UK); CSIC, MICINN and JA (Spain); Tekes, AoF and CSC (Finland); DLR and MPG (Germany); CSA (Canada); DTU Space (Denmark); SER/SSO (Switzerland); RCN (Norway); SFI (Ireland); FCT/MCTES (Portugal); and DEISA (EU).
A description of the \Planck\ Collaboration and a list of its members with the technical or scientific activities they have been involved into, can be found at
\url{http://www.rssd.esa.int/index.php?project=PLANCK&page=Planck_Collaboration}

\end{acknowledgements}


\begin{thebibliography}{38}
\expandafter\ifx\csname natexlab\endcsname\relax\def\natexlab#1{#1}\fi

\bibitem[Arnaud et al.(2010)]{arnaud:2010} Arnaud, M., Pratt, G.~W., Piffaretti, R., B{\"o}hringer, H., Crost
on, J.~H., \& Pointecouteau, E.\ 2010, \aap, 517, A92

\bibitem[Aatrokoski et al.(2010)]{Aatrokoski2010} Aatrokoski, J., 
L{\"a}hteenm{\"a}ki, A., Tornikoski, M., Valtaoja, E., Maino, D., Galeotta, 
S., Zacchei, A., \& Pasian, F.\ 2010, \mnras, 401, 597

\bibitem[Beichman et al.(1988)]{beichman} Beichman, C.~A.,
Neugebauer, G., Habing, H.~J., Clegg, P.~E., 
\& Chester, T.~J.\ 1988, Infrared astronomical satellite (IRAS) catalogs and atlases.~Volume 1: Explanatory s
upplement, 1

\bibitem[Bennett et al.(2003)]{Bennett2003} Bennett, C.~L., et al.\ 
2003, \apjs, 148, 1 

\bibitem[{{Bersanelli} {et~al.}(2010){Bersanelli}, {Mandolesi}, {Butler},
  {Mennella}, {Villa}, {Aja}, {Artal}, {Artina}, {Baccigalupi}, {Balasini},
  {Baldan}, {Banday}, {Bastia}, {Battaglia}, {Bernardino}, {Blackhurst},
  {Boschini}, {Burigana}, {Cafagna}, {Cappellini}, {Cavaliere}, {Colombo},
  {Crone}, {Cuttaia}, {D'Arcangelo}, {Danese}, {Davies}, {Davis}, {de Angelis},
  {de Gasperis}, {de La Fuente}, {de Rosa}, {de Zotti}, {Falvella}, {Ferrari},
  {Ferretti}, {Figini}, {Fogliani}, {Franceschet}, {Franceschi}, {Gaier},
  {Garavaglia}, {Gomez}, {Gorski}, {Gregorio}, {Guzzi}, {Herreros},
  {Hildebrandt}, {Hoyland}, {Hughes}, {Janssen}, {Jukkala}, {Kettle},
  {Kilpi{\"a}}, {Laaninen}, {Lapolla}, {Lawrence}, {Lawson}, {Leahy},
  {Leonardi}, {Leutenegger}, {Levin}, {Lilje}, {Lowe}, {Lubin}, {Maino},
  {Malaspina}, {Maris}, {Marti-Canales}, {Martinez-Gonzalez}, {Mediavilla},
  {Meinhold}, {Miccolis}, {Morgante}, {Natoli}, {Nesti}, {Pagan}, {Paine},
  {Partridge}, {Pascual}, {Pasian}, {Pearson}, {Pecora}, {Perrotta},
  {Platania}, {Pospieszalski}, {Poutanen}, {Prina}, {Rebolo}, {Roddis},
  {Rubi{\~n}o-Martin}, {Salmon}, {Sandri}, {Seiffert}, {Silvestri},
  {Simonetto}, {Sjoman}, {Smoot}, {Sozzi}, {Stringhetti}, {Taddei}, {Tauber},
  {Terenzi}, {Tomasi}, {Tuovinen}, {Valenziano}, {Varis}, {Vittorio}, {Wade},
  {Wilkinson}, {Winder}, {Zacchei}, \& {Zonca}}]{Bersanelli2010}
{Bersanelli}, M., {Mandolesi}, N., {Butler}, R.~C., {et~al.} 2010, \aap, 520,
  A4+

\bibitem[Bertin \& Arnouts(1996)]{ber96}
Bertin, E., \& Arnouts, S. 1996, \aaps, 117, 393

\bibitem[Carlstrom {et~al.}(2002)]{carl02} 
Carlstrom, J., {et~al.}, 2002, ARA\&A, 40, 643

\bibitem[Carvalho et al.(2009)]{car09} Carvalho, P., Rocha,
G., \& Hobson, M.~P.\ 2009, \mnras, 393, 681

\bibitem[Chary {et~al.}(2004)]{cha}
Chary, R., {et~al.}, 2004, \apj, 154, 80

\bibitem[{{Dupac} \& {Tauber}(2005)}]{dupac2005}
{Dupac}, X. \& {Tauber}, J. 2005, \aap, 430, 363

\bibitem[{{Gold} {et~al.}(2011){Gold}, {Odegard}, {Weiland}, {Hill}, {Kogut},
  {Bennett}, {Hinshaw}, {Chen}, {Dunkley}, {Halpern}, {Jarosik}, {Komatsu},
  {Larson}, {Limon}, {Meyer}, {Nolta}, {Page}, {Smith}, {Spergel}, {Tucker},
  {Wollack}, \& {Wright}}]{gold2010}
{Gold}, B., {Odegard}, N., {Weiland}, J.~L., {et~al.} 2011, \apjs, 192, 15

\bibitem[G\'{o}rski {et~al.}(2005)]{gor05}
G\'{o}rski, K.~M., Hivon, E., Banday, A.~J., Wandelt, B.~D., Hansen, F.~K., Reinecke, M., \& Bartelmann, M.
2005, \apj, 622, 759

\bibitem[Gregory {et~al.}(1996)]{greg} 
Gregory, P.~C., Scott, W.~K., Douglas, K., \& Condon, J.~J.\ 1996, \apjs, 103, 427

\bibitem[Griffith et al.(1995)]{griffith} Griffith, M.~R., 
Wright, A.~E., Burke, B.~F., \& Ekers, R.~D.\ 1995, \apjs, 97, 347

\bibitem[Healey {et~al.}(2007)]{hea}
Healey, S.~E., Romani, R.~W., Taylor, G.~B., Sadler, E.~M., Ricci, R., Murphy, T., Ulvestad, J.~S., \& Winn, 
J.~N.\ 2007, \apjs, 171, 61

\bibitem[Hobson \& McLachlan(2003)]{hob03} Hobson, M.~P., \&
McLachlan, C.\ 2003, \mnras, 338, 765

\bibitem[Kron(1980)]{kron1980} Kron, R.~G.\ 1980, \apjs, 43, 305

\bibitem[{{Lamarre} {et~al.}(2010){Lamarre}, {Puget}, {Ade}, {Bouchet},
  {Guyot}, {Lange}, {Pajot}, {Arondel}, {Benabed}, {Beney}, {Beno{\^i}t},
  {Bernard}, {Bhatia}, {Blanc}, {Bock}, {Br\'{e}elle}, {Bradshaw}, {Camus},
  {Catalano}, {Charra}, {Charra}, {Church}, {Couchot}, {Coulais}, {Crill},
  {Crook}, {Dassas}, {de Bernardis}, {Delabrouille}, {de Marcillac}, {Delouis},
  {D\'{e}sert}, {Dumesnil}, {Dupac}, {Efstathiou}, {Eng}, {Evesque},
  {Fourmond}, {Ganga}, {Giard}, {Gispert}, {Guglielmi}, {Haissinski},
  {Henrot-Versill\'{e}}, {Hivon}, {Holmes}, {Jones}, {Koch}, {Lagard{\`e}re},
  {Lami}, {Land\'{e}}, {Leriche}, {Leroy}, {Longval},
  {Mac{\'{\i}}as-P\'{e}rez}, {Maciaszek}, {Maffei}, {Mansoux}, {Marty}, {Masi},
  {Mercier}, {Miville-Desch{\^e}nes}, {Moneti}, {Montier}, {Murphy},
  {Narbonne}, {Nexon}, {Paine}, {Pahn}, {Perdereau}, {Piacentini}, {Piat},
  {Plaszczynski}, {Pointecouteau}, {Pons}, {Ponthieu}, {Prunet}, {Rambaud},
  {Recouvreur}, {Renault}, {Ristorcelli}, {Rosset}, {Santos}, {Savini},
  {Serra}, {Stassi}, {Sudiwala}, {Sygnet}, {Tauber}, {Torre}, {Tristram},
  {Vibert}, {Woodcraft}, {Yurchenko}, \& {Yvon}}]{Lamarre2010}
{Lamarre}, J., {Puget}, J., {Ade}, P.~A.~R., {et~al.} 2010, \aap, 520, A9+

\bibitem[{{Leahy} {et~al.}(2010){Leahy}, {Bersanelli}, {D'Arcangelo}, {Ganga},
  {Leach}, {Moss}, {Keih{\"a}nen}, {Keskitalo}, {Kurki-Suonio}, {Poutanen},
  {Sandri}, {Scott}, {Tauber}, {Valenziano}, {Villa}, {Wilkinson}, {Zonca},
  {Baccigalupi}, {Borrill}, {Butler}, {Cuttaia}, {Davis}, {Frailis},
  {Francheschi}, {Galeotta}, {Gregorio}, {Leonardi}, {Mandolesi}, {Maris},
  {Meinhold}, {Mendes}, {Mennella}, {Morgante}, {Prezeau}, {Rocha},
  {Stringhetti}, {Terenzi}, \& {Tomasi}}]{Leahy2010}
{Leahy}, J.~P., {Bersanelli}, M., {D'Arcangelo}, O., {et~al.} 2010, \aap, 520,
  A8+

\bibitem[L\'{o}pez-Caniego {et~al.}(2007)]{lop07}
L\'{o}pez-Caniego, M., Gonz\'{a}lez-Nuevo, J., Herranz, D., Massardi, M.,
Sanz, J.~L., De Zotti, G., Toffolatti, L., \& Arg{\"u}eso, F. 007, \apjs,
170, 108

\bibitem[{{Mandolesi} {et~al.}(2010){Mandolesi}, {Bersanelli}, {Butler},
  {Artal}, {Baccigalupi}, {Balbi}, {Banday}, {Barreiro}, {Bartelmann},
  {Bennett}, {Bhandari}, {Bonaldi}, {Borrill}, {Bremer}, {Burigana}, {Bowman},
  {Cabella}, {Cantalupo}, {Cappellini}, {Courvoisier}, {Crone}, {Cuttaia},
  {Danese}, {D'Arcangelo}, {Davies}, {Davis}, {de Angelis}, {de Gasperis}, {de
  Rosa}, {de Troia}, {de Zotti}, {Dick}, {Dickinson}, {Diego}, {Donzelli},
  {D{\"o}rl}, {Dupac}, {En{\ss}lin}, {Eriksen}, {Falvella}, {Finelli},
  {Frailis}, {Franceschi}, {Gaier}, {Galeotta}, {Gasparo}, {Giardino}, {Gomez},
  {Gonzalez-Nuevo}, {G\'{o}rski}, {Gregorio}, {Gruppuso}, {Hansen}, {Hell},
  {Herranz}, {Herreros}, {Hildebrandt}, {Hovest}, {Hoyland}, {Huffenberger},
  {Janssen}, {Jaffe}, {Keih{\"a}nen}, {Keskitalo}, {Kisner}, {Kurki-Suonio},
  {L{\"a}hteenm{\"a}ki}, {Lawrence}, {Leach}, {Leahy}, {Leonardi}, {Levin},
  {Lilje}, {L\'{o}pez-Caniego}, {Lowe}, {Lubin}, {Maino}, {Malaspina}, {Maris},
  {Marti-Canales}, {Martinez-Gonzalez}, {Massardi}, {Matarrese}, {Matthai},
  {Meinhold}, {Melchiorri}, {Mendes}, {Mennella}, {Morgante}, {Morigi},
  {Morisset}, {Moss}, {Nash}, {Natoli}, {Nesti}, {Paine}, {Partridge},
  {Pasian}, {Passvogel}, {Pearson}, {P\'{e}rez-Cuevas}, {Perrotta}, {Polenta},
  {Popa}, {Poutanen}, {Prezeau}, {Prina}, {Rachen}, {Rebolo}, {Reinecke},
  {Ricciardi}, {Riller}, {Rocha}, {Roddis}, {Rohlfs}, {Rubi{\~n}o-Martin},
  {Salerno}, {Sandri}, {Scott}, {Seiffert}, {Silk}, {Simonetto}, {Smoot},
  {Sozzi}, {Sternberg}, {Stivoli}, {Stringhetti}, {Tauber}, {Terenzi},
  {Tomasi}, {Tuovinen}, {T{\"u}rler}, {Valenziano}, {Varis}, {Vielva}, {Villa},
  {Vittorio}, {Wade}, {White}, {White}, {Wilkinson}, {Zacchei}, \&
  {Zonca}}]{Mandolesi2010}
{Mandolesi}, N., {Bersanelli}, M., {Butler}, R.~C., {et~al.} 2010, \aap, 520,
  A3+

\bibitem[Massardi et al.(2009)]{massardi} Massardi, M., 
L\'{o}pez-Caniego, M., Gonz\'{a}lez-Nuevo, J., Herranz, D., de Zotti, G.,
\& Sanz, J.~L.\ 2009, \mnras, 392, 733

\bibitem[{{Melin} {et~al.}(2006){Melin}, {Bartlett}, \&
  {Delabrouille}}]{melin2006}
{Melin}, J., {Bartlett}, J.~G., \& {Delabrouille}, J. 2006, \aap, 459, 341

\bibitem[Melin {et~al.}(2010)]{melin:2010}
Melin, J.-B., et al.\ 2010, \aap, submitted

\bibitem[{{Mennella et al.}(2011)}]{planck2011-1.4}
{Mennella et al.} 2011, {Planck early results 03: First assessment of the Low
  Frequency Instrument in-flight performance} ({Submitted to \aap,
  [arXiv:astro-ph/1101.2038]})

\bibitem[{{Mitra} {et~al.}(2011){Mitra}, {Rocha}, {G\'{o}rski}, {Huffenberger},
  {Eriksen}, {Ashdown}, \& {Lawrence}}]{Mitra2010}
{Mitra}, S., {Rocha}, G., {G\'{o}rski}, K.~M., {et~al.} 2011, \apjs, 193, 5

\bibitem[{{Montier} {et~al.}(2010){Montier}, {Pelkonen}, {Juvela},
  {Ristorcelli}, \& {Marshall}}]{Montier2010}
{Montier}, L.~A., {Pelkonen}, V.-M., {Juvela}, M., {Ristorcelli}, I., \&
  {Marshall}, D.~J. 2010, \aap, 522, A83+

\bibitem[Murakami et al.(2007)]{Murakami2007} Murakami, H., et al.\ 
2007, \pasj, 59, 369 

\bibitem[Murphy {et~al.}(2010)]{mur}
Murphy, T., {et~al.} 2010, \mnras, 402, 2403

\bibitem[Piffaretti {et al.}(2010)]{Piff2010} Piffaretti, R., Arnaud, M., Pratt, G.~W., 
Pointecouteau, E.,
\& Melin, J.-B.\ 2010, \aap, submitted, arXiv:1007.1916

\bibitem[{{Planck Collaboration}(2011{\natexlab{a}})}]{planck2011-1.1}
{Planck Collaboration}. 2011{\natexlab{a}}, {Planck early results 01: The
  Planck mission} ({Submitted to \aap, [arXiv:astro-ph/1101.2022]})

\bibitem[{{Planck Collaboration}(2011{\natexlab{b}})}]{planck2011-1.3}
{Planck Collaboration}. 2011{\natexlab{b}}, {Planck early results 02: The
  thermal performance of Planck} ({Submitted to \aap,
  [arXiv:astro-ph/1101.2023]})

\bibitem[{{Planck Collaboration}(2011{\natexlab{c}})}]{planck2011-1.10}
{Planck Collaboration}. 2011{\natexlab{c}}, {Planck early results 07: The Early
  Release Compact Source Catalogue} ({Submitted to \aap,
  [arXiv:astro-ph/1101.2041]})

\bibitem[{{Planck Collaboration}(2011{\natexlab{d}})}]{planck2011-5.1a}
{Planck Collaboration}. 2011{\natexlab{d}}, {Planck early results 08: The
  all-sky early Sunyaev-Zeldovich cluster sample} ({Submitted to \aap,
  [arXiv:astro-ph/1101.2024]})

\bibitem[{{Planck Collaboration}(2011{\natexlab{e}})}]{planck2011-5.1b}
{Planck Collaboration}. 2011{\natexlab{e}}, {Planck early results 09:
  XMM-Newton follow-up for validation of Planck cluster candidates} ({Submitted
  to \aap, [arXiv:astro-ph/1101.2025]})

\bibitem[{{Planck Collaboration}(2011{\natexlab{f}})}]{planck2011-5.2a}
{Planck Collaboration}. 2011{\natexlab{f}}, {Planck early results 10:
  Statistical analysis of Sunyaev-Zeldovich scaling relations for X-ray galaxy
  clusters} ({Submitted to \aap, [arXiv:astro-ph/1101.2043]})

\bibitem[{{Planck Collaboration}(2011{\natexlab{g}})}]{planck2011-5.2b}
{Planck Collaboration}. 2011{\natexlab{g}}, {Planck early results 11:
  Calibration of the local galaxy cluster Sunyaev-Zeldovich scaling relations}
  ({Submitted to \aap, [arXiv:astro-ph/1101.2026]})

\bibitem[{{Planck Collaboration}(2011{\natexlab{h}})}]{planck2011-5.2c}
{Planck Collaboration}. 2011{\natexlab{h}}, {Planck early results 12: Cluster
  Sunyaev-Zeldovich optical Scaling relations} ({Submitted to \aap,
  [arXiv:astro-ph/1101.2027]})

\bibitem[{{Planck Collaboration}(2011{\natexlab{i}})}]{planck2011-6.1}
{Planck Collaboration}. 2011{\natexlab{i}}, {Planck early results 13:
  Statistical properties of extragalactic radio sources in the Planck Early
  Release Compact Source Catalogue} ({Submitted to \aap,
  [arXiv:astro-ph/1101.2044]})

\bibitem[{{Planck Collaboration}(2011{\natexlab{j}})}]{planck2011-6.2}
{Planck Collaboration}. 2011{\natexlab{j}}, {Planck early results 14: Early
  Release Compact Source Catalogue validation and extreme radio sources}
  ({Submitted to \aap, [arXiv:astro-ph/1101.1721]})

\bibitem[{{Planck Collaboration}(2011{\natexlab{k}})}]{planck2011-6.3a}
{Planck Collaboration}. 2011{\natexlab{k}}, {Planck early results 15: Spectral
  energy distributions and radio continuum spectra of northern extragalactic
  radio sources} ({Submitted to \aap, [arXiv:astro-ph/1101.2047]})

\bibitem[{{Planck Collaboration}(2011{\natexlab{l}})}]{planck2011-6.4a}
{Planck Collaboration}. 2011{\natexlab{l}}, {Planck early results 16: The
  Planck view of nearby galaxies} ({Submitted to \aap,
  [arXiv:astro-ph/1101.2045]})

\bibitem[{{Planck Collaboration}(2011{\natexlab{m}})}]{planck2011-6.4b}
{Planck Collaboration}. 2011{\natexlab{m}}, {Planck early results 17: Origin of
  the submillimetre excess dust emission in the Magellanic Clouds} ({Submitted
  to \aap, [arXiv:astro-ph/1101.2046]})

\bibitem[{{Planck Collaboration}(2011{\natexlab{n}})}]{planck2011-6.6}
{Planck Collaboration}. 2011{\natexlab{n}}, {Planck early results 18: The power
  spectrum of cosmic infrared background anisotropies} ({Submitted to \aap,
  [arXiv:astro-ph/1101.2028]})

\bibitem[{{Planck Collaboration}(2011{\natexlab{o}})}]{planck2011-7.0}
{Planck Collaboration}. 2011{\natexlab{o}}, {Planck early results 19: All-sky
  temperature and dust optical depth from Planck and IRAS --- constraints on
  the ``dark gas" in our Galaxy} ({Submitted to \aap,
  [arXiv:astro-ph/1101.2029]})

\bibitem[{{Planck Collaboration}(2011{\natexlab{p}})}]{planck2011-7.2}
{Planck Collaboration}. 2011{\natexlab{p}}, {Planck early results 20: New light
  on anomalous microwave emission from spinning dust grains} ({Submitted to
  \aap, [arXiv:astro-ph/1101.2031]})

\bibitem[{{Planck Collaboration}(2011{\natexlab{q}})}]{planck2011-7.3}
{Planck Collaboration}. 2011{\natexlab{q}}, {Planck early results 21:
  Properties of the interstellar medium in the Galactic plane} ({Submitted to
  \aap, [arXiv:astro-ph/1101.2032]})

\bibitem[{{Planck Collaboration}(2011{\natexlab{r}})}]{planck2011-7.7a}
{Planck Collaboration}. 2011{\natexlab{r}}, {Planck early results 22: The
  submillimetre properties of a sample of Galactic cold clumps} ({Submitted to
  \aap, [arXiv:astro-ph/1101.2034]})

\bibitem[{{Planck Collaboration}(2011{\natexlab{s}})}]{planck2011-7.7b}
{Planck Collaboration}. 2011{\natexlab{s}}, {Planck early results 23: The
  Galactic cold core population revealed by the first all-sky survey}
  ({Submitted to \aap, [arXiv:astro-ph/1101.2035]})

\bibitem[{{Planck Collaboration}(2011{\natexlab{t}})}]{planck2011-7.12}
{Planck Collaboration}. 2011{\natexlab{t}}, {Planck early results 24: Dust in
  the diffuse interstellar medium and the Galactic halo} ({Submitted to \aap,
  [arXiv:astro-ph/1101.2036]})

\bibitem[{{Planck Collaboration}(2011{\natexlab{u}})}]{planck2011-7.13}
{Planck Collaboration}. 2011{\natexlab{u}}, {Planck early results 25: Thermal
  dust in nearby molecular clouds} ({Submitted to \aap,
  [arXiv:astro-ph/1101.2037]})

\bibitem[{{Planck Collaboration}(2011{\natexlab{v}})}]{planck2011-1.10sup}
{Planck Collaboration}. 2011{\natexlab{v}}, {The Explanatory Supplement to the
  Planck Early Release Compact Source Catalogue} ({ESA})

\bibitem[{{Planck HFI Core Team}(2011{\natexlab{a}})}]{planck2011-1.5}
{Planck HFI Core Team}. 2011{\natexlab{a}}, {Planck early results 04: First
  assessment of the High Frequency Instrument in-flight performance}
  ({Submitted to \aap, [arXiv:astro-ph/1101.2039]})

\bibitem[{{Planck HFI Core Team}(2011{\natexlab{b}})}]{planck2011-1.7}
{Planck HFI Core Team}. 2011{\natexlab{b}}, {Planck early results 06: The High
  Frequency Instrument data processing} ({Submitted to \aap,
  [arXiv:astro-ph/1101.2048]})

\bibitem[{{Rosset} {et~al.}(2010){Rosset}, {Tristram}, {Ponthieu}, {Ade},
  {Aumont}, {Catalano}, {Conversi}, {Couchot}, {Crill}, {D\'{e}sert}, {Ganga},
  {Giard}, {Giraud-H\'{e}raud}, {Ha{\"i}ssinski}, {Henrot-Versill\'{e}},
  {Holmes}, {Jones}, {Lamarre}, {Lange}, {Leroy}, {Mac{\'{\i}}as-P\'{e}rez},
  {Maffei}, {de Marcillac}, {Miville-Desch{\^e}nes}, {Montier}, {Noviello},
  {Pajot}, {Perdereau}, {Piacentini}, {Piat}, {Plaszczynski}, {Pointecouteau},
  {Puget}, {Ristorcelli}, {Savini}, {Sudiwala}, {Veneziani}, \&
  {Yvon}}]{Rosset2010}
{Rosset}, C., {Tristram}, M., {Ponthieu}, N., {et~al.} 2010, \aap, 520, A13+

\bibitem[Savage
\& Oliver(2007)]{savage07} Savage, R.~S., \& Oliver, S.\ 2007, \apj, 661, 1339

\bibitem[Sunyaev \& Zel'dovich(1972)]{sz72} 
Sunyaev, R. \& Zeldovich, Ya. (1972), A\&A, 20, 189

\bibitem[{{Tauber} {et~al.}(2010){Tauber}, {Mandolesi}, {Puget}, {Banos},
  {Bersanelli}, {Bouchet}, {Butler}, {Charra}, {Crone}, {Dodsworth}, \&
  et~al.}]{tauber2010a}
{Tauber}, J.~A., {Mandolesi}, N., {Puget}, J., {et~al.} 2010, \aap, 520, A1+

\bibitem[Vieira et al.(2010)]{Vieira2010} Vieira, J.~D., et al.\
2010, \apj, 719, 763

\bibitem[Ward-Thompson et al.(2002)]{WardThompson} Ward-Thompson,
D., Andr\'{e}, P., \& Kirk, J.~M.\ 2002, \mnras, 329, 257

\bibitem[{{Wright} {et~al.}(2009){Wright}, {Chen}, {Odegard}, {Bennett},
  {Hill}, {Hinshaw}, {Jarosik}, {Komatsu}, {Nolta}, {Page}, {Spergel},
  {Weiland}, {Wollack}, {Dunkley}, {Gold}, {Halpern}, {Kogut}, {Larson},
  {Limon}, {Meyer}, \& {Tucker}}]{wright2009}
{Wright}, E.~L., {Chen}, X., {Odegard}, N., {et~al.} 2009, \apjs, 180, 283

\bibitem[Wright et al.(2010)]{Wright2010} Wright, E.~L., et al.\ 
2010, \aj, 140, 1868 

\bibitem[{{Zacchei et al.}(2011)}]{planck2011-1.6}
{Zacchei et al.} 2011, {Planck early results 05: The Low Frequency Instrument
  data processing} ({Submitted to \aap, [arXiv:astro-ph/1101.2040]})

\end{thebibliography}
\end{document}